\documentclass[11pt,reqno]{amsproc}    
\usepackage[german,english]{babel}  
\usepackage[latin1]{inputenc} 
\usepackage{graphicx}
\usepackage{amssymb}
\usepackage{times} 
\numberwithin{equation}{section}
\newtheorem{theorem}{Theorem}[section]
\newtheorem{proposition}[theorem]{Proposition}
\newtheorem{lemma}[theorem]{Lemma}
\newtheorem{corollary}[theorem]{Corollary}
\newtheorem{definition}[theorem]{Definition}
\newtheorem{result}[theorem]{Result}
\theoremstyle{definition}
\newtheorem{remark}[theorem]{Remark}
\newtheorem{remarks}[theorem]{Remarks}

\newtheoremstyle{break}
  {}
  {}
  {}
  {}
  {}
  {}
  {\newline}
  {}

\theoremstyle{break}

\def\supp{\mathop{\hbox{supp}}}
\def\tr{\mathop{\textnormal{tr}}}
\newcommand{\R}{\mathbb{R}}
\newcommand{\Rd}{\R^d}
\newcommand{\C}{\mathbb{C}}
\newcommand{\Z}{\mathbb{Z}}
\newcommand{\Q}{\mathbb{Q}}
\renewcommand{\P}{\mathbb{P}}
\newcommand{\E}{\mathbb{E}}
\newcommand{\N}{\mathbb{N}}

\newcommand{\F}{\mathcal{F}}
\renewcommand{\H}{\mathcal{H}}

\newcommand{\Zd}{\mathbb{Z}^{d}}
\newcommand{\LL}{\Lambda_L}
\newcommand{\Ll}{\Lambda_l}

\newcommand{\eh}{\frac{1}{2}}

\newcommand{\ua}[1]{\subsection{#1} \quad\par\vspace{0.25cm}}
\newcommand{\ut}{e^{-itH}}
\newcommand{\utr}{e^{-itH_\omega}}
\newcommand{\sn}[1]{||\,#1||_\infty}
\newcommand{\no}[1]{||\,#1||}
\newcommand{\nor}[1]{|\,#1|}
\newcommand{\ns}[1]{||\,#1||_1}
\newcommand{\platz}{\quad\\ \vspace{-0.45cm}}
\newcommand{\oexists}{\operatorname*{\displaystyle\exists}}
\newcommand{\es}{\quad\par\vspace{0.25cm}}
\newcommand{\inorm}[1]{||\,#1||_\infty}
\newcommand{\norm}[1]{||\,#1||}
\newcommand{\gnorm}[1]{||\,#1||_{1}}
\renewcommand{\L}{\Lambda}
\renewenvironment{proof}{\par\vspace{0.2cm}{\scshape Proof:\quad}}{\hfill\qed\\}
\newenvironment{nproof}[1]{\par\vspace{0.2cm}{\scshape Proof}\; \textsl{(#1)}\,:\quad}{\hfill\qed\\}
\makeatletter
\def\@tocpagenum#1{\hss{\mdseries #1}}
\def\@tocwrite#1{\@xp\@tocwriteb\csname toc#1\endcsname{#1}}
\def\@tocwriteb#1#2#3{%
  \begingroup
    \def\@tocline##1##2##3##4##5##6{%
      \ifnum##1>\c@tocdepth
      \else \sbox\z@{##5\let\indentlabel\@tochangmeasure##6}\fi}%
    \csname l@#2\endcsname{#1{\csname#2name\endcsname}{\@secnumber}{}}%
  \endgroup
  \addcontentsline{toc}{#2}%
    {\protect#1{\csname#2name\endcsname}{\@secnumber}{#3}}}
\def\l@section{\@tocline{1}{0pt}{0.5pc}{}{}}
\renewcommand{\tocsection}[3]{%
  \indentlabel{\@ifnotempty{#2}{\ignorespaces#1 #2.\quad}}#3}
\def\l@subsection{\@tocline{2}{0pt}{2.5pc}{5pc}{}}

\def\l@subsubsection{\@tocline{3}{0pt}{1pc}{7pc}{}}

\def\l@part{\@tocline{-1}{12pt plus2pt}{0pt}{}{\bfseries}}

\def\l@chapter{\@tocline{0}{8pt plus1pt}{0pt}{}{}}

 \makeatother

\newcommand{\mindex}[1]{#1 \index{#1}}
\newcommand{\dindex}[1]{\emph{#1}\index{#1}}
\newcommand{\mathdex}[1]{#1 \index{$#1$}}
\newcommand{\tn}[1]{\textnormal{#1}}
\newcommand{\ve}{\varepsilon}
\newcommand{\NeueSeite}{\cleardoublepage}
 \makeindex

\begin{document}
\bibliographystyle{alpha}


\quad

\vspace{1cm}

{\Huge{\centerline{An Invitation to }}} \vspace{0.5cm}
{\Huge{\centerline{Random Schrödinger operators}}} \vspace{2cm}

{\LARGE{\centerline{Werner Kirsch}}} \vspace{0.3cm}
{\LARGE{\centerline{Institut für Mathematik}}}

{\LARGE{\centerline{Ruhr-Universität Bochum}}}

{\LARGE{\centerline{D-44780 Bochum, Germany}}} \vspace{1cm}
{\large{\centerline{email: werner.kirsch@rub.de}}} \vspace{2.5cm}
\centerline{\textbf{Abstract}}
\noindent\\[3mm]
This review is an extended version of my mini course at the États de
la recherche: Opérateurs de Schr\"odinger aléatoires at the
Université Paris 13 in June 2002, a summer school organized by
Frédéric
Klopp.\\[2mm]
These lecture notes try to give some of the basics of random
Schrödinger operators. They are meant for nonspecialists and require
only minor previous knowledge about functional analysis and
probability theory. Nevertheless this survey includes complete
proofs of Lifshitz tails and Anderson localization.

\vfill \centerline{Copyright by the author. Copying for academic
purposes is permitted.}

\NeueSeite
\setlength{\parskip}{0.2ex}
\tableofcontents 

\NeueSeite


\section{Preface}

In these lecture notes I try to give an introduction to (some part
of) the basic theory of random Schr\"odinger operators. I intend to
present the field in a rather self contained and elementary way. It
is my hope that the text will serve as an introduction to random
Schr\"odinger operators for students, graduate students and
researchers who have not studied this topic before. If some scholars
who are already acquainted with random Schr\"odinger operators might
find the
text useful as well I will be even more satisfied.\\[0.2cm]
Only a basic knowledge in Hilbert space theory and some basics from
probability theory are required to understand the text (see the
Notes below). I have restricted the considerations in this text
almost exclusively to the Anderson model, i.e. to random operators
on the Hilbert space $\ell^2(\Z^d)$. By doing so I tried to avoid
many of the technical difficulties that are necessary to deal with
in the continuous case (i.e. on $L^2(\R^d)$). Through such technical
problems sometimes the main ideas become obscured and less
transparent.

The theory I present is still not exactly easy staff. Following
Einstein's advice, I tried to make
things as easy as possible, but not easier.\\[1mm]

The author has to thank many persons. The number of colleagues and
friends I have learned from about mathematical physics and
especially disordered systems is so large that it is impossible to
mention a few without doing injustice to many others. A lot of the
names can be found as authors in the list of references. Without
these persons the writing of this review would have been impossible.

A colleague and friend I \emph{have} to mention though is Frédéric
Klopp who organized a summer school on Random Schr\"odinger
operators in Paris in 2002. My lectures there were the starting
point for this review. I have to thank Frédéric especially for his
enormous patience when I did not obey the third, forth, \ldots,
deadline for delivering the manuscript.

It is a great pleasure to thank Bernd Metzger for his advice, for
many helpful discussions, for proofreading the manuscript, for
helping me with the text and especially with the references and for
many other things.

Last, not least I would like to thank Jessica Langner, Riccardo
Catalano and Hendrik Meier for the skillful typing of the
manuscript, for proofreading and for
their patience with the author.\\[2mm]

{\bf Notes and Remarks }\\[2mm]
For the spectral theory needed in this work we recommend \cite{rs1}
or \cite{weid}. We will also need the min-max theorem (see
\cite{rs4}).

The probabilistic background we need can be found e.g. in
\cite{lamp1} and \cite{lamp2}.

For further reading on random Schr\"odinger operators we recommend
\cite{KleinSurv} for the state of the art in multiscale analysis. We
also recommend the textbook \cite{stoll1}. A modern survey on the
density of states is \cite{KirMetz}. \cleardoublepage


\section{Introduction: Why random Schrödinger operators ?}
\subsection{The setting of quantum mechanics}\es

A quantum mechanical particle moving in d-dimensional space is
described by a vector $\psi$ in the Hilbert space $L^2(\R^d)$. The
time evolution of the state $\psi$ is determined by the
\dindex{Schr\"odinger operator}

\begin{equation}
H=H_0+V
\end{equation}

acting on $L^2(\R^d)$. The operator $H_0$ is called the \dindex{free
operator}. It represents the kinetic energy of the particle. In the
absence of magnetic fields it is given by the Laplacian

\begin{equation}
H_0=-\frac{\hbar^2}{2m}\,\Delta=-\frac{\hbar^2}{2m}\,\sum_{\nu=1}^{d}\,\frac{\partial^2}{{\partial
x_\nu}^2} \ .
\end{equation}

The physics of the system is encoded in the potential $V$ which is
the multiplication operator with the function $V(x)$ in the Hilbert
space $L^2(\R^d)$. The function $V(x)$ is the (classical) potential
energy. Consequently, the forces are given by

$$F(x)
=-\nabla{V}(x)\ .$$

In the following we choose physical units in such a way that
$\frac{\hbar^2}{2m}=1$ since we are not interested in the explicit
dependence of quantities on $\hbar$ or $m$. The time evolution of
the state $\psi$ is obtained from the time dependent Schrödinger
equation

\begin{equation}\label{eq:3.0}
i\,\frac{\partial}{\partial t}\,\psi \;= \;H\,\psi \ .
\end{equation}

By the spectral theorem for self adjoint operators equation
(\ref{eq:3.0}) can be solved by

\begin{equation}\label{eq:3.1}
\psi(t)=e^{-itH}\psi_0
\end{equation}

where $\psi_0$ is the state of the system at time $t=0$.

To extract valuable information from (\ref{eq:3.1}) we have to know
as much as possible about the spectral theory of the operator $H$
and this is what we try to do in this text.

\subsection{Random Potentials}\es

In this review we are interested in random Schrödinger operators.
These operators model disordered solids. Solids occur in nature in
various forms. Sometimes they are (almost) totally ordered. In
crystals the atoms or nuclei are distributed on a periodic lattice
(say the lattice $\Z^d$ for simplicity) in a completely regular way.
Let us assume that a particle (electron) at the point $x\in\R^d$
feels a potential of the form $q\, f(x-i) $ due to an atom (or ion
or nucleus) located at the point $i\in\Z^d$. Here, the constant $q$,
the charge or coupling constant in physical terms, could be absorbed
into the function $f$. However, since we are going to vary this
quantity from atom to atom later on, it is useful to write the
potential in the above way. Then, in a regular crystal our particle
is exposed to a total potential

\begin{eqnarray}\label{eq:4.0}\label{alloy}
V(x)=\sum_{i\in\Z^d}\;q\,f(x-i) \ .
\end{eqnarray}

We call the function $f$ the \emph{\mindex{single site potential}}
to distinguish it from the total potential $V$. The potential $V$ in
(\ref{eq:4.0}) is periodic with respect to the lattice $\Z^d$, i.~e.
$V(x-i)=V(x)$ for all $x\in\R^d$ and $i\in\Z^d$. The mathematical
theory of Schrödinger operators with periodic potentials is well
developed (see e.g. \cite{eastham}, \cite{rs4} ). It is based on a
thorough analysis of the symmetry properties of periodic operators.
For example, it is known that such operators have a spectrum with
band structure, i.e. $\sigma(H)=\bigcup_{n=0}^\infty [a_n, b_n]$
with $a_n<b_n\le a_{n+1}$. This spectrum is also known to be
absolutely continuous.

Most solids do not constitute ideal crystals. The positions of the
atoms may deviate from the ideal lattice positions in a non regular
way due to imperfections in the crystallization process. Or the
positions of the atoms may be completely disordered as is the case
in amorphous or glassy materials. The solid may also be a mixture of
various materials which is the case for example for alloys or doped
semiconductors. In all these cases it seems reasonable to look upon
the potential as a random quantity.

For example, if the material is a pure one, but the positions of the
atoms deviate from the ideal lattice positions randomly, we may
consider a random potential of the form

\begin{equation}
V_\omega(x)\,=\;\sum_{i\in\Z^d}\,q\,f\left(x-i-\xi_i(\omega)\right).
\end{equation}

Here the $\xi_i$ are random variables which describe the deviation
of the `$i^{\textnormal{th}}$' atom from the lattice position $i$.
One may, for example assume that the random variables $\xi_i$ are
independent and identically distributed. We have added a subscript
$\omega$ to the potential $V$ to make clear that $V_\omega$ depends
on (unknown) random parameters.

To model an amorphous material like glass or rubber we assume that
the atoms of the material are located at completely random points
$\eta_i$ in space. Such a random potential may formally be written
as
\begin{equation}\label{eq:amorphous}
V_\omega(x)=\sum_{i\in\Z^d}\,q\,f(x-\eta_i).
\end{equation}

To write the potential (\ref{eq:amorphous}) as a sum over the
lattice $\Z^d$ is somewhat misleading, since there is, in general,
no natural association of the  $\eta_i$ with a lattice point $i$. It
is more appropriate to think of a collection of random points in
$\R^d$ as a random point measure. This representation emphasizes
that any ordering of the $\eta_i$ is completely artificial.

A \emph{\mindex{counting measure}} is a Borel measure on $\R^d$ of
the form $\nu=\sum_{x\in M}\,\delta_x$ with a countable set $M$
without (finite) accumulation points. By a \emph{\mindex{random
point measure}} we mean a mapping $\omega \mapsto \mu_\omega$, such
that $\mu_\omega$ is a counting measure with the property that the
function $\omega \mapsto \mu_\omega(A)$ is measurable for any
bounded Borel set $A$. If $\nu=\nu_\omega$ is the random point
measure $\nu=\sum_i\,\delta_{\eta_i}$ then (\ref{eq:amorphous}) can
be written as
\begin{eqnarray}\label{poisson}
V_\omega(x)=\int_{\R^d}\,q\,f(x-\eta)\;\,d\nu(\eta) \ .
\end{eqnarray}

The most frequently used example of a random point measure and the
most important one is the \emph{\mindex{Poisson random measure}}
$\mu_\omega$. Let us set $n_A = \mu_\omega(A)$, the number of random
points in the set A. The Poisson random measure can be characterized
by the following specifications
\begin{itemize}
\item The random variables $n_A$ and $n_B$ are independent for disjoint (measurable) sets $A$ and $B$.
\item The probability that $n_A = k$ is equal to $\frac{{|A|}^{k}}{k!}\;e^{-|A|}$, where $|A|$ is the Lebesgue
measure of $A$.
\end{itemize}

A random potential of the form (\ref{poisson}) with the Poisson
random measure is called the \emph{\mindex{Poisson model}}.

\vspace{0.1cm} The most popular model of a disordered solid and the
best understood one as well is the \emph{\mindex{alloy-type
potential}} (see (\ref{eq:alloy}) below). It models an unordered
alloy, i.e. a mixture of several materials the atoms of which are
located at lattice positions. The type of atom at the lattice point
$i$ is assumed to be random. In the model we consider here the
different materials are described by different charges (or coupling
constants) $q_i$. The total potential $V$ is then given by

\begin{equation}\label{eq:alloy}
V_\omega(x)=\sum_{i\in\Z^d}\;q_i(\omega)\,f(x-i) \ . \end{equation}

 The $q_i$ are random
variables which we assume to be independent and identically
distributed. Their range describes the possible values the coupling
constant can assume in the considered alloy. The physical model
suggests that there are only finitely many values the random
variables can assume. However, in the proofs of some results we have
to assume that the distribution of the random variables $q_i$ is
continuous (even absolutely continuous) due to limitations of the
mathematical techniques. One might argue that such an assumption is
acceptable as a purely technical one. On the other hand one could
say we have not understood the problem as long as we can not handle
the physically relevant cases.

For a given $\omega$ the potential $V_\omega(x)$ is a pretty
complicated `normal' function. So, one may ask: What is the
advantage of `making it random'?

With the introduction of random variables we implicitly change our
point of view. From now on we are hardly interested in  properties
of $H_\omega$ for a single given $\omega$. Rather, we look at
`typical' properties of $H_\omega$. In mathematical terms, we are
interested in results of the form: The set of all $\omega$ such that
$H_\omega$ has the property $\mathcal{P}$ has probability one. In
short: $\mathcal{P}$ holds for $\P$-almost all\index{$\P$-almost
all} $\omega$ (or $\P$-almost surely\index{$\P$-almost surely}).
Here $\P$ is the probability measure on the underlying probability
space.

In this course we will encounter a number of such properties. For
example we will see that (under weak assumptions on $V_\omega$),
there is a closed, nonrandom (!) subset $\Sigma$ of the real line
such that $\Sigma=\sigma(H_\omega)$, the spectrum of the operator
$H_\omega$, $\P$-almost surely.

This and many other results can be proven for various types of
random Schrödinger operators. In this lecture we will restrict
ourselves to a relatively simple system known as the Anderson model.
Here the Hilbert space is the sequence space $\ell^2(\Z^d)$ instead
of $L^2(\R^d)$ and the free operator $H_0$ is a finite-difference
operator rather than the Laplacian. We will call this setting the
\emph{\mindex{discrete case}} in contrast to Schrödinger operators
on $L^2(\R^d)$ which we refer to as the \emph{\mindex{continuous
case}}. In the references the reader may find papers which extend
results we prove here to the continuous setting.

\subsection{The one body approximation}\es

In the above setting we have implicitly assumed that we describe a
single particle moving in a static exterior potential. This is at
best a caricature of what we find in nature. First of all there are
many electrons moving in a solid and they interact with each other.
The exterior potential originates in nuclei or ions which are
themselves influenced both by the other nuclei and by the electrons.
In the above discussion we have also implicitly assumed that the
solid we consider extends to infinity in all directions, (i.e. fills
the universe). Consequently, we ought to consider infinitely many
interacting particles. It is obvious that such a task is out of
range of the methods available today. As a first approximation it
seems quite reasonable to separate the motion of the nuclei from the
system and to take the nuclei into account only via an exterior
potential. Indeed, the masses of the nuclei are much larger than
those of the electrons.

The second approximation is to neglect the electron-electron
interaction. It is \textsl{not at all clear} that this approximation
gives a qualitatively correct picture. In fact, there is physical
evidence that the interaction between the electrons is fundamental
for a number of phenomena.

Interacting particle systems in condensed matter are an object of
intensive research in theoretical physics. In mathematics, however,
this field of research is still in its infancy despite of an
increasing interest in the subject.

If we neglect the interactions between the electrons we are left
with a system of noninteracting electrons in an exterior potential.
It is not hard to see that such a system (and the corresponding
Hamiltonian) separates, i.e. the eigenvalues are just sums of the
one-body eigenvalues and the eigenfunctions have product form. So,
if $\psi_1, \psi_2, \ldots, \psi_N$ are eigenfunctions of the
one-body system corresponding to eigenvalues $E_1, E_2, \ldots, E_N$
respectively, then

\begin{equation}
\Psi(x_1, x_2, \ldots, x_N)=\psi_1(x_1) \cdot\psi_2(x_2)\cdot
\ldots\cdot \psi_N(x_N) \ . \label{eq:prodstate}
\end{equation}

is an eigenfunction of the full system with eigenvalue $E_1+E_2+
\ldots+E_N$.

However, there is a subtlety to obey here, which is typical to many
particle Quantum Mechanics. The electrons in the solid are
indistinguishable, since we are unable to `follow their
trajectories'. The corresponding Hamiltonian is invariant under
permutation of the particles. As a consequence, the $N$-particle
Hilbert space consists either of totally {\em symmetric} or of
totally {\em antisymmetric} functions of the particle positions
$x_1, x_2, \ldots, x_N$. It turns out that for particles with
integer spin the symmetric subspace is the correct one. Such
particles, like photons, phonons or mesons, are called Bosons.

Electrons, like protons and neutrons, are Fermions, particles with
half integer spin. The Hilbert space for Fermions consists of
totally antisymmetric functions, i.e.: if\;$x_1,x_2, \ldots, x_N\in
\R^d$ are the coordinates of $N$ electrons, then any state $\psi$ of
the system satisfies $\psi(x_1, x_2, x_3, \ldots, x_N)=-\psi(x_2,
x_1, x_3, \ldots, x_N)$ and similarly for interchanging any other
pair of particles.

It follows, that the product in (\ref{eq:prodstate}) is {\em not} a
vector of the (correct) Hilbert space (of antisymmetric functions).
Only its anti-symmetrization is

\begin{equation}
\Psi_f(x_1, x_2, \ldots, x_N):=\sum_{\pi\in S_N}\;(-1)^\pi \,
\psi_1(x_{\pi_1}) \psi_2(x_{\pi_2}) \ldots,
\psi_N(x_{\pi_N})\label{eq:antisym} \ .
\end{equation}

Here, the symbol $S_N$ stands for the permutation group and
$(-1)^\pi$ equals $1$ for even permutations (i.e. products of an
even number of exchanges), it equals $-1$ for odd permutations.

The anti-symmetrization (\ref{eq:antisym}) is non zero only if  the
functions $\psi_j$ are pairwise different. Consequently, the
eigenvalues of the multi-particle system are given as sums $E_1 +
E_2 + \ldots + E_N$ of the eigenvalues $E_j$ of the one-particle
system where the eigenvalues $E_j$ are all different. (We count
multiplicity, i.e. an eigenvalue of multiplicity two may occur twice
in the above sum). This rule is known as the Pauli-principle.

\goodbreak
The ground state energy of a system of $N$ identical, {\em
noninteracting} Fermions is therefore given by

$$E_1 + E_2 + \ldots + E_N$$
\goodbreak
where the $E_n$ are the eigenvalues of the single particle system in
increasing order, $E_1 \le E_2 \le \ldots $ counted according to
multiplicity.

It is not at all obvious how we can implement the above rules for
the systems considered here. Their spectra tend to consist of whole
intervals rather than being discrete, moreover, since the systems
extend to infinity they ought to have infinitely many electrons.

To circumvent this difficulty we will introduce a procedure  known
as the `thermodynamic limit': We first restrict the system to a
finite but large box (of length $L$ say), then we define quantities
of interest in this system, for example the number of states in a
given energy region {\em per unit volume}. Finally, we let the box
grow indefinitely (i.e. send L to infinity) and hope (or better
prove) that the quantity under consideration has a limit as L goes
to infinity. In the case of the number of states per unit volume
this limit, in deed, exists. It is called the density of states
measure and will play a major role in what follows. We will discuss
this issue in detail in chapter \ref{ch:dos}.
\\[5mm]
\goodbreak

{\bf Notes and Remarks }\\[2mm]
Standard references for  mathematical methods of quantum mechanics
are    \cite{K}, \cite{rs1}, \cite{rs2}, \cite{rs3}, \cite{rs4}  and
\cite{weid},  \cite{cfks}.

Most of the necessary prerequisites from spectral theory can be
found in \cite{rs1} or \cite{weid}. A good source for the
probabilistic background is \cite{lamp1} and \cite{lamp2}.

The physical theory of random Schrödinger operators  is described in
\cite{an}, \cite{LifGP}, \cite{thou1} and \cite{thou2}. References
for the mathematical approach to random Schrödinger operators are
\cite{cl}, \cite{cfks}, \cite{lang}, \cite{ksonder}, \cite{pf} and
\cite{stoll1}.
 \NeueSeite


\section{Setup: The Anderson model}

\subsection{Discrete Schr\"odinger operators}\es

In the Anderson model the Hilbert space $L^2(\R^d)$ is replaced by
the sequence space
\begin{align}
\ell^2(\Z^d)&=\{(u_i)_{i\in\Z^d}|\sum_{i\in\Z^d} |u_i|^2 < \infty\}\\
&=\{u:\Z^d\rightarrow \C\,\big|\sum_{n\in\Z^d} |u(n)|^2 < \infty\} \
.
\end{align}
\index{$\ell^2(\Z^d)$}

We denote the norm on $\ell^2(\Z^d)$ by

\begin{align}
\norm{u}~=~\Big(\sum_{n\in\Z^d} |u(n)|^2\,\Big)^\eh
\end{align}\index{$\norm{u}$}

Here, we think of a particle moving on the lattice $\Z^d$, so that
in the case $\|u\|=1$ the probability to find the particle at the
point $n\in\Z^d$ is given by $|u(n)|^2$. Note, that we may think of
$u$ either as a function $u(n)$ on $\Z^d$ or as a sequence $u_n$
indexed by $\Z^d$.

It will be convenient to equip $\Z^d$ with two different norms. The
first one is

\begin{equation}
\inorm{n\,} := \sup_{\nu=1, \dots, d} |\,n_\nu| \ .
\end{equation}
\index{$\inorm{n}$}

This norm respects the cubic structure of the lattice $\Z^d$. For
example, it is convenient to define the cubes ($n_0\in\Z^d$,
$L\in\N$)

\begin{equation}\label{eq:cube}
    \Lambda_L(n_0):=\{n\in\Z^d; \sn{n-n_0}\le L\} \ .
\end{equation}
\index{$\Lambda_L(n_0)$}

$\Lambda_L(n_0)$ is the cube of side length $2L+1$ centered at
$n_0$. It contains
    $|\Lambda_L(n_0)\,|:=\,(2L+1)^d\,$
points. Sometimes we call $|\Lambda_L(n_0)\,|$ the volume of
$\Lambda_L(n_0)$. In general, we denote by $|A|$ \index{$"|A"|$}
the number of elements of the set $A$.
 To shorten notation we write \mindex{$\Lambda_L$} for
$\Lambda_L(0)$. The other norm we use on $\Z^d$ is
\begin{equation}\label{eq:grnorm}
    \gnorm{n}:=\;\sum_{\nu=1}^d\,|\,n_\nu| \ .
\end{equation}
\index{$\ns{n}$}

This norm reflects the graph structure of $\Z^d$. Two vertices $n$
and $m$ of the graph $\Z^d$ are connected by an edge, if they are
nearest neighbors, i. e. if $\ns{n-m}=1$. For arbitrary $n,
m\in\Z^d$ the norm $\ns{n-m}$ gives the length of the shortest path
between $n$ and $m$.

The kinetic energy operator \mindex{$H_0$} is a discrete analogue of
the (negative) Laplacian, namely

\begin{equation} (H_0\,u)(n) = -\sum_{\gnorm{m-n}=1}\,(u(m)-u(n)) \ .
\label{def:H0}
\end{equation}

This operator is also known as the \dindex{graph Laplacian} for the
graph $\Z^d$ or the \emph{discrete Laplacian}. \index{Laplacian,
discrete} Its quadratic form is given by 
\begin{equation}\label{eq:dirichletform}
 \langle u, H_0\,v\rangle = \eh\sum_{n\in\Z^d}\;\sum_{\ns{m-n}=1}\;\overline{\left(u\left(n\right)-u\left(m\right)\right)}
 \left({v(n)-v(m)}\right) \ .
\end{equation}

We call this sesquilinear form a Dirichlet form because of its
similarity to the classical Dirichlet form

\begin{displaymath}
    \langle u, -\triangle\,v\rangle = \int_{\R^d}\;\overline{\nabla u(x)}\cdot\nabla
    v(x)\,dx \ .
\end{displaymath}
The operator $H_0$ is easily seen to be symmetric and bounded, in
fact

\begin{align}
\|H_0\,u \|  &=  \Big(\sum_{n\in\Z^d} \;  \Big(
\sum_{\gnorm{j\,}=1} \big(u(n+j)-u(n)\big)\;\Big)^2
\Big)^{\frac{1}{2}}\label{triangle1}\\
&\le   \sum_{\gnorm{j\,}=1}\,\Big( \, \sum_{n\in\Z^d} \,|\,u(n+j)-u(n)\,|^{\,2}\Big) ^{\frac{1}{2}}\label{trianglen}\\
&\le \sum_{\gnorm{j}=1}\,\Big( \, \sum_{n\in\Z^d} \big(|\,u(n+j)\,|\,+\,|\,u(n)\,|\big)^{\,2}\Big) ^{\frac{1}{2}}\\
&\le \sum_{\gnorm{j}=1}\,\bigg(\,\Big( \, \sum_{n\in\Z^d} \;|\,u(n+j)\,|^{\,2}\,\Big)^{\frac{1}{2}}
+\Big(\,\sum_{n\in\Z^d}\;|\,u(n)\,|^{\,2}\Big)^{\frac{1}{2}}\bigg)\label{triangle2}\\
&\le \quad 4\,d\;\|\,u\,\|\label{2d} \ .
\end{align}


From line (\ref{triangle1}) to (\ref{trianglen}) we applied the
triangle inequality for $\ell^2$ to the functions $\sum\,f_j(n)$
with $f_j(n)=u(n+j)-u(n)$. In (\ref{triangle2}) and (\ref{2d}) we
used the triangle inequality and the fact that any lattice point in
$\Z^d$ has $2d$ neighbors.

Let us define the Fourier transform \index{Fourier transform}
\index{$\mathcal{F}$} from $\ell^2(\Z^d)$ to $L^2([0,2\pi]^d)$ by

\begin{equation}
(\mathcal{F}u)(k)=\;\hat{u}(k)=\sum_n \;u_n\, e^{-i\,n\cdot k} \ .
\end{equation}

$\mathcal{F}$ is a unitary operator. Under $\mathcal{F}$ the
discrete Laplacian $H_0$ transforms to the multiplication operator
with the function $h_0(k)= 2\,\sum_{\,\nu=1}^{\,d}
\left(1-\cos(k_\nu)\right)$, i.e. $\mathcal{F}H_0\mathcal{F}^{-1}$
is the multiplication operator on $L^2([0,2\pi]^d)$ with the
function $h_0$. This shows that the spectrum $\sigma(H_0)$ equals
$[0,4d]$ (the range of the function $h_0$) and that $H_0$ has purely
absolutely continuous spectrum.

It is very convenient that the `discrete Dirac function'
\mindex{$\delta_i$} defined by $(\delta_i)_j=0$ for $i\neq j$ and
$(\delta_i)_i=1$ is an `honest' $\ell^2$-vector, in fact the
collection $\{\delta_i\}_{i\in\Z^d}$ is an orthonormal basis of
$\ell^2(\Z^d)$. This allows us to define \emph{matrix
entries}\index{matrix entry} or a \dindex{kernel} for every (say
bounded) operator $A$ on $\ell^2(\Z^d)$ by
\begin{equation}
    A(i,j)=\langle\delta_i,A\delta_j\rangle \ .
\end{equation}
\index{$A(i,j)$}

We have $(Au)(i)=\sum_{j\in\Z^d}A(i,j)u(j)$. So, the $A(i,j)$ define
the operator $A$ uniquely.

In this representation the multiplication operator $V$ is diagonal,
while
\begin{equation}\label{eq:H0matrix}
    H_0(i,j)=\left\{ \begin{array}{cl}
        -1 & \quad\mbox{if} \quad \gnorm{i-j}=1, \\ 2d & \quad\mbox{if}
        \quad i=j, \\ 0 & \quad\mbox{otherwise.}
        \end{array}\right.
\end{equation}

In many texts  the diagonal term in $H_0$ is dropped and absorbed
into the potential $V$. Moreover, one can also neglect the $-$-sign
in the offdiagonal terms of (\ref{eq:H0matrix}). The corresponding
operator is up to a constant equivalent to $H_0$ and has spectrum
$[-2d,2d]$.

In this setting the potential $V$  is a multiplication operator with
a function $V(n)$ on $\Z^d$. The simplest form to make this random
is to take $V(n)=V_\omega(n)$ itself as independent, identically
distributed random variables (see Section \ref{sec:randpot}), so we
have

\begin{displaymath}
H_\omega=H_0+V_\omega \ .
\end{displaymath}

We call this random operator the \emph{\mindex{Anderson model}}. For
most of this course we will be concerned with this operator.

\subsection{Spectral calculus\label{sec:speccal}}\es

One of the most important tools of spectral theory is the functional
calculus (spectral theorem) for self adjoint operators. We discuss
this topic here by giving a brief sketch of the theory and establish
notations. Details about functional calculus can be found in
\cite{rs1}. (For an alternative approach see \cite{Davies}).

Throughout this section let $A$ denote a self adjoint operator with
domain $D(A)$ on a (separable) Hilbert space $\mathcal{H}$. We will
try to define functions $f(A)$ of $A$ for a huge class of functions
$f$. Some elementary functions of $A$ can be defined in an obvious
way. One example is the resolvent which we consider first.

For any $z\in\C$ the operator $A-z=A-\,z\,\textnormal{id}\;$ is
defined by $(A-z)\varphi=A\varphi-z\varphi$. The resolvent set
\index{resolvent set} $\rho(A)$ \index{$\rho(A)$} of $A$ is the set
of all $z\in\C$ for which $A-z$ is a bijective mapping from $D(A)$
to $\mathcal{H}$. The spectrum \index{spectrum} \index{$\sigma(A)$}
$\sigma(A)$ of $A$ is defined by $\sigma(A)=\C\setminus\rho(A)$. For
self adjoint $A$ we have $\sigma(A)\subset\R$. The spectrum is
always a closed set. If $A$ is bounded, $\sigma(A)$ is compact.

For $z\in\rho(A)$ we can invert $A-z$. The inverse operator
$(A-z)^{-1}$ is called the resolvent \index{resolvent} of $A$. For
self adjoint $A$ the $(A-z)^{-1}$ is a bounded operator for all
$z\in\rho(A)$.

Resolvents observe the following important identities, known as the
\dindex{resolvent equations}

\begin{align}
(A-z_1)^{-1}\,-\,(A-z_2)^{-1}~&=~(z_1-z_2)\,(A-z_1)^{-1}\,(A-z_2)^{-1}\\
&=~(z_1-z_2)\,(A-z_2)^{-1}\,(A-z_1)^{-1}\label{eq:res}\\
\intertext{and,\; if $D(A)=D(B)$,}
(A-z)^{-1}\,-\,(B-z)^{-1}~&=~(A-z)^{-1}\,(B-A)\,(B-z)^{-1}\\
&=~(B-z)^{-1}\,(B-A)\,(A-z)^{-1}
\end{align}

For $z\in\C$ and $M\subset\C$ we define
\begin{equation}\label{def:dist1}
\tn{dist}(z,M)~=~\inf \{|z-\zeta|; \zeta\in M\}
\end{equation}\index{dist}
It is not hard to see that for any self adjoint operator $A$ and any
$z\in\rho(A)$ the operator norm $\|(A-z)^{-1}\|$ of the resolvent is
given by
\begin{equation}\label{eq:resnorm}
\|(A-z)^{-1}\| ~=~\frac{1}{\;\tn{dist}(z,\sigma(A))\;}\ .
\end{equation}
In particular, for a self adjoint operator $A$ and
$z\in\C\setminus\R$
\begin{equation}\label{est:resnorm}
\|(A-z)^{-1}\| ~\leq~\frac{1}{\;\tn{Im}\, z\;}\ .
\end{equation}

For the rest of this section we assume that the operator $A$ is
bounded. In this case, polynomials of the operator $A$ can be
defined straightforwardly
\begin{align}A^2\,\varphi~&=~A\big(A(\varphi)\big)\\
A^3\,\varphi~&=~A\Big(A\big(A(A \varphi)\big)\Big)\quad \tn{etc.}
\end{align}
More generally, if $P$ is a complex valued polynomial in one real
variable, $P(\lambda)=\sum_{j=0}^n a_n \lambda^j$ then
\begin{equation}
P(A)~=~\sum_{j=0}^n a_n\,A^j\ .
\end{equation}

It is a key observation that
\begin{equation}\label{eq:norm}
\|P(A)\|~=~\sup_{\lambda\in\sigma(A)}\;|P(\lambda)|
\end{equation}

Let now $f$ be a function in $C\big(\sigma(A)\big)$ the
complex-valued \index{$C(K)$} continuous functions on (the compact
set) $\sigma(A)$. The Weierstraß approximation theorem tells us,
that on $\sigma(A)$ the function $f$ can be uniformly approximated
by polynomials. Thus using (\ref{eq:norm}) we can define the
operator $f(A)$ as a norm limit of polynomials $P_n(A)$. These
operators satisfy

\begin{align}
(\alpha f + \beta g)\,(A)~&=~\alpha f(A) + \beta g(A)\label{proper1}\\
f\cdot g\,(A)~&=~f(A)\,g(A)\label{proper2}\\
\overline f\,(A)~&=~f(A)^*\label{proper3}\\
\textnormal{If }f\geq 0 \quad\textnormal{then}\quad &\langle
\varphi, f(A) \varphi\rangle\geq 0 \;\textnormal{ for all
}\;\varphi\in\mathcal{H} \label{proper4}
\end{align}

By the Riesz-representation theorem it follows, that for each
$\varphi\in\mathcal{H}$ there is a positive and bounded measure
$\mu_{\varphi,\varphi}$ on $\sigma(A)$ such that for all $f\in
C\big(\sigma(A)\big)$

\begin{equation}
\langle\varphi,
f(A)\varphi\rangle~=~\int\;f(\lambda)\;d\mu_{\varphi,\varphi}(\lambda)\
.
\end{equation}

For $\varphi, \psi \in\mathcal{H}$, using the polarization identity,
we find complex-valued measures $\mu_{\varphi,\psi}$ such that

\begin{equation}\label{def:specmeas}
\langle\varphi,
f(A)\varphi\rangle~=~\int\;f(\lambda)\;d\mu_{\varphi,\psi}(\lambda)\
.
\end{equation}

Equation (\ref{def:specmeas}) can be used to define the operator
$f(A)$ for bounded \emph{measurable} functions. The operators
$f(A)$, $g(A)$ satisfy (\ref{proper1})--(\ref{proper4}) for bounded
measurable functions as well, moreover we have:

\begin{equation}
\|f(A)\|~\leq~\sup_{\lambda\in\sigma(A)}\,|f(\lambda)|
\end{equation}
with equality for continuous $f$.

For any Borel set $M\subset\R$ we denote by $\chi_M$
\index{$\chi_M$} the \emph{characteristic function}
\index{characteristic function} of $M$ defined by

\begin{equation}
\chi_M(\lambda)~=~\left\{ \begin{array}{@{\quad}r l}
1&\quad\textnormal{if } \lambda\in M\\
0&\quad\textnormal{otherwise.}\end{array} \right.
\end{equation}

The operators $\mu(A)=\chi_M(A)$ play a special role. It is not hard
to check that they satisfy the following conditions:

\begin{align}
\mu(A)\quad \tn{is an}&\tn{ orthogonal projection.}\label{def:pm1}\\
\mu(\emptyset)~=~0 \quad&\textnormal{and\quad}\mu\big(\sigma(A)\big)~=~1\label{def:pm2}\\
\mu(M\cap N)~&=~\mu(M)\,\mu(N)\label{def:pm3}\\
\intertext{If the Borel sets $M_n$ are pairwise disjoint, then for
each $\varphi\in\mathcal{H}$}
\mu\big(\bigcup_{n=1}^\infty\;M_n\big)\,\varphi~&=~
\sum_{n=1}^\infty \;\mu(M_n)\,\varphi\label{def:pm4}
\end{align}

Since $\mu(M)=\chi_M(A)$ satisfies (\ref{def:pm1})--(\ref{def:pm4})
it is called the \dindex{projection valued measure} associated to
the operator $A$ or the \emph{projection valued spectral measure}
\index{spectral measure!projection valued} of $A$. We have

\begin{equation}
\langle\varphi, \mu(M)\psi\rangle~=~\mu_{\varphi,\psi}(M)
\end{equation}
The functional calculus can be implemented for \emph{unbounded self
adjoint} operators as well. For such operators the spectrum is
always a closed set. It is compact only for bounded operators.

We will use the functional calculus virtually everywhere throughout
this paper. For example, it gives meaning to the operator $e^{-itH}$
used in (\ref{eq:3.1}). We will look at the projection valued
measures $\chi_M(A)$ more closely in chapter \ref{ch:spec}.

\subsection{Some more functional analysis\label{sec:morefun}}\es

In this section we recall a few results from functional analysis and
spectral theory and establish notations at the same time. In
particular, we discuss the min-max principle and the
Stone-Weierstraß theorem.

Let $A$ be a selfadjoint (not necessarily bounded) operator on the
(separable) Hilbert space  $\mathcal{H}$ with domain $D(A)$. We
denote the set of eigenvalues of $A$ by $\ve(A)$. \index{$\ve(H)$}
Obviously, any eigenvalue of $A$ belongs to the spectrum
$\sigma(A)$. The \dindex{multiplicity} of an eigenvalue $\lambda$ of
$A$ is the dimension of the eigenspace $\{\varphi\in D(A);
A\varphi=\lambda\varphi\}$  associated to $\lambda$. If $\mu$ is the
projection valued spectral measure of $A$, then the multiplicity of
$\lambda$ equals $\,\tr\,\mu(\{\lambda\})$. An eigenvalue is called
\dindex{simple} or \dindex{non degenerate} if its multiplicity is
one, it is called \dindex{finitely degenerate} if its eigenspace is
finite dimensional. An eigenvalue $\lambda$ is called
\dindex{isolated} if there is an $\ve>0$ such that
$\sigma(A)\cap(\lambda-\ve,\lambda+\ve)=\{\lambda\}$. Any isolated
point in the spectrum is always an eigenvalue. The \dindex{discrete
spectrum} $\sigma_{dis}(A)$ \index{$\sigma_{dis}(A)$} is the set of
all isolated eigenvalues of finite multiplicity. The
\dindex{essential spectrum} $\sigma_{ess}(A)$
\index{$\sigma_{ess}(A)$} is defined by
$\sigma_{ess}(A)=\sigma(A)\setminus\sigma_{dis}(A)$.

The operator $A$ is called \emph{positive}\index{positive operator}
if $\langle\phi,A\phi\rangle\geq 0$ for all $\phi$ in the domain
$D(A)$, $A$ is called \dindex{bounded below} if
$\langle\phi,A\phi\rangle\geq -M \langle\phi,\phi\rangle$ for some
$M$ and all $\phi\in D(A)$.

We define
\begin{align}
\mu_0(A)~&=~\inf\,\{\langle\phi,A\phi\rangle\,;\ \phi\in D(A), \|\phi\|=1\}\\
\intertext{and for $k\geq 1$}
\mu_k(A)~&=~\sup_{\psi_1,\dots,\psi_k\in\mathcal{H}}\;\inf\,\{\langle\phi,A\phi\rangle\,;\
\phi\in D(A), \|\phi\|=1, \phi\perp\psi_1,\dots,\psi_k\}
\end{align}

The operator $A$ is bounded below iff $\mu_0(A)>-\infty$ and
$\mu_0(A)$ is the infimum of the spectrum of $A$.

If $A$ is bounded below and has purely discrete spectrum (i.e.
$\sigma_{ess}(A)=\emptyset$), we can order the eigenvalues of $A$ in
increasing order and repeat them according to their multiplicity,
namely
\begin{equation}\label{eigvord}
E_0(A)\leq E_1(A) \leq E_2(A) \leq \dots\ .
\end{equation}
If an eigenvalue $E$ of $A$ has multiplicity $m$ it occurs in
(\ref{eigvord}) exactly $m$ times.

The min-max principle \index{min-max principle} relates the $E_k(A)$
with the $\mu_k(A)$.

\begin{theorem}[Min-max principle] \label{th:minmax} If the self adjoint operator $A$ has purely discrete spectrum and is
bounded below, then
\begin{equation} E_k(A)~=~\mu_k(A)\qquad \tn{for all $k\geq 0$}\ .
\end{equation}
\end{theorem}

A proof of this important result can be found in \cite{rs4}. The
formulation there contains various refinements of our version. In
particular \cite{rs4} deals also with discrete spectrum below the
infimum of the essential spectrum.

We state an application of Theorem \ref{th:minmax}. By $A\leq B$ we
mean that the domain $D(B)$ is a subset of the domain $D(A)$ and
$\langle\phi,A\phi\rangle\leq\langle\phi,B\phi\rangle$ for all
$\phi\in D(B)$.\index{$A\leq B$}
\begin{corollary} Let $A$ and $B$ are self adjoint operators which are bounded below and have purely discrete spectrum.
If $A\leq B$ then $E_k(A)\leq E_k(B)$ for all $k$.
\end{corollary}
The Corollary follows directly from Theorem \ref{th:minmax}.\\

We end this section with a short discussion of the Stone-Weierstraß
Theorem \index{Stone-Weierstraß Theorem} in the context of spectral
theory. The Stone-Weierstraß Theorem deals with subsets of the space
$C_\infty(\R)$, the set of all (complex valued) continuous functions
on $\R$ which vanish at infinity.\index{$C_\infty(\R)$}.

A subset $\mathcal{D}$ is called an involutative subalgebra of
$C_\infty(\R)$, if it is a linear subspace and if for
$f,g\in\mathcal{D}$ both the product $f\cdot g$ and the complex
conjugate $\overline{\,f\,}$ belong to \;$\mathcal{D}$. We say that
$\mathcal{D}$ \emph{seperates points} if for $x,y\in\R$ there is a
function $f\in\mathcal{D}$ such that $f(x)\not=f(y)$ and both $f(x)$
and $f(y)$ are non zero.

\begin{theorem}[Stone-Weierstraß]\label{th:StoneW} If $\mathcal{D}$ is an involutative subalgebra of $C_\infty(\R)$ which seperates points,
then $\mathcal{D}$ is dense in $C_\infty(\R)$ with respect to the
topology of uniform convergence.
\end{theorem}

A proof of this theorem is contained e.g. in \cite{rs1}. Theorem
\ref{th:StoneW} can be used to prove some assertion $\mathcal{P}(f)$
for the operators $f(A)$ for \emph{all} $f\in C_\infty(\R)$ if we
know $\mathcal{P}(f)$ for \emph{some} $f$. Suppose we know
$\mathcal{P}(f)$ for all $f\in\mathcal{D}_0$. If we can show that
$\mathcal{D}_0$ seperates points and that the set of all $f$
satisfying $\mathcal{P}(f)$ is a closed involutative subalgebra of
$C_\infty(\R)$, then the Stone-Weierstraß theorem tells us that
$\mathcal{P}(f)$ holds for all $f\in C_\infty(\R)$.

Theorem \ref{th:StoneW} is especially useful in connection with
resolvents. Suppose a property $\mathcal{P}(f)$ holds for all
functions $f$ in $\mathcal{R}$, the set of linear combinations of
the functions $f_\zeta(x)=\frac{1}{\,x-\zeta\,}$ for all
$\zeta\in\C\setminus\R$,  so for resolvents of $A$ and their linear
combinations. The resolvent equations (or rather basic algebra of
$\R$) tell us that $\mathcal{R}$ is actually an involutative
algebra. So, if the property $\mathcal{P}(f)$ survives uniform
limits, we can conclude that $\mathcal{P}(f)$ is valid for all $f\in
C_\infty(\R)$. The above procedure was dubbed the `Stone-Weierstraß
Gavotte' in \cite{cfks}. More details can be found there.

\subsection{Random potentials\label{sec:randpot}}\es

\begin{definition}
A \dindex{random variable} is a real valued measurable function on a
probability space $(\Omega,\mathcal{F},\P)$.

If $X$ is a random variable we call the probability measure $P_0$ on
$\R$ defined by
\begin{align}
P_0(A)~=~\P\,\big(\{\omega\,|\,X(\omega)\in A\,\}\big)\qquad
\textnormal{\em for any Borel set } A
\end{align}
the \dindex{distribution} of $X$. If the distributions of the random
variables $X$ and $Y$ agree we say that $X$ and $Y$ are
\dindex{identically distributed}. We also say that $X$ and $Y$ have
a common distribution in this case.

A family $\{X_i\}_{i\in I}$ of random variables is called
\dindex{independent} if for any finite subset $\{i_1, \dots, i_n\}$
of $I$
\begin{align}\nonumber
&\P\,\Big(\big\{\omega |\;X_{i_1}(\omega)\in [a_1,b_1],
X_{i_2}(\omega)\in [a_2,b_2],\ldots\; X_{i_n}(\omega)\in
[a_n,b_n]\,\big\}\Big)\\\label{eq:independent}
&=\;\P\,\Big(\big\{\omega |\;X_{i_1}(\omega)\in
[a_1,b_1]\,\big\}\Big) \;\cdot\,\ldots\,\cdot
\;\P\,\Big(\big\{\omega |\;X_{i_n}(\omega)\in
[a_n,b_n]\,\big\}\Big)\ .
\end{align}
\end{definition}
\begin{remark}
If $X_i$ are independent and identically distributed (iid)
\index{iid} with common distribution  \mindex{$P_0$} then
\begin{eqnarray}\nonumber
&&\P\,\Big(\big\{\omega |\;X_{i_1}(\omega)\in [a_1,b_1],
X_{i_2}(\omega)\in [a_2,b_2],\ldots\; X_{i_n}(\omega)\in
[a_n,b_n]\,\big\}\Big)\\\nonumber &&=\; P_0([a_1,b_1])\cdot
P_0([a_2,b_2])\cdot\;\ldots\;\cdot P_0([a_n,b_n])\ .
\end{eqnarray}
\end{remark}
For the reader's convenience we state a very useful result of
elementary probability theory which we will need a number of times
in this text.

\begin{theorem}[Borel-Cantelli lemma]\label{th:BorelCantelli}\index{Borel-Cantelli lemma}
Let $(\Omega,\mathcal{F},\P)$ be a probability space and
$\{A_n\}_{n\in\N}$ be a sequence of set in $\mathcal{F}$. Denote by
$A_\infty$ the set
\begin{equation}
A_\infty~=~\{\omega\in\Omega\,|\;\omega\in A_n \tn{ \;for infinitely
many \;}n\}
\end{equation}
\begin{enumerate}
\item If \;
$\sum_{n=1}^\infty \P(A_n)<\infty$, then
 \;$\P(A_\infty)=0$\vspace{2mm}
\item If the sets $\{A_n\}$ are independent\\ and \,
$\sum_{n=1}^\infty \P(A_n)=\infty$, then
 \;$\P(A_\infty)=1$
\end{enumerate}
\end{theorem}

\begin{remark}\es\vspace{-0.2cm}
\begin{enumerate}
\item We recall that a sequence $\{A_n\}$ of events \index{event} (i.e. of sets from $\mathcal{F}$) is called
\emph{independent} \index{independent} if for any finite subsequence
$\{A_{n_j}\}_{j=1,\dots,M}$
\begin{equation}
\P\,\big(\bigcap_{j=1}^M\;
A_{n_j}\big)~=~\prod_{j=1}^M\;\P\,(A_{n_j})
\end{equation}
\item The set $A_\infty$ can be written as $A_\infty=\bigcap_{N}\,\bigcup_{n\geq N} A_n$.
\end{enumerate}
\end{remark}

For the proof of Theorem \ref{th:BorelCantelli} see e.g. \cite{Bau}
or \cite{lamp1}. \vspace{0.5cm}

From now on we assume that the random variables
$\{V_\omega(n)\}_{n\in\Z^d}$ are independent and identically
distributed with common distribution $P_0$.

By \mindex{$\supp P_0$} we denote the \dindex{support} of the
measure $P_0$, i.e.
\begin{equation}
\supp
P_0~=~\{x\in\R\,|\,P_0\big(\,(x-\varepsilon,x+\varepsilon)\,\big)>0
\textnormal{ for all } \varepsilon>0 \}\ .
\end{equation}

If $\supp P_0$ is compact then the operator $H_\omega = H_0 +
V_\omega$ is bounded. In fact, if $\supp P_0 \subset [-M , M]$, with
probability one

\begin{displaymath}
\sup_{j \in \Z^d} |V_\omega (j)| \leq M \ .
\end{displaymath}

Even if $\supp P_0$ is not compact the multiplication operator
$V_\omega$ is selfadjoint on $D = \{\varphi \in \ell^2 | V_\omega
\varphi \in \ell^2\}$. It is essentially selfadjoint on

\begin{displaymath}
{\ell_0^2(\Z^d)} = \{\varphi \in \ell^2(\Z^d)\; |\; \varphi(i) = 0
\; \textnormal{for all but finitely many points } i\} \ .
\end{displaymath}
\index{$\ell_0^2(\Z^d)$}

Since $H_0$ is bounded it is a fortiori Kato bounded with respect to
$V_\omega$. By the Kato-Rellich theorem it follows that $H_\omega =
H_0 + V_\omega$ is essentially selfadjoint on $\ell_0^2(\Z^d)$ as
well (see \cite{rs2} for details).

In a first result about the Anderson model we are now going to
determine its spectrum (as a set). In particular we will see that
the spectrum $\sigma(H_\omega)$ is ($\P$-almost surely) a fixed non
random set. First we prove a proposition which while easy is very
useful in the following. Roughly speaking, this proposition tells
us:

Whatever \emph{can} happen, \emph{will} happen, in fact infinitely
often.

\begin{proposition}{\label{pr:1}}
There is a set $\Omega_0$ of probability one such that the following
is true: For any $\omega \in \Omega_0$, any finite set $\Lambda
\subset \Z^d$, any sequence $\{q_i\}_{i \in \Lambda}$, $q_i \in
\supp P_0$ and any $\varepsilon > 0$, there exists a sequence
$\{j_n\}$ in $\Z^d$ with $\inorm{j_n} \rightarrow \infty$ such that

\begin{displaymath}
\sup_{i \in \Lambda} |\, q_i - V_\omega(i + j_n)\, | < \varepsilon \
.
\end{displaymath}
\end{proposition}

\begin{proof}
Fix a finite set $\Lambda$, a sequence $\{q_i\}_{i \in \Lambda}$,
$q_i \in \supp P_0$ and $\varepsilon > 0$. Then, by the definition
of $\supp$ and the independence of the $q_i$ we have for $A=\{\omega
| \sup_{i \in \Lambda} |V_\omega(i) - q_i| < \varepsilon\}$

\begin{displaymath}
\P(A) > 0 \ .
\end{displaymath}

Pick a sequence $\ell_n \in \Z^d$, such that the distance between
any $\ell_n$, $\ell_m \; (n \neq m)$ is bigger than twice the
diameter of $\Lambda$. Then, the events

\begin{displaymath}
A_n =A_n(\Lambda,\{q_i\}_{i\in\Lambda},\varepsilon)= \{\omega |
\sup_{i \in \Lambda} |V_\omega(i + \ell_n) - q_i| < \varepsilon\}
\end{displaymath}

are independent and $\P(A_n) = \P(A) > 0$. Consequently, the
Borel-Cantelli lemma (see Theorem \ref{th:BorelCantelli}) tells us
that

\begin{displaymath}
\Omega_{\Lambda, \{q_i\}, \varepsilon} = \{\,\omega\; |\; \omega \in
A_n\, \textnormal{ for infinitely many}\ n\}
\end{displaymath}

has probability one.

The set $\supp P_0$ contains a countable dense set $R_0$. Moreover,
the system $\Xi$ of all finite subsets of $\Z^d$ is countable. Thus
the set

\begin{displaymath}
\Omega_0 := \bigcap_{\Lambda\, \in\; \Xi, \atop \{q_i\} \in R_0,
n\in \N} \!\!\Omega_{\Lambda, \{q_i\}, \frac{1}{n}}
\end{displaymath}

has probability one. It is a countable intersection of sets of
probability one.

By its definition, $\Omega_0$ satisfies the requirements of the
assertion.

\end{proof}

We now turn to the announced theorem

\begin{theorem}\label{th:spAnd}
For\; $\P$-almost all $\omega$ we have $\sigma(H_\omega)= [0,4d\,] +
\tn{supp}\,P_0$.
\end{theorem}

\begin{proof}
The spectrum $\sigma(V)$ of the multiplication operator with $V(n)$
is given by the closure of the set $R(V) = \{V(n) | n \in \Z^d\}$.
Hence $\sigma(V_\omega) = \supp P_0$ almost surely. Since $ 0 \leq
H_0  \leq 4d$ we have

\begin{eqnarray}\nonumber
\sigma (H_0 + V_\omega) &\subset& \sigma(V_\omega) + [0,
\|H_0\|]\\\nonumber &=& \supp P_0 + [0, 4d] \ .
\end{eqnarray}

Let us prove the converse. We use the \dindex{Weyl criterion} (see
\cite{rs1} or \cite{weid}):

\begin{displaymath}
\lambda \in \sigma(H_\omega) \quad\Leftrightarrow \quad\exists\
\varphi_n \in D_0, || \varphi_n || = 1:\quad ||(H_\omega - \lambda)
\varphi_n|| \rightarrow 0 \ ,
\end{displaymath}

where $D_0$ is any vector space such that $H_\omega$ is essentially
selfadjoint on $D_0$. The sequence $\varphi_n$ is called a
\dindex{Weyl sequence}. In a sense, $\varphi_n$ is an `approximate
eigenfunction'.

Let $\lambda \in [0, 4d] + \supp P_0$, say $\lambda = \lambda_0 +
\lambda_1$, \ $\lambda_0 \in \sigma(H_0) = [0, 4d], \ \lambda_1 \in
\supp P_0$. Take a Weyl sequence $\varphi_n$ for $H_0$ and
$\lambda_0$, i. e. $|| (H_0 - \lambda_0)\varphi_n || \rightarrow 0,
||\varphi_n|| = 1$. Since $H_0$ is essentially selfadjoint on
$D_0=\ell_0^2(\Z^d)$ (in fact $H_0$ is bounded), we may suppose
$\varphi_n \in D_0$. Setting $\varphi^{(j)}(i) = \varphi(i-j)$, we
easily see

\begin{displaymath}
H_0 \varphi^{(j)} = (H_0 \varphi)^{(j)} \ .
\end{displaymath}

Due to Proposition \ref{pr:1} there is (with probability one) a
sequence $\{j_n\}$, \hbox{$\inorm{j_n} \rightarrow \infty$} such
that
\begin{eqnarray} \sup_{i \in \supp \varphi_n} |V_\omega(i + j_n) -
\lambda_1| < \frac{1}{n} \ .
\end{eqnarray}

Define $\psi_n = \varphi_n^{j_n}$. Then $\psi_n$ is a Weyl sequence
for $H_\omega$ and $\lambda = \lambda_0 + \lambda_1$. This proves
the theorem.
\end{proof}

The above result tells us in particular that the spectrum
$\sigma(H_\omega)$ is (almost surely) non random. Moreover, an
inspection of the proof shows that there is no discrete spectrum
(almost surely), as the constructed Weyl sequence tends to zero
weakly, in fact can be chosen to be orthonormal. Both results are
valid in much bigger generality. They are due to ergodicity
properties of the potential $V_\omega$. We will discuss this topic
in the following chapter.\\[2mm]
{\bf Notes and Remarks }\\[2mm]
For further information see  \cite{cl} and \cite{cfks} or consult
\cite{kuso} and \cite{km2,km3}. \NeueSeite


\section{Ergodicity properties}

\subsection{Ergodic stochastic processes}\es

Some of the basic questions about random Schr\"odinger can be
formulated and answered most conveniently within the framework of
`ergodic operators'. This class of operators comprises many random
operators, such as the Anderson model and its continuous analogs,
the Poisson model, as well as random acoustic operators. Moreover,
also operators with almost periodic potentials can be viewed as
ergodic operators.

In these notes we only briefly touch the topic of ergodic operators.
We just collect a few definitions and results we will need in the
following chapters. We recommend
the references cited in the notes at the end of this chapter for further reading.\\[2mm]
Ergodic stochastic processes are a certain generalization of
independent, identically distributed random variables. The
assumption that the random variables $X_i$ and $X_j$ are independent
for $|i-j|>0$ is replaced by the requirement that $X_i$ and $X_j$
are `almost independent' if $|i-j|$ is large (see the discussion
below, especially (\ref{eq:2}), for a precise statement). The most
important result about ergodic processes is the ergodic theorem (see
Theorem \ref{th:Birkhoff} below), which says that the strong law of
large numbers, one of the basic results about independent,
identically distributed random variables, extends to ergodic
processes.

At a number a places in these notes we will have to deal with
ergodic processes. Certain important quantities connected with
random operators are ergodic but not independent even if the
potential $V_\omega$ \emph{is }\,a sequence of independent random
variables.

A family $\{X_i\}_{i\in\Z^d}$ of random variables is called a
\dindex{stochastic process} (with index set $\Z^d$). This means that
there is a probability space $(\Omega,\mathcal{F},\P)$
($\mathcal{F}$ a $\sigma$-algebra on $\Omega$ and $\P$ a probability
measure on $(\Omega,\F)$) such that the $X_i$ are real valued,
measurable functions on $(\Omega,\F)$.

The quantities of interest are the probabilities of events that can
be expressed through the random variables $X_i$, like
\begin{equation*}
\{\omega | \lim_{N\rightarrow\infty}
\frac{1}{|\L_N|}\sum_{\,\inorm{i}\leq N} X_i(\omega)=0\}\ .
\end{equation*}

The special way $\Omega$ is constructed is irrelevant. For example,
one may take the set $\R^{\Z^d}$ as $\Omega$. The corresponding
$\sigma$-algebra $\mathcal{F}$ is generated by
\emph{\mindex{cylinder sets}} of the form
\begin{equation}\label{eq:1}
    \{\omega\; |\; \omega_{i_1}\in A_1,\dots,\omega_{i_n}\in A_n\}
\end{equation}
where $A_1,\dots,A_n$ are Borel subsets of $\R$. On $\Omega$ the
random variables $X_i$ can be realized by $X_i(\omega)=\omega_i$.

This choice of $(\Omega,\F)$ is called the \emph{\mindex{canonical
probability space}}. For details in connection with random operators
see e.g. \cite{ksonder, km2}. Given a probability space
$(\Omega,\F,\P)$ we call a measurable mapping $T:
\Omega\rightarrow\Omega$ a \emph{\mindex{measure preserving
transformation}} if~~ $\P(T^{-1}A)=\P(A)$~~ for all $A\in\F$. If
$\{T_i\}_{i\in\Z^d}$ is a family of measure preserving
transformations we call a set $A\in\F$ \emph{\mindex{invariant}}
(under $\{T_i\}$) if $T_i^{-1}A=A$ for all $i\in\Z^d$.

A family $\{T_i\}$ of measure preserving transformations on a
probability space $(\Omega,\F,\P)$ is called
\emph{\mindex{ergodic}}(with respect to the probability measure
$\P$) if any invariant $A\in\F$ has probability zero or one. A
stochastic process $\{X_i\}_{i\in \Z^d}$ is called \emph{ergodic},
if there exists an ergodic family of measure preserving
transformations $\{T_i\}_{i\in \Z^d}$ such that $X_i(T_j\,
\omega)=X_{i-j}(\omega)$.

Our main example of an ergodic stochastic process is given by
independent, identically distributed random variables
$X_i(\omega)=V_\omega(i)$ (a random potential on $\Z^d$). Due to the
independence of the random variables the probability measure $\P$
(on $\Omega=\R^{\Z^d})$ is just the infinite product measure of the
probability measure $P_0$ on $\R$ given by
$P_0(M)=\P(V_\omega(0)\in\ M)$. $P_0$ is the distribution of
$V_\omega(0)$.

It is easy to see that the shift operators
\begin{displaymath}
(T_i\, \omega)_j = \omega_{j-i}
\end{displaymath}
form a family of measure preserving transformations on $\R^{\Z^d}$
in this case.

It is not hard to see that the family of shift operators is ergodic
with respect to the product measure $\P$. One way to prove this is
to show that
\begin{equation}\label{eq:2}
\P(T_i^{-1}A\cap B)\rightarrow \P(A)\ \P(B)
\end{equation}
as $\inorm{i} \rightarrow \infty$ for all $A,\ B\ \in\ \F$. This is
obvious if both $A$ and $B$ are of the form (\ref{eq:1}). Moreover,
the system of sets $A,\ B$ for which (\ref{eq:2}) holds is a
$\sigma$-algebra, thus (\ref{eq:2}) is true for the $\sigma$-algebra
generated by sets of the form (\ref{eq:1}), i.e. on $\F$.

Now let $M$ be an invariant set. Then (\ref{eq:2}) (with $A=B=M$)
gives
\begin{displaymath}
\P(M)=\P(M\cap M)=\P(T_i^{-1}M\cap M) \rightarrow \P(M)^2
\end{displaymath}
proving that $M$ has probability zero or one.

\vspace{0.2cm} We will need two more results on ergodicity.

\begin{proposition}\label{prop:const}
Let $\{T_i\}_{i\in \Z^d}$ be an ergodic family of measure preserving
transformations on a probability space $(\Omega, \F, P)$. If a
random variable $Y$ is invariant under $\{T_i\}$ (i.e. $Y(T_i
\omega)=Y(\omega)$ for all $i\in \Z^d$) then $Y$ is almost surely
constant, i.e. there is a $c \in \R$, such that $\P(Y=c)=1$.
\end{proposition}

We may allow the values $\pm \infty$ for $Y$ (and hence for $c$) in
the above result.The proof is not difficult (see e.g. \cite{cfks}).

The final result is the celebrated ergodic theorem by Birkhoff. It
generalizes the strong law of large number to ergodic processes.

We denote by $\E(\cdot)$ the expectation with respect to the
probability measure $\P$.

\begin{theorem}\label{th:Birkhoff}
If $\{X_i\}_{i \in \Z^d}$ is an ergodic process and $\E(|X_0|) <
\infty$ then
\begin{displaymath}
\lim_{L \rightarrow \infty} \frac{1}{(2L+1)^d} \sum_{i \in
\Lambda_L}X_i \rightarrow \E(X_0)
\end{displaymath}
for $\P$-almost all $\omega$.
\end{theorem}

For a proof of this fundamental result see e.g. \cite{lamp2}. We
remark that the ergodic theorem has important extensions in various
directions (see \cite{krengel}).

\bigskip

\subsection{Ergodic operators}\es

Let $V_\omega(n)$, $n \in \Z^d$ be an ergodic process (for example,
one may think of independent identically distributed $V_\omega(n)$).

Then there exist measure preserving transformations $\{T_i\}$ on
$\Omega$ such that
\begin{enumerate}
    \item $V_\omega(n)$ satisfies
    \begin{equation}\label{eq:3}
        V_{T_i \omega}(n)=V_\omega(n-i) \ .
        \end{equation}
    \item Any measurable subset of $\Omega$ which is invariant
    under the $\{T_i\}$ has trivial probability (i.e. $\P(A)=0$ or
    $\P(A)=1$) \ .
\end{enumerate}

We define translation operators $\{U_i\}_{i \in
\mathbb{Z}^d}$\index{$U_i$} on $\ell^2(\Z^d)$ by
\begin{equation}\label{eq:4}(U_i
\varphi)_m = \varphi_{m-i} \quad , \varphi \in \ell^2(\Z^d)  \ .
\end{equation}

It is clear that the operators $U_i$ are unitary. Moreover, if we
denote the multiplication operators with the function $V$ by
$\underline{V}$ then

\begin{equation}\label{eq:5}
\underline{V_{T_i \omega}}=U_i \underline{V_\omega} U_i^*.
\end{equation}

The free Hamiltonian $H_0$ of the Anderson model (\ref{def:H0})
commutes with $U_i$, thus (\ref{eq:5}) implies

\begin{equation}\label{eq:6}
H_{T_i \omega}=U_i H_\omega U_i^* \ .
\end{equation}

i.e. $H_{T_i \omega}$ and $H_\omega$ are unitarily equivalent.

Operators satisfying (\ref{eq:6}) (with ergodic $T_i$ and unitary
$U_i$) are called \emph{\mindex{ergodic operators}}.

The following result is basic to the theory of ergodic operators.

\begin{theorem} (Pastur)\label{th:specconst}
If $H_\omega$ is an ergodic family of selfadjoint operators, then
there is a (closed, nonrandom) subset $\Sigma$ of $\R$, such that
\begin{verse}
$\sigma(H_\omega)=\Sigma$ \quad for $\P$-almost all $\omega$.
\end{verse}
Moreover, there are sets $\Sigma_{ac}$, $\Sigma_{sc}$, $\Sigma_{pp}$
such that
\begin{verse}
$\sigma_{ac}(H_\omega)=\Sigma_{ac}$, \quad
$\sigma_{sc}(H_\omega)=\Sigma_{sc}$, \quad
$\sigma_{pp}(H_\omega)=\Sigma_{pp}$ \\ \qquad\qquad\qquad for
$\P$-almost all $\omega$.
\end{verse}
\end{theorem}

\begin{remark}
\mbox{}
\begin{enumerate}
    \item The theorem in its original form is due to Pastur  \cite{p1}. It was extended in
    \cite{kuso} and \cite{km2}.
    \item We have been sloppy about the measurability properties
    of $H_\omega$ which have to be defined and checked carefully. They are satisfied in our case (i.e. for the Anderson
    model). For a precise formulation and proofs see \cite{km2}.
    \item We denote by $\sigma_{ac}(H), \sigma_{sc}(H), \sigma_{pp}(H)$ the absolutely continuous (resp.
    singularly continuous, resp. pure point) spectrum of the operator $H$. For a definition and basic poperties
    we refer to Sections \ref{sec:measspec}and \ref{sec:RAGE}.
\end{enumerate}
\end{remark}

\begin{nproof}{Sketch}
If $H_\omega$ is ergodic and $f$ is a bounded (measurable) function
then $f(H_\omega)$ is ergodic as well, i.e.
\begin{displaymath}
f(H_{T_i w})=U_i f(H_\omega) U_i^* \ .
\end{displaymath}
(see Lemma \ref{lem:UAU}).

We have $(\lambda, \mu) \cap\ \sigma(H_\omega) \neq \emptyset$ \quad
if and only if \quad $\chi_{(\lambda, \mu)}\ (H_\omega) \neq 0$.

This is equivalent to $Y_{\lambda, \mu}(\omega):= \tr
\chi_{(\lambda, \mu)}\ (H_\omega) \neq 0$.

Since $\chi_{(\lambda, \mu)}\ (H_\omega)$ is ergodic, $Y_{\lambda,
\mu}$ is an invariant random variable and consequently, by
Proposition \ref{prop:const}\; $Y_{\lambda, \mu}=c_{\lambda, \mu}$
for all $\omega \in \Omega_{\lambda, \mu}$ with $P(\Omega_{\lambda,
\mu})=1$.

Set
\begin{displaymath}
\Omega_0 = \bigcap_{\lambda, \mu \in \Q,\ \lambda \leq \mu}
\Omega_{\lambda, \mu} \ .
\end{displaymath}
Since $\Omega_0$ is a \emph{countable} intersection of sets of full
measure, it follows that $P(\Omega_0)=1$. Hence we can set
\begin{displaymath}
\Sigma = \{E\ |\ c_{\lambda, \mu} \not= 0 \textnormal{ for all }
\lambda < E < \mu, \quad \lambda, \mu \in \Q\} \ .
\end{displaymath}

To prove the assertions on $\sigma_{ac}$  we need that the
projection onto $\H_{ac}$, the absolutely continuous subspace with
respect to $H_\omega$ is measurable, the rest is as above. The same
is true for $\sigma_{sc}$ and $\sigma_{pp}$\,.

We omit the measurability proof and refer to  \cite{km2}  or
\cite{cl}.
\end{nproof}

Above we used the following results

\begin{lemma}\label{lem:UAU}
Let $A$ be a self adjoint operators and $U$ a unitary operator, then
for any bounded measurable function $f$ we have
\begin{equation}\label{eq:UAU}
 f\big(UAU^*\big)~=~U\,f(A)\,U^*\ .
\end{equation}
\end{lemma}

\begin{proof}
For resolvents, i.e. for $f_z(\lambda)=\frac{1}{\lambda-z}$ with
$z\in\C\setminus\R$ equation (\ref{eq:UAU}) can be checked directly.
Linear combinations of the $f_z$ are dense in
$C_\infty(\R)$,\index{$C_\infty(\R)$} the continuous functions
vanishing at infinity, by the Stone-Weierstraß theorem (see Section
\ref{sec:morefun}). Thus (\ref{eq:UAU}) is true for $f\in
C_\infty(\R)$.

If $\mu$ and $\nu$ are the projection valued measures for $A$ and
$B=UAU^*$ respectively, we have therefore for all $f\in
C_\infty(\R)$

\begin{align}\label{eq:UAUint}
\int f(\lambda)\;d\,{\nu}_{\varphi,\psi}(\lambda)~=&
\quad\langle \varphi, f(B) \,\psi \rangle \notag\\
=&\quad\langle \varphi, U f(A)\, U^* \psi \rangle\notag\\
=&\quad\langle U^*\varphi, f(A)\, U^*\psi \rangle\notag\\
~=&~\int f(\lambda)\;d\,{\mu}_{\,U^*\varphi,U^*\psi}(\lambda)
\end{align}
holds for all . Thus the measures ${\mu}_{\varphi,\psi}$ and
${\nu}_{U^*\varphi,U^*\psi}$ agree. Therefore (\ref{eq:UAUint})
holds for all bounded measurable $f$.
\end{proof}
$\;$\\[2mm]
{\bf Notes and Remarks }\\[2mm]
For further information see  \cite{cl}, \cite{ksonder}, \cite{km2},
\cite{km3},  \cite{kuso},
 \cite{p1} and \cite{pf}. An recent extensive review on ergodic operators can be found in \cite{Jitomir}.


\cleardoublepage


\section{The density of states \label{ch:dos}}

\subsection{Definition and existence}\label{sec:defdos}\es

Here, as in the rest of the paper we consider the Anderson model,
i.e.\linebreak\hbox{$H_\omega=H_0+V_\omega$} on $\ell^2(\Z^d)$ with
independent random variables $V_\omega(n)$ with a common
distribution $P_0$.

In this section we define a quantity of fundamental importance for
models in condensed matter physics: the density of states. The
density of states measure $\nu([E_1,E_2])$ gives the `number of
states per unit volume' with energy between $E_1$ and $E_2$. Since
the spectrum of our Hamiltonian $H_\omega$ is not discrete we can
not simply count eigenvalues within the interval $[E_1,E_2]$ or,
what is the same, take the dimension of the corresponding spectral
projection. In fact, the dimension of any spectral projection of
$H_\omega$ is either zero or infinite.
 Instead we restrict the spectral projection to the finite cube $\Lambda_L$ (see \ref{eq:cube})
 in $\Z^d$, take the dimension of its range and
divide by $|\,\Lambda_L|=(2L+1)^d$ \index{$"|\Lambda_L"|$} the
number of points in $\Lambda_L$. Finally, we send the parameter $L$
to infinity. This procedure is sometimes called the thermodynamic
limit.\index{thermodynamic limit}

For any bounded measurable function $\varphi$ on the real line we
define  the quantity
\begin{equation}\label{def:nul}
\mathdex{\nu_L}(\varphi)\;=\;\frac{1}{|\Lambda_L|}\,\tr\left(\,\chi_{\Lambda_L}\,\varphi(H_\omega)\,\chi_{\Lambda_L}\right)\;=
\;\frac{1}{|\Lambda_L|}\,\tr\left(\,\varphi(H_\omega)\,\chi_{\Lambda_L}\right)
\ .
\end{equation}

Here \mindex{$\chi_\Lambda$} denotes the \dindex{characteristic
function} of the set $\Lambda$, (i.e. $\chi_\Lambda(x)=1$ for
$x\in\Lambda$ and $=0$ otherwise). The operators $\varphi(H_\omega)$
are defined via the spectral theorem (see Section
\ref{sec:speccal}). In equation (\ref{def:nul}) we used the
cyclicity of the trace, (i.e.: $\tr(AB)=\tr(BA)$) and the fact that
${\chi_\Lambda}^2=\chi_\Lambda$.

 Since $\nu_L$ is a positive linear functional on the
bounded continuous functions, by Riesz representation theorem,
 it comes from a measure which we also
call $\nu_L$, i.e.

\begin{equation}
\nu_L(\varphi)= \;\int_\R \,\varphi(\lambda)\;d\nu_L(\lambda).
\end{equation}

We will show in the following that the measures $\nu_L$ converge to
a limit measure $\nu$ as $L\to\infty$ in the sense of vague
convergence of measures for $\P$-almost all $\omega$.

\begin{definition}\label{def:vagcon}
A series $\nu_n$ of Borel measures on $\R$ is said to \emph{converge
vaguely}\index{vague convergence} to a Borel measure $\nu$ if
\begin{equation}\notag
\int\,\varphi(x)\;d\,\nu_n(x) \to \int\,\varphi(x)\;d\,\nu(x)
\end{equation}
for all function $\varphi\in C_0(\mathbb{R})$,
\index{$C_0(\mathbb{R})$}the set of continuous functions with
compact support.
\end{definition}

We start with a proposition which establishes the almost sure
convergence of the integral of $\nu_L$ over a \emph{given} function.
\begin{proposition}\label{P1}
 If $\varphi$ is a bounded measurable function, then for $\P$-almost all $\omega$
 \begin{equation}\label{eq:b1}
    \lim_{L\rightarrow\infty}\frac{1}{|\Lambda_L|}\;\tr\big(\varphi(H_\omega)\,\chi_{\Lambda_L}\big)=
    \E\,\big(\,\langle\delta_0,\varphi(H_\omega)\delta_0\rangle\,\big)\quad \tn{}.
 \end{equation}
\end{proposition}

\begin{remark} The right hand side of (\ref{eq:b1}) defines a positive measure $\nu$ by
\begin{displaymath}
    \int{\varphi(\lambda)\,d\nu(\lambda)}=\E(\langle\delta_0,\varphi(H_\omega)\delta_0\rangle) \ .
\end{displaymath}
\end{remark}
This measure satisfies $\nu(\R)=1$, hence it is a probability measure (just insert $\varphi(\lambda)\equiv 1$).\\
\begin{definition}
The measure $\nu$, defined by
\begin{equation}
\nu(A)=\E\,\big(\langle\delta_0,\chi_A(H_\omega)\,\delta_0\rangle\big)
\qquad\textnormal{for $A$ a Borel set in $\R$}
\end{equation}
is called the \emph{\mindex{density of states measure}}.

The distribution function $N$ of  $\nu$, defined by
\begin{equation}
    \mathdex{N(E)}=\nu\big((-\infty,E]\big)
\end{equation}
is known as the \emph{\mindex{integrated density of states}}.
\end{definition}

\begin{nproof}{Proposition}
\begin{eqnarray}
    &&\frac{1}{|\Lambda_L|}\;\tr\,(\varphi(H_\omega)\chi_{\Lambda_L})\nonumber \\
    &=&\frac{1}{(2L+1)^d}\sum_{i\in\Lambda_L}\langle\delta_i,\varphi(H_\omega)\delta_i\rangle
\end{eqnarray}
The random variables
$X_i=\langle\delta_i,\varphi(H_\omega)\delta_i\rangle$ form an
ergodic stochastic process since the shift operators $\{T_i\}$ are
ergodic and since
\begin{eqnarray}
    X_i(T_j\omega)&=&\langle\delta_i,\varphi(H_{T_j\omega})\,\delta_i\rangle\nonumber\\
        &=&\langle\delta_i,U_j\,\varphi(H_\omega)\,U_j^*\,\delta_i\rangle\nonumber\\
        &=&\langle U_j^*\,\delta_i,\varphi(H_\omega)\,U_j^*\,\delta_i\rangle\nonumber\\
        &=&\langle\delta_{i-j},\varphi(H_\omega)\,\delta_{i-j}\rangle\nonumber\\&=&X_{i-j}(\omega) \ .
\end{eqnarray}
We used that $U_j^*\delta_i(n) = \delta_i(n+j) = \delta_{i-j}(n)$.

Since $|X_i|\leq ||\varphi||_\infty$, the $X_i$ are integrable (with
respect to $\P$). Thus we may apply the ergodic theorem
(\ref{th:Birkhoff}) to obtain
\begin{align}
    &\frac{1}{|\Lambda_L|}\tr(\varphi(H_\omega)\chi_{\Lambda_L})\;
    =\;\frac{1}{(2L+1)^d}\sum_{i\in\Lambda_L}\,X_i\;\\[4mm]
    \;\longrightarrow\quad&
    \E(X_0)\;=\;\E(\langle\delta_0,\varphi(H_\omega)\delta_0\rangle) \ .
\end{align}

\end{nproof}

We have proven that (\ref{eq:b1}) holds for \emph{fixed} $\varphi$
on a set of full probability. This set, let's call it
$\Omega_\varphi$, may (and will) depend on $\varphi$. We can
conclude that (\ref{eq:b1}) holds for \emph{all} $\varphi$ for
$\omega\in\bigcap_{\varphi}\Omega_\varphi$. However, this is an
\emph{uncountable} intersection of sets of probability one. We do
\emph{not} know whether this intersection has full measure, in fact
we even don't know whether this set is measurable.

\begin{theorem}\label{T1}
 The measures $\nu_L$ converge vaguely to the measure $\nu$ $\P$-almost surely, i.e. there is a set \,$\Omega_0$\, of
 probability one, such that
 \begin{equation}\label{eq:b2}
  \int\varphi(\lambda)\,d\nu_L(\lambda)\rightarrow \int\varphi(\lambda)\,d\nu(\lambda)
 \end{equation}
 for all $\varphi\in C_0(\R)$ and all $\omega\in\Omega_0$ \ .
 \end{theorem}

 \begin{remark}The measure $\nu$ is \emph{non random} by definition.\end{remark}
\begin{proof}
 Take a countable dense set $D_0$ in $C_0(\R)$ in the uniform topology.
 With $\Omega_\varphi$ being the set of full measure for which
 (\ref{eq:b2}) holds, we set
 \begin{displaymath}
  \Omega_0=\bigcap_{\varphi\in D_0}\Omega_\varphi \ .
 \end{displaymath}
 Since $\Omega_0$ is a \emph{countable} intersection of sets of full measure, $\Omega_0$ has probability one.

 For $\omega\in\Omega_0$ the convergence (\ref{eq:b2}) holds for all $\varphi\in D_0$.

 By assumption on $D_0$, if $\varphi\in C_0(\R)$ there is a sequence $\varphi_n\in D_0$ with
 $\varphi_n\rightarrow\varphi$ uniformly. It follows
\begin{eqnarray}
 &&|\int\varphi(\lambda)\,d\nu(\lambda)-\int\varphi(\lambda)\,d\nu_L(\lambda)|\nonumber\\
 \leq&&|\int\varphi(\lambda)\,d\nu(\lambda)-\int\varphi_n(\lambda)\,d\nu(\lambda)|\nonumber\\
 &+&|\int\varphi_n(\lambda)\,d\nu(\lambda)-\int\varphi_n(\lambda)\,d\nu_L(\lambda)|\nonumber\\
 &+&|\int\varphi_n(\lambda)\,d\nu_L(\lambda)-\int\varphi(\lambda)\,d\nu_L(\lambda)|\nonumber\\[2mm]
 \leq&&||\,\varphi-\varphi_n||_\infty\cdot\nu(\R)\;+\;||\,\varphi-\varphi_n||_\infty\cdot\nu_L(\R)\nonumber\\
 &+&|\int\varphi_n(\lambda)\,d\nu(\lambda)-\int\varphi_n(\lambda)\,d\nu_L(\lambda)| \ .
\end{eqnarray}
Since both $\nu(\R)$ and $\nu_L(\R)$ are bounded by $1$ (in fact are
equal to one) the first two terms can be made small by taking $n$
large enough. We make the third term small by taking $L$ large.
\end{proof}
\goodbreak
\begin{remarks}\noindent
\begin{enumerate}
\item
As we remarked already in the above proof both $\nu_L$ and $\nu$ are
probability measures. Consequently, the measures $\nu_L$ converge
even weakly\index{weak convergence} to $\nu$, i.~e. when integrated
against a bounded continuous function (see e.~g. \cite{Bau}).
Observe that the space of bounded continuous functions
$C_b(\mathbb{R})$ \index{$C_b(\mathbb{R})$} does \emph{not} contain
a countable dense set, so the above \emph{proof} does not work for
$C_b$ directly.
\item
In the continuous case the density of states measure is unbounded,
even for the free Hamiltonian. So, in the continuous case, it does
not make sense even to talk about weak convergence, we have to
restrict ourselves to vague convergence in this case.

\item
Given a countable set $D$ of bounded measurable functions we can
find a set $\Omega_1$ of probability one such that
\begin{displaymath}
 \int\varphi(\lambda)d\nu_L(\lambda)\rightarrow\int\varphi(\lambda)d\nu(\lambda)
\end{displaymath}
for all $\varphi\in D \cup C_b(\R)$ and all $\omega\in\Omega_1$.
\end{enumerate}
\end{remarks}
\goodbreak

\begin{corollary}\label{cor:limN}
For\, $\P$-almost all $\omega$ the following is true:\\
For all $E\in\R$

\begin{equation}
N(E)=\lim_{L\rightarrow\infty}\;\nu_L\big((-\infty,E]\big)
\label{eq:limN} \ .
\end{equation}

\end{corollary}

\begin{remarks}\noindent
It is an immediate consequence of Proposition \ref{P1} that for
\emph{fixed} $E$ the convergence in (\ref{eq:limN}) holds for almost
all $\omega$, with the set of exceptional $\omega$ being
$E$-dependent. The statement of Corollary \ref{cor:limN} is
stronger: It claims the existence of an $E$-\emph{independent} set
of $\omega$ such that \ref{eq:limN} is true for all $E$.
\end{remarks}

\begin{proof}
We will prove (\ref{eq:limN}) first for energies $E$ where $N$ is
continuous.

Since $N$ is monotone increasing the set of \emph{discontinuity}
points of $N$ is at most countable (see Lemma \ref{lem:dp} below).
Consequently, there is a countable set $S$ of continuity points of
$N$ which is dense in $\R$. By Proposition \ref{P1} there is a set
of full $\P$-measure such that
$$ \int\;\chi_{(-\infty,E]}(\lambda)\,d\nu_L(\lambda)\;\to\; N(E)$$
for all $E\in S$.

Take $\varepsilon>0$. Suppose $E$ is an arbitrary continuity point
of $N$. Then, we find $E_+, E_- \in S$ with $E_-\le E\le E_+$ such
that $N(E_+) - N(E_-)<\frac{\varepsilon}{2}$.

We estimate ($N$ is monotone increasing)
\begin{eqnarray}
\lefteqn{N(E) - \int\;\chi_{(-\infty,E]}(\lambda)\,d\nu_L(\lambda) }\\
&\le& N(E_+)
- \int\;\chi_{(-\infty,E_-]}(\lambda)\,d\nu_L(\lambda) \\
  &\le& N(E_+) - N(E_-) +\big|{N(E_-) - \int\;\chi_{(-\infty,E_-]}(\lambda)\,d\nu_L(\lambda)}\big| \\
  &\le& \varepsilon
  \end{eqnarray}
for $L$ large enough.

Analogously we get
\begin{eqnarray}
\lefteqn{N(E) - \int\;\chi_{(-\infty,E]}(\lambda)\,d\nu_L(\lambda) }\\
  &\ge& N(E_-) - N(E_+) - \big|{N(E_+) - \int\;\chi_{(-\infty,E_+]}(\lambda)\,d\nu_L(\lambda)}\big| \\
  &\ge& -\varepsilon \ .
  \end{eqnarray}
Hence
\begin{equation*}
\big|\,{N(E) - \int\;\chi_{(-\infty,E]}(\lambda)\,d\nu_L(\lambda)
}\;\big|\to 0
\end{equation*}
This proves (\ref{eq:limN}) for continuity points. Since there are
at most countably many  points of discontinuity for $N$ another
application of Proposition \ref{P1} proves the result for all $E$.
\end{proof}
Above we used the following Lemma.
\begin{lemma}\label{lem:dp}
If the function $F:\R \rightarrow \R$ is monotone increasing then
$F$ has at most countably many
 points of discontinuity.
\end{lemma}
\begin{proof}
Since $F$ is monotone both $F(t-)=\lim_{s\nearrow t}F(s)$ and
$F(t+)=\lim_{s\searrow t}F(s)$ exist. If $F$ is discontinuous at
$t\in\R$ then $F(t+)-F(t-)~>0$. Set
$$D_n=\{t\in\R\; |\; F(t+)-F(t-)>\frac{1}{n}\}$$
then the set $D$ of discontinuity points of $F$ is given by
$\bigcup_{n\in\N} D_n$.

Let us assume that $D$ is uncountable. Then also one of the $D_n$
must be uncountable.

Since $F$ is monotone and defined on all of $\R$ it must be bounded
on any bounded interval. Thus we conclude that $D_n\cap [-M,M]$ is
finite for any $M$. It follows that $D_n=\bigcup_{M\in\N}
\big(D_n\cap [-M,M]\big)$ is countable. This is a contradiction to
the conclusion above.
\end{proof}
\begin{remark}\quad\\
The proof of Corollary \ref{cor:limN} shows that we also have
\begin{eqnarray}
N(E-)\;&=&\;\sup_{\varepsilon>0}\,N(E-\varepsilon)\notag\\
&=&\;\int\,\chi_{(-\infty, E)}(\lambda)\,d\nu(\lambda)\notag\\
&=&\; \lim_{L\to\infty}\;\int\,\chi_{(-\infty,
E)}(\lambda)\,d\nu_L(\lambda)
\end{eqnarray}
for \emph{all} $E$ and $\P$-almost all $\omega$ (with an
$E$-independent set of $\omega$).

Consequently, we also have
$\nu(\{E\})=\lim_{L\to\infty}\,\nu_L(\{E\})$.
\end{remark}
\vspace{1mm}
\begin{proposition}
 $\supp(\nu)=\Sigma\quad(=\sigma(H_\omega))$.
\end{proposition}
\begin{proof}
 If $\lambda\notin\Sigma$ then there is an $\epsilon > 0$ such that
 $\chi_{(\lambda-\epsilon,\lambda+\epsilon)}(H_\omega)=0$ \\
 $\P$-almost surely, hence
 \begin{displaymath}
   \nu\big((\lambda-\epsilon,\lambda+\epsilon)\big)=E\big(\chi_{(\lambda-\epsilon,\lambda+\epsilon)}(H_\omega)(0,0)\big)=0 \ .
 \end{displaymath}
If $\lambda\in\Sigma\;$ then
$\;\chi_{(\lambda-\epsilon,\lambda+\epsilon)}(H_\omega)\not= 0\;$
$\P$-almost surely for any $\epsilon >0$.

Since $\chi_{(\lambda-\epsilon,\lambda+\epsilon)}(H_\omega)$ is a
projection, it follows that for some $j\in\Z^d$
\begin{eqnarray}
  0 &\not=&\; \E\big(\chi_{(\lambda-\epsilon,\lambda+\epsilon)}(H_\omega)(j,j)\big)\nonumber\\
    &=&\;
    \E\big(\chi_{(\lambda-\epsilon,\lambda+\epsilon)}(H_\omega)(0,0)\big)\notag\\
    &=&\;\nu\big((\lambda-\epsilon,\lambda+\epsilon)\big)
    \ .
\end{eqnarray}

Here, we used that by Lemma \ref{lem:UAU}
\begin{displaymath}
  f(H_\omega)(j,j)=f(H_{T_j\,\omega})(0,0)
\end{displaymath}
and the assumption that $T_j$ is measure preserving.
\end{proof}

It is not hard to see that the integrated density of states
$N(\lambda)$ is a continuous function, which is equivalent to the
assertion that $\nu$ has no atoms, i.e. $\nu(\{\lambda\})=0$ for all
$\lambda$. We note, that an analogous result for the continuous case
(i.e. Schr\"odinger operators on $L^2(\R^d)$) is unknown in this
generality.

We first state
\begin{lemma}\label{L1}
 Let $\mathcal{V}_\lambda$ be the eigenspace of $H_\omega$ with respect to the eigenvalue $\lambda$ then
 $\textnormal{dim}\,(\,\chi_{\Lambda_L}(\mathcal{V}_\lambda))\leq C L^{d-1}.$
\end{lemma}

From this we deduce

\begin{theorem}\label{T2}
 For any $\lambda\in\R\qquad\nu(\{\lambda\})=0$.
\end{theorem}

\begin{nproof}{of the Theorem assuming the Lemma}\\
 By Proposition~\ref{P1} and Theorem~\ref{T1} we have
 \begin{equation}\label{eq:*}
 \nu(\{\lambda\})=\lim_{L\rightarrow\infty}\frac{1}{(2L+1)^d}\,\tr\,(\,\chi_{\Lambda_L}\;\chi_{\{\lambda\}}(H_\omega)).
 \end{equation}
 If $f_i$ is an orthonormal basis of $\chi_{\Lambda_L}(\mathcal{V}_\lambda)$ and $g_j$ an orthonormal basis of
 $\chi_{\Lambda_L}(\mathcal{V}_\lambda)^\bot$ we have,
 noting that $\chi_{\Lambda_L}(\mathcal{V}_\lambda)$ is finite dimensional,
\begin{eqnarray}
  && \tr\,\big(\,\chi_{\Lambda_L}\,\chi_{\{\lambda\}}(H_\omega)\,\big)\nonumber\\
  &=& \sum_{i}\;\langle f_i,\chi_{\Lambda_L}\,\chi_{\{\lambda\}}(H_\omega)f_i\rangle
  + \sum_{j}\;\langle g_j,\chi_{\Lambda_L}\,\chi_{\{\lambda\}}(H_\omega)g_j\rangle \nonumber\\
  &=& \sum_{i}\;\langle f_i,\chi_{\Lambda_L}\,\chi_{\{\lambda\}}(H_\omega)f_i\rangle\nonumber\\
  &\leq& \dim\,\chi_{\Lambda_L}(\mathcal{V}_\lambda) \leq C L^{d-1}
\end{eqnarray}
hence (\ref{eq:*}) converges to zero. Thus
$\nu\big(\{\lambda\}\big)=0$.
\end{nproof}

\begin{nproof}{Lemma}\\
We define
$\widetilde{\Lambda}_L=\{i\in\Lambda_L|\;\, (L-1)\le \inorm{i} \le \,L \;\}$\\
$\widetilde{\Lambda}_L$ consists of the two outermost layers of
$\Lambda_L$.

The values $u(n)$ of an eigenfunction $u$ of $H_\omega$ with
$H_\omega u=\lambda u$ can be computed from the eigenvalue equation
for all $n\in\LL$ once we know its values on
$\widetilde{\Lambda}_L$. So, the dimension of
$\chi_{\Lambda_L}(\mathcal{V}_\lambda)$ is at most the number of
points in $\widetilde{\Lambda}_L$.
\end{nproof}


\subsection{Boundary conditions}\es

Boundary conditions are used to define differential operators on
sets $M$ with a boundary. A rigorous treatment of boundary
conditions for differential operators is most conveniently based on
a quadratic form approach (see \cite{rs4}) and is out of the scope
of this review. Roughly speaking boundary conditions restrict the
domain of a differential operator $D$ by requiring that functions in
the domain of $D$ have a certain behavior at the boundary of $M$. In
particular, Dirichlet boundary conditions force the functions $f$ in
the domain to vanish at $\partial M$. Neumann boundary conditions
require the normal derivative to vanish at the boundary. Let us
denote by $-\Delta_M^D$ and $-\Delta_M^N$ the Laplacian on $M$ with
Dirichlet and Neumann boundary condition respectively.

The definition of boundary conditions for the discrete case are
somewhat easier then in the continuous case. However, they are
presumably less familiar to the reader and may look somewhat
technical at a first glance. The reader might therefore skip the
details for the first reading and concentrate of `simple' boundary
conditions defined below. Neumann and Dirichlet boundary conditions
will be needed for this text only in chapter \ref{ch:Lifshitz} in
the proof of Lifshitz tails.

For our purpose the most important feature of Neumann and Dirichlet
boundary conditions is the so called \mindex{Dirichlet-Neumann
bracketing}. Suppose $M_1$ and $M_2$ are disjoint open sets in
$\R^d$ and $M=(\overline{M_1\cup M_2}\,)^{\,\circ}$, ($^\circ$
denoting the interior) then
\begin{eqnarray}\label{NDbrackC}
-\Delta_{M_1}^N \oplus -\Delta_{M_2}^N \leq -\Delta_{M}^N \leq
-\Delta_{M}^D \leq -\Delta_{M_1}^D \oplus -\Delta_{M_2}^D \ .
\end{eqnarray}

in the sense of quadratic forms. In particular the eigenvalues of
the operators in (\ref{NDbrackC}) are increasing from left to right.

We recall that a bounded operator $A$ on a Hilbert space
$\mathcal{H}$ is called \emph{positive}\index{positive operator} (or
positive definite or $A\,\geq\,0$) if

\begin{equation}\label{eq:positiveOp}
\langle\varphi,A\,\varphi\rangle~\ge~0\quad\quad \textnormal{for all
}\varphi\in\mathcal{H} \ .
\end{equation}
For unbounded $A$ the validity of equation \ref{eq:positiveOp} is
required for the (form-)domain of $A$ only.

By \mindex{$A\,\leq\,B$} for two operators $A$ and $B$ we mean $B-A\,\geq\,0$.\\[1.5mm]

For the lattice case we introduce boundary conditions which we call
Dirichlet and Neumann conditions as well. Our choice is guided by
the chain of inequalities(\ref{NDbrackC}).

The easiest version of boundary conditions for the lattice is given
by the following procedure

\begin{definition}\label{def:simpleBC}
The Laplacian with \emph{\mindex{simple boundary conditions}} on
$\Lambda \subset \mathbb{Z}^d$ is the operator on $\ell^2(\Lambda)$
defined by
\begin{equation}\label{simp}
(H_0)_\Lambda(n,m)= \langle \delta_n ,H_0\, \delta_m \rangle
\end{equation}
whenever both $n$ and $m$ belong to $\Lambda$. We also set
$H_\L=(H_0)_\L+V$\index{$H_\L$}.

In particular, if $\Lambda$ is finite, the operator $H_{\Lambda}$
acts on a finite dimensional space, i.e. is a matrix.
\end{definition}

We are going to use simple boundary conditions frequently in this
work. At a first glance simple boundary conditions seem to be a
reasonable analog of Dirichlet boundary conditions. However, they do
not satisfy (\ref{NDbrackC}) as we will see later. Thus, we will
have to search for other boundary conditions.

Let us define

\begin{eqnarray}\nonumber
\partial \Lambda &=&\big\{(n,m) \in \mathbb{Z}^d \times
\mathbb{Z}^d\;\; \big| \  \ \gnorm{n-m} =1 \ \textnormal{and}\\
&&\quad\textnormal{either}\ n \in \Lambda ,m \not \in \Lambda \
\textnormal{or} \ n \not \in \Lambda ,m \in \Lambda \big\} \ .
\end{eqnarray}
\index{$\partial \Lambda$}

 The set $\partial \Lambda$ is the
\dindex{boundary} of $\Lambda$. It consists of the \emph{edges}
connecting points in $\Lambda$ with points outside $\Lambda$.
 We also define the \dindex{inner boundary} of $\Lambda$ by

\begin{equation}\label{eq:inbound}
\partial^- \Lambda
 = \big\{\,n \in \mathbb{Z}^d\; \big| \ n \in \Lambda, \ \exists \ m \not \in
\Lambda \ \ (n,m) \in  \partial \Lambda \big\}
\end{equation}
\index{$\partial^- \Lambda$}

 and the \dindex{outer boundary} by

\begin{equation}
\partial^+ \Lambda
 = \big\{\,m \in \mathbb{Z}^d\; \big| \ m \not \in \Lambda, \ \exists \ n \in
\Lambda \ \ (n,m) \in  \partial \Lambda \big\} \ .
\end{equation}
\index{$\partial^+ \Lambda$}

Hence $\partial^+ \Lambda=\partial^- (\complement \Lambda)$ and the
boundary $\partial\Lambda$ consists of edges between
$\partial^-\Lambda$ and $\partial^+\Lambda$.

For any set $\Lambda$ we define the boundary operator
\mindex{$\Gamma_\Lambda$} by

\begin{equation} \label{def:Gamma}
\Gamma_\Lambda(n,m) = \left\{
\begin{array}{rl} -1 & \textnormal{ if}\ (n,m) \in
\partial \Lambda , \\ 0 & \textnormal{ otherwise.}\end{array}\right.
\end{equation}

Thus for the Hamilitonian $H=H_0 +V$ we have the important relation

\begin{eqnarray}\label{eq:splitH}
H=H_\Lambda \oplus H_{\complement \Lambda} + \Gamma_\Lambda \ .
\end{eqnarray}

In this equation we identified $\ell^2(\mathbb{Z}^d)$ with
$\ell^2(\Lambda) \oplus \ell^2(\complement \Lambda)$.
\newline More precisely

\begin{equation}
(H_\Lambda \oplus H_{\complement \Lambda})(n,m)=\left\{
\begin{array}{ll} H_\Lambda(n,m) & \textnormal{if } n,m \in\Lambda,
\\ H_{\complement \Lambda}(n,m) & \textnormal{if } n,m \not\in\Lambda, \\ 0 & \textnormal{otherwise.}
\end{array}\right.
\end{equation}

In other words $H_\Lambda \oplus H_{\complement \Lambda}$ is a block
diagonal matrix and $\Gamma_\Lambda$ is the part of $H$ which
connects these blocks.

It is easy to see, that $\Gamma_\Lambda$ is neither negative nor
positive definite. Consequently, the operator $H_\Lambda$ will not
satisfy any inequality of the type (\ref{NDbrackC}).

To obtain analogs to Dirichlet and Neumann boundary conditions we
should substitute the operator $\Gamma_\Lambda$ in (\ref{eq:splitH})
by a negative definite resp. positive definite operator and
$H_\Lambda \oplus H_{\complement\Lambda}$ by an appropriate block
diagonal matrix.

For the operator $H_0$ the diagonal term $H_0(i,i)=2d$ gives the
number of sites $j$ to which $i$ is connected (namely the $2d$
neighbors in $\mathbb{Z}^d$). This number is called the
\dindex{coordination number} of the graph $\Z^d$. In the matrix
$H_\Lambda$ the edges to $\complement\Lambda$ are removed but the
diagonal still contains the `old' number of adjacent edges. Let us
set $n_\Lambda(i)= |\, \{ j \in\Lambda |\; \gnorm{j-i}=1
\}|$\index{$n_\Lambda(i)$} to be the number of sites adjacent to $i$
{\em in} $\Lambda$, the \dindex{coordination number} for the graph
$\Lambda$. $n_\Lambda(i)=2d$ as long as $i \in \Lambda \backslash
\partial^-\Lambda$ but $n_\Lambda(i)<2d$ at the boundary.  We also
define the \emph{\mindex{adjacency matrix}} on $\Lambda$ by

\begin{equation}
A_\Lambda(i,j)=\left\{
\begin{array}{rl} -1 &\textnormal{ if } i,j\in\Lambda,\  \gnorm{i-j}=1
\\ 0 & \textnormal{ otherwise.}
\end{array} \right.
\end{equation}
\index{$A_\Lambda(i,j)$}

The operator $(H_0)_\Lambda$ on $\ell^2(\Lambda)$ is given by
\begin{equation}
(H_0)_\Lambda\quad=\quad 2d \; + \; A_\Lambda
\end{equation}
where $2d$ denotes a multiple of the identity.
\begin{definition}\label{def:Neum}
The \emph{\mindex{Neumann Laplacian}} on $\Lambda \subset
\mathbb{Z}^d$ is the operator on $\ell^2(\Lambda)$ defined by
\begin{equation}
(H_0)^N_\Lambda \quad=\quad n_\Lambda\; +\; A_\Lambda \ .
\end{equation}
\end{definition}
Above $n_\Lambda$ stands for the multiplication operator with the
function $n_\Lambda(i)$ on $\ell^2(\Lambda)$.

\goodbreak
\begin{remark}\noindent
\begin{enumerate}
\item In $(H_0)_\Lambda$ the off diagonal term `connecting'
$\Lambda$ to $\Z^d\backslash\Lambda$ are removed. However, through
the diagonal term $2d$ the operator still `remembers' there were
$2d$ neighbors originally.
\item
The Neumann Laplacian ${H_0}^N$ on $\Lambda$ is also called the
\dindex{graph Laplacian}. It is the canonical and intrinsic
Laplacian with respect to the graph structure of $\Lambda$. It
`forgets' completely that the set $\Lambda$ is imbedded in $\Z^d$.
\item
The quadratic form corresponding to $(H_0)^N_\Lambda$ is given by
\begin{displaymath}
\langle u, (H_0)_\Lambda^N v \rangle = \frac{1}{2} \sum_{n,m
\in\Lambda \atop \gnorm{n-m} =1} \overline{(u(n)-u(m))}(v(n)-v(m)) \
.
\end{displaymath}
\end{enumerate}
\end{remark}

\begin{definition}\label{def:Diri}
The \emph{\mindex{Dirichlet Laplacian}} on $\Lambda$ is the operator
on $\ell^2(\Lambda)$ defined by
\begin{displaymath}
(H_0)_\Lambda^D = 2d + (2d-n_\Lambda) + A_\Lambda \ .
\end{displaymath}
\end{definition}

\begin{remark}\indent
\begin{enumerate}
\item The definition of the Dirichlet Laplacian may look a bit strange at the first
glance. The main motivation for this definition is to preserve the
properties (\ref{NDbrackC}) of the continuous analog.
\item The Dirichlet Laplacian not only remembers that there were $2d$ neighboring
sites before introducing boundary conditions, it even increases the
diagonal entry by one for each adjacent edge which was removed. Very
loosely speaking, one might say that the points at the boundary get
an additional connection for every `missing' link to points outside
$\Lambda$.
\end{enumerate}
\end{remark}

It is not hard to see, that
\begin{eqnarray}\label{NDbrack}
(H_0)_\Lambda^N \leq (H_0)_\Lambda \leq (H_0)_\Lambda^D
\end{eqnarray}

and
\begin{eqnarray}\label{eq:splitHN}
H_0 & = &(H_0)_\Lambda^N \oplus (H_0)_{\complement \Lambda}^N +
\Gamma_\Lambda^N
\\  & = & (H_0)_\Lambda^D \oplus (H_0)_{\complement \Lambda}^D +
\Gamma_\Lambda^D\label{eq:splitHD}
\end{eqnarray}

with
\begin{equation} \mathdex{\Gamma_\Lambda^N} (i,j)=\left\{
\begin{array}{ll} 2d-n_\Lambda(i) & \textnormal{if } i=j, i
\in\Lambda,
\\ 2d-n_{ \complement\Lambda}(i) & \textnormal{if } i=j, i
\in\complement\Lambda,
\\ -1 & \textnormal{if } (i,j) \in\partial\Lambda,
\\ 0 & \textnormal{otherwise} \end{array} \right.
\end{equation}

and
\begin{equation}\mathdex{\Gamma_\Lambda^D} (i,j)=\left\{
\begin{array}{ll} n_\Lambda(i)-2d & \textnormal{if } i=j, i
\in\Lambda,
\\ n_{ \complement\Lambda}(i)-2d & \textnormal{if } i=j, i
\in\complement\Lambda,
\\ -1 & \textnormal{if } (i,j) \in\partial\Lambda,
\\ 0 & \textnormal{otherwise.} \end{array} \right.
\end{equation}

The operator $\Gamma_\Lambda^N$ is positive definite as

\begin{displaymath}
\langle u, \Gamma_\Lambda^N v \rangle = \frac{1}{2}
\sum_{(i,j)\in\partial\Lambda}
\overline{\big(u(i)-u(j)\big)}\big(v(i)-v(j)\big)
\end{displaymath}

is its quadratic form. In a similar way, we see that
$\Gamma_\Lambda^D$ is negative definite, since

\begin{displaymath}
\langle u, \Gamma_\Lambda^D v \rangle = - \frac{1}{2}
\sum_{(i,j)\in\partial\Lambda}
\overline{\big(u(i)+u(j)\big)}\big(v(i)+v(j)\big)
\end{displaymath}

 Hence we have in analogy to (\ref{NDbrackC})
\begin{equation}\label{NDbrackdis1}
(H_0)_\Lambda^N \oplus (H_0)_{\complement \Lambda}^N \leq H_0 \leq
(H_0)_\Lambda^D \oplus (H_0)_{\complement \Lambda}^D \ .
\end{equation}

If $H=H_0+V$ we define $H_\Lambda=(H_0)_\Lambda +V$, where in the
latter expression $V$ stands for the multiplication with the
function $V$ restricted to $\Lambda$. Similarly,
$H_\Lambda^N=(H_0)_\Lambda^N +V$\index{$H_\Lambda^N$} and
$H_\Lambda^D=(H_0)_\Lambda^D +V$\index{$H_\Lambda^D$}.

These operators satisfy
\begin{equation}\label{NDbrackdis2}
H_\Lambda^N \oplus H_{\complement \Lambda}^N \leq H \leq H_\Lambda^D
\oplus H_{\complement \Lambda}^D \ .
\end{equation}
\vspace{2mm}

For $\Lambda_1 \subset \Lambda \subset \mathbb{Z}^d$ we have analogs
of the `splitting' formulae (\ref{eq:splitH}), (\ref{eq:splitHN})
and (\ref{eq:splitHD}). To formulate them it will be useful to
define the `relative' boundary
\mindex{$\partial_{\Lambda_2}\Lambda_1$} of $\L_1\subset\L_2$ in
$\L_2$.
\begin{align}
&\partial_{\Lambda_2}\Lambda_1~=~\partial\Lambda_1\cap(\Lambda_2\times\Lambda_2)~
=~\partial\Lambda_1\setminus\partial\Lambda_2\\\notag
&=~\big\{\,(i,j)\,\big|\,\gnorm{i-j}=1 \textnormal{ and }\,i\in\L_1,
j\in\L_2\setminus\L_1 \textnormal{ or }i\in\L_2\setminus\L_1,
j\in\L_1\,\big\}
\end{align}

The analogs of the splitting formulae are

\begin{align}
H_{\L_2}~=&~H_{\L_1} \oplus H_{\L_2\setminus\L_1}\;+ \Gamma_{\L_1}^{\L_2}\label{eq:splitHL}\\
H^N_{\L_2}~=&~H^N_{\L_1} \oplus H^N_{\L_2\setminus\L_1}\;+ {\Gamma_{\L_1}^{\L_2\,N}}\label{eq:splitHLN}\\
H^D_{\L_2}~=&~H^D_{\L_1} \oplus H^D_{\L_2\setminus\L_1}\;+
{\Gamma_{\L_1}^{\L_2\,D}}\label{eq:splitHLD}
\end{align}
with

\begin{align} \label{def:GammaL}
\Gamma^{\L_2}_{\L_1}(i,j) =& \left\{ \begin{array}{rl}
\hspace{29mm}0 & \tn{if } i=j \textnormal{ and } i\in\L_1\\
0 & \tn{if } i=j \textnormal{ and } i\in\L_2\setminus\L_1\\
-1 & \textnormal{if } (i,j)\in \partial_{\Lambda_2}\Lambda_1\\
0 & \textnormal{otherwise.}
\end{array}\right.\\
\Gamma^{\L_2\,N}_{\L_1}(i,j) =& \left\{ \begin{array}{rl}
n_{\L_2}(i)-n_{\L_1}(i) & \tn{if } i=j \textnormal{ and } i\in\L_1\\
n_{\L_2}(i)-n_{\L_2\setminus\L_1}(i) & \tn{if } i=j \textnormal{ and } i\in\L_2\setminus\L_1\\
-1 & \textnormal{if } (i,j)\in \partial_{\Lambda_2}\Lambda_1\\
0 & \textnormal{otherwise.}
\end{array}\right.\\
\Gamma^{\L_2\,D}_{\L_1}(i,j) =& \left\{ \begin{array}{rl}
n_{\L_1}(i)-n_{\L_2}(i) & \tn{if } i=j \textnormal{ and } i\in\L_1\\
n_{\L_2\setminus\L_1}(i)-n_{\L_2}(i) & \tn{if } i=j \textnormal{ and } i\in\L_2\setminus\L_1\\
-1 & \textnormal{if } (i,j)\in \partial_{\Lambda_2}\Lambda_1\\
0 & \textnormal{otherwise.}
\end{array}\right.
\end{align}

In particular, for $\Lambda=\Lambda_1\cup\Lambda_2$ with disjoint
sets $\Lambda_1$ and $\Lambda_2$ we have

\begin{equation}\label{NDbrackdis}
H_{\Lambda_1}^N \oplus H_{\Lambda_2}^N~\leq~
H_\Lambda^N~\leq~H_\Lambda^D~\leq ~H_{\Lambda_1}^D \oplus
H_{\Lambda_2}^D  .
\end{equation}

since $ \Gamma^{\L\,N}_{\L_1}\geq 0 $ and $  \Gamma^{\L\,D}_{\L_1}
\leq 0  $.


\subsection{The geometric resolvent equation}\es

The equations (\ref{eq:splitHL}), (\ref{eq:splitHLN})and
(\ref{eq:splitHLD}) allow us to prove the so called geometric
resolvent equation. It expresses the resolvent of an operator on a
larger set in terms of operators on smaller sets. This equality is a
central tool of multiscale analysis.

We do the calculations for simple boundary conditions (\ref{simp})
but the results are valid for Neumann and Dirichlet boundary
conditions with the obvious changes.

We start from equation (\ref{eq:splitHL}) for
$\Lambda_1\subset\Lambda_2\subset\Z^d$.

For $z\in\C\setminus\R$ this equation and the resolvent equation
(\ref{eq:res}) imply
\begin{align}\nonumber
&(H_{\Lambda_2}-z)^{-1}\\ \nonumber =~~&(H_{\Lambda_1}\oplus
H_{\Lambda_2\backslash\Lambda_1}-z)^{-1} - (H_{\Lambda_1}\oplus
H_{\Lambda_2\backslash\Lambda_1}-z)^{-1}\Gamma_{\Lambda_1}^{\Lambda_2}(H_{\Lambda_2}-z)^{-1}\\
\nonumber=~~ &(H_{\Lambda_1}\oplus
H_{\Lambda_2\backslash\Lambda_1}-z)^{-1}
-(H_{\Lambda_2}-z)^{-1}\Gamma_{\Lambda_1}^{\Lambda_2}(H_{\Lambda_1}\oplus
H_{\Lambda_2\backslash\Lambda_1}-z)^{-1}\ .\label{res} \\
\end{align}

In fact, (\ref{res}) holds for $z\not\in
\sigma(H_{\Lambda_1})\cup\,\sigma(H_{\Lambda_2})\cup\,\sigma(H_{\Lambda_2\setminus\L_1})$.

For $n\in\Lambda_1$, $m\in\Lambda_2\backslash\Lambda_1$ we have

\begin{displaymath}
H_{\Lambda_1}\oplus H_{\Lambda_2\backslash\Lambda_1}(n,m)=0
\end{displaymath}

hence

\begin{displaymath}
(H_{\Lambda_1}\oplus H_{\Lambda_2\backslash\Lambda_1}-z)^{-1}(n,m)=0
\ .
\end{displaymath}

Note that $(H_{\Lambda_1}\oplus
H_{\Lambda_2\backslash\Lambda_1}-z)^{-1}=(H_{\Lambda_2}-z)^{-1}\oplus(H_{\Lambda_2\backslash\Lambda_1}-z)^{-1}$.

\goodbreak
Thus (\ref{res}) gives (for $n\in\Lambda_1$,
$m\in\Lambda_2\backslash\Lambda_1$)

\begin{eqnarray}\nonumber
&&(H_{\Lambda_2}-z)^{-1}(n,m)\\ \nonumber &=&
-\sum_{k,k'\in\Lambda_2}(H_{\Lambda_1}\oplus
H_{\Lambda_2}-z)^{-1}(n,k)\;\;\Gamma_{\Lambda_1}^{\Lambda_2}(k,k')\;\;(H_{\Lambda_2}-z)^{-1}(k',m)\\
&=&\sum_{(k,k')\in\partial\Lambda_1 \atop k\in\Lambda_1,\
k'\in\Lambda_2}(H_{\Lambda_1}-z)^{-1}(n,k)\;\;(H_{\Lambda_2}-z)^{-1}(k',m)
\ .
\end{eqnarray}

We summarize in the following theorem
\goodbreak
\begin{theorem}[Geometric resolvent equation]\es \label{th:geoRes}
If $\;\L_1\subset\L_2$ and $n\in\L_1, \;m\in\L_2\setminus\L_1$ and
if $\;z\not\in\big(\sigma(H_{\L_1})\cup\sigma(H_{\L_2})\big)$, then
\begin{align}
&\quad(H_{\Lambda_2}-z)^{-1}(n,m)\notag\\
~=~&\sum_{(k,k')\in\partial\Lambda_1 \atop k\in\Lambda_1,\
k'\in\Lambda_2}(H_{\Lambda_1}-z)^{-1}(n,k)\;(H_{\Lambda_2}-z)^{-1}(k',m)\label{geoRes}
\ .
\end{align}
\end{theorem}

Equation (\ref{geoRes}) is the \dindex{geometric resolvent
equation}. It expresses the resolvent on a large set $(\Lambda_2)$
in terms of the resolvent on a smaller set $(\Lambda_1)$. Of course,
the right hand side still contains the resolvent on the large set.

\begin{remark}
Above we derived (\ref{geoRes}) only for
$z\not\in\sigma(H_{\Lambda_2\setminus\L_1})$. However, both sides of
(\ref{geoRes}) exist and are analytic outside
$\sigma(H_{\L_1})\cup\sigma(H_{\L_2})$,\; so the formula is valid
for all $z$ outside $\sigma(H_{\L_1})\cup\sigma(H_{\L_2})$
\end{remark}
We introduce a short-hand notation for the matrix elements of
resolvents

\begin{equation}\label{def:Green}
\mathdex{G_z^\Lambda}(n,m)=(H_\Lambda-z)^{-1}(n,m).
\end{equation}

The functions $G_z^\Lambda$ are called \emph{\mindex{Green's
functions}}.

With this notation the geometric resolvent equation reads
\begin{align}
G_z^{\L_2}(n,m)~=~\sum_{(k,k')\in\partial\Lambda_1 \atop
k\in\Lambda_1,\ k'\in\Lambda_2}
G_z^{\Lambda_1}(n,k)G_z^{\Lambda_2}(k',m) \ .
\end{align}

There are analogous equations to (\ref{geoRes}) for Dirichlet or
Neumann boundary conditions which can be derived from
(\ref{eq:splitHLD}) and (\ref{eq:splitHLD}) in the same way as
above.


\subsection{An alternative approach to the density of states}\es

In this section we present an alternative definition of the density
of states measure. Perhaps, this is the more traditional one. We
prove its equivalence to the definition given above.

In section \ref{sec:defdos} we defined the density of states measure
by starting with a function $\varphi$ of the Hamiltonian, taking its
trace restricted to a cube $\Lambda_L$ and normalizing this trace.
In the second approach we first restrict the \emph{Hamiltonian}
$H_\omega$ to $\Lambda_L$ with appropriate boundary conditions,
apply the function $\varphi$ to the restricted Hamiltonian and then
take the normalized trace.

For any $\Lambda$ let \mindex{$H_\Lambda^X$} be either $H_\Lambda$
or $H_\Lambda^N$ or $H_\Lambda^D$. We define the measures
\mindex{$\tilde{\nu}_L^X$} (\,i.e.  $\tilde{\nu}_L$,\;
$\tilde{\nu}_L^N$,\; $\tilde{\nu}_L^D$)
\index{$\tilde{\nu}_L$}\index{$\tilde{\nu}_L^N$}\index{$\tilde{\nu}_L^D$}
by

\begin{equation}\label{dosalt}
\int\varphi(\lambda)\;d\tilde{\nu}_L^X(\lambda)=\frac{1}{|\Lambda_L|}\tr\varphi(H_{\LL}^X)
\ .
\end{equation}

Note that the operators $H_{\LL}^X$ act on the finite dimensional
Hilbert space $\ell^2(\LL)$, so their spectra consist of eigenvalues
\mindex{$E_n(H_{\LL}^X)$} which we enumerate in increasing order

\begin{displaymath}
E_0(H_{\LL}^X)\leq E_1(H_{\LL}^X)\leq\dots\quad \ .
\end{displaymath}

In this enumeration we repeat each eigenvalue according to its
multiplicity (see also (\ref{eigvord}).

With this notation (\ref{dosalt}) reads

\begin{displaymath}
\int\varphi(\lambda)\,d\tilde{\nu}_L^X(\lambda)=\frac{1}{|\Lambda_L|}\sum_n\varphi(E_n(H_{\LL}^X))
\ .
\end{displaymath}

The measure $\tilde{\nu}^X_L$ is concentrated on the eigenvalues of
$H^X_{\LL}$. If $E$ is an eigenvalue of $H^X_{\LL}$ then
$\tilde{\nu}^X_L(\{E\})$ is equal to the dimension of the eigenspace
corresponding to $E$.

We define the eigenvalue counting function by
\begin{equation}\label{def:NLambda}
\mathdex{N(H_\L^X,E)}=\big|\{n\,|\, E_n(H_\L^X)<E\}\big|
\end{equation}
(where $|M|$ is the number of elements of $M$). Then
$\frac{1}{|\LL|}\,N(H_{\LL}^X,E)$ is the distribution function of
the measure $\tilde{\nu}_L^X$, i.~e.

\begin{equation}
    \frac{1}{|\LL|}\,N(H_{\LL}^X,E)~=~\int\;\chi_{(-\infty,
    E)}(\lambda)\;d\,\tilde{\nu}_L^X(\lambda) \ .
\end{equation}

\begin{theorem}
The measures \mindex{$\tilde{\nu}_L $}, \mindex{$\tilde{\nu}_L^D$}
and \mindex{$\tilde{\nu}_L^N$} converge $\P$-almost surely vaguely
to the density of states measure $\nu$.
\end{theorem}

\begin{proof}
We give the proof for $\tilde{\nu}_L$. An easy modification gives
the result for $\tilde{\nu}_L^D$ and $\tilde{\nu}_L^N$ as well. To
prove that $\tilde{\nu}_L$ converges vaguely to $\nu$ it suffices to
prove
\begin{displaymath}
 \int\varphi(\lambda)\,d\tilde{\nu}_L(\lambda)\rightarrow\int\varphi(\lambda)\,d\nu(\lambda)
\end{displaymath}

for all $\varphi$ of the form

\begin{displaymath}
 \varphi(x)=r_z(x)=\frac{1}{x-z} \quad\textnormal{for } z\in\C\setminus\R
 \end{displaymath}

because linear combination of these functions are dense in
$\C_\infty(\R)$ by the Stone-Weierstraß Theorem (see Section
\ref{sec:morefun}). (\,$C_\infty(\R)$ \index{$C_\infty(\R)$} are the
continuous functions vanishing at infinity.)
\goodbreak
We have

\begin{displaymath}
\int
r_z(\lambda)\,d\tilde{\nu}_L(\lambda)=\frac{1}{(2L+1)^d}\,\tr\,\big((H_{\LL}-z)^{-1}\big)
=\frac{1}{(2L+1)^d}\,\sum_{n\in\Lambda_L} (H_{\LL}-z)^{-1}(n,n)
\end{displaymath}

and

\begin{displaymath}
 \int r_z(\lambda)\,d\nu_L(\lambda)=\frac{1}{(2L+1)^d}\,\tr\,\big(\chi_{\Lambda_L}(H-z)^{-1}\big)
 =\frac{1}{(2L+1)^d}\,\sum_{n\in\Lambda_L} (H-z)^{-1}(n,n)
\end{displaymath}
\vspace{1mm}

We use the resolvent equation in the form (\ref{res}) for $n\in\LL$:

\begin{eqnarray}\nonumber
&&\big|\sum_{n\in\LL}(H_{\LL}-z)^{-1}(n,n)-(H-z)^{-1}(n,n)\,\big|\\\nonumber
&=&\big|\sum_{n\in\LL}\sum_{(k,k')\in\partial\LL \atop k\in\LL,\
k'\in\complement\LL}(H_{\LL}-z)^{-1}(n,k)\;(H-z)^{-1}(k',n)\,\big|\\\nonumber
&\leq&\sum_{(k,k')\in\partial\LL \atop k\in\LL,\
k'\in\complement\LL}\big(\sum_n|(H_{\LL}-z)^{-1}(n,k)|^2\big)^{\frac{1}{2}}
\cdot\big(\sum_n|(H-z)^{-1}(k',n)|^2\big)^{\frac{1}{2}}\\\nonumber
&=&\sum_{(k,k')\in\partial\LL \atop k\in\LL,\
k'\in\complement\LL}||(H_{\LL}-z)^{-1}\delta_k||\cdot||(H-z)^{-1}\delta_{k'}||\\\nonumber
&\leq& c\,L^{d-1}\,||(H_{\LL}-z)^{-1}||\cdot||(H-z)^{-1}||\\
&\leq&\frac{c}{(\textnormal{Im}\ z)^2}\;L^{d-1} \ .
\end{eqnarray}

Hence

\begin{eqnarray}\nonumber
&&|\int r_z(\lambda)\,d\tilde{\nu}_L(\lambda)-\int
r_z(\lambda)\,d\nu_L(\lambda)|\leq \frac{c'}{(\textnormal{Im}\
z)^2}\cdot\frac{1}{L}~~\to~~0\qquad\textnormal{as }L\to\infty \ .
\end{eqnarray}

\end{proof}


\subsection{The Wegner estimate\label{sec:Wegner}}\es

We continue with the celebrated `Wegner estimate'. This result due
to Wegner \cite{wegner} shows not only the regularity of the density
of states, it is also a key ingredient to prove Anderson
localization. We set $\mathdex{N_\Lambda(E)}:=N(H_\L, E)$.

\begin{theorem}
(\mindex{Wegner estimate}{\label{th:1}\label{th:Wegner}}) Suppose
the measure $P_0$ has a bounded density $g$, (i.e. $P_0(A)=\int_A
g(\lambda)d\lambda, \ ||g||_\infty < \infty)$ then

\begin{equation}
\E\,
\big(N_\Lambda(E+\varepsilon)-N_\Lambda(E-\varepsilon)\,\big)\leq
C\,\|\,g\,\|_\infty \;|\Lambda|\;\varepsilon \ .
\end{equation}

\end{theorem}

Before we prove this estimate we note two important consequences.

\begin{corollary}{\label{co:1}}
Under the assumption of Theorem \ref{th:1} the integrated density of
states is absolutely continuous with a bounded density $n(E)$.
\end{corollary}

Thus $N(E)=\int_{-\infty}^E n(\lambda)\,d\lambda$. We call
$n(\lambda)$\index{$n(\lambda)$} the \dindex{density of states}.
Sometimes, we also call $N$ the density of states, which, we admit,
is an abuse of language.

\begin{corollary} {\label{co:2}}
Under the assumptions of Theorem \ref{th:1} we have for any $E$ and
$\Lambda$

\begin{equation}
\P\,\big(\textnormal{dist}(E,\sigma(H_\Lambda))<\varepsilon\,\big)\leq
\;C\,\|\,g\,\|_\infty \; \varepsilon\; |\Lambda| \ .
\end{equation}

\end{corollary}

\begin{nproof}{Corollary \ref{co:2}} By the Chebycheff inequality we get
\begin{eqnarray}
&&\P\,\big(\textnormal{dist}(E, \sigma(H_\Lambda))<\varepsilon\,\big)\nonumber \\
&&= \P\,\big(N_\Lambda (E+\varepsilon)-N_\Lambda(E-\varepsilon)\geq 1\,\big)\nonumber \\
&&\leq \E\,\big(N_\Lambda(E+\varepsilon)-N_\Lambda(E-\varepsilon)\,\big) \nonumber\\
&&\leq C\;\|\,g\,\|_\infty \; \varepsilon\; |\Lambda| \qquad
\textnormal{by Theorem (\ref{th:1}).}
\end{eqnarray}
\end{nproof}

\begin{nproof}{Corollary \ref{co:1}} By Theorem \ref{th:1} we have
\begin{align}
N(E+\varepsilon)- N(E-\varepsilon)
&=\lim_{|\Lambda|\rightarrow\infty}\;
\frac{1}{|\Lambda|} \;\E\,\big(N_\Lambda(E+\varepsilon)-N_\Lambda(E-\varepsilon)\,\big)\nonumber\\[2mm]
&\leq C\; \|\,g\,\|_\infty \;\varepsilon \ .
\end{align}
\end{nproof}
\goodbreak
We turn to the proof of the theorem.

\begin{nproof}{Wegner estimate}
Let $\varrho$ be a non decreasing $C^\infty$-function with\\
$\varrho(\lambda)=1$ for $\lambda \geq \varepsilon$,
$\varrho(\lambda)=0$ for $\lambda\leq -\varepsilon$ and consequently
$0\leq \varrho(\lambda)\leq 1$. Then
\begin{align}\nonumber
0 \;\;\leq\quad & \chi_{(-\infty,
E+\varepsilon)}(\lambda)-\chi_{(-\infty,
E-\varepsilon)}(\lambda)\\\nonumber \leq\quad &
\varrho(\lambda-E+2\varepsilon)-\varrho(\lambda-E-2\varepsilon)
\intertext{hence} \nonumber 0 \;\;\leq\quad & \chi_{(-\infty,
E+\varepsilon)}(H_\Lambda)-\chi_{(-\infty,
E-\varepsilon)}(H_\Lambda)\\\nonumber \leq\quad &
\varrho(H_\Lambda-E+2\varepsilon)-\varrho(H_\Lambda-E-2\varepsilon).
\intertext{Consequently,} \nonumber
&N_\L(E+\varepsilon)-N_\L(E-\varepsilon)\\\nonumber =\quad &
\textnormal{tr} \;\big(\; \chi_{(-\infty,
E+\varepsilon)}(H_\Lambda)-\chi_{(-\infty,
E-\varepsilon)}(H_\Lambda)\,\big)\\
\leq\quad & \textnormal{tr} \ \varrho(H_\Lambda-E+2\varepsilon)-
\textnormal{tr} \ \varrho(H_\Lambda-E-2\varepsilon)\label{eq:WBew} \
.
\end{align}

To compute the expectation of (\ref{eq:WBew}) we look upon the
operators $H_\Lambda$ (and their eigenvalues $E_n(H_\Lambda)$) as
functions of the values
$V_{\L}=\{V_i\}_{i\in\Lambda}$\index{$V_{\L}$} of the potential
inside $\L$. More precisely, we view the mapping
\begin{displaymath}
V_\L \rightarrow H_\Lambda=H_{\L}(V_\L)
\end{displaymath}
\index{$H_{\L}(V_\L)$} as a matrix-valued function on $\R^{|\L|}$.
This function is differentiable and

\begin{equation}
\left( \frac{\partial H_\Lambda}{\partial V_i} \right)_{\ell m} =
\delta_{\ell m} \delta_{\ell i} \ .
\end{equation}

The function
\begin{displaymath}
(E, V_\L) \rightarrow \tr \ \varrho \big(H_\Lambda(V_\L)-E\big) \ .
\end{displaymath}
is differentiable as well. Furthermore, since

\begin{align}
H_\L(V_\L)-E\,&=\,H_\L(V_\L-E)\label{eq:diffW} \ .
\end{align}

it follows

\begin{equation}
\tr \ \varrho
\big(H_\Lambda(V_\L)-E\big)=F\big(\{V_i-E\}_{i\in\L}\big)
\end{equation}

and consequently

\begin{equation}
\frac{\partial}{\partial E}\;\Big(\tr \ \varrho
\big(H_\Lambda(V_\L)-E\big)\Big)
~=~-\,\sum_{i\in\L}\;\frac{\partial}{\partial V_i}\;\Big(\tr \
\varrho \big(H_\Lambda(V_\L)-E\big)\Big)
\end{equation}

Therefore, with (\ref{eq:WBew})
\begin{eqnarray}\nonumber
 N_\L(E+\varepsilon)-N_\L(E-\varepsilon)&\leq&\textnormal{tr} \ \varrho\,(H_\Lambda-E+2\varepsilon)-
\textnormal{tr} \ \varrho\,(H_\Lambda-E-2\varepsilon)\\[2mm]
&=& - \big( \textnormal{tr} \ \varrho\,(H_\Lambda-(E+2\varepsilon))-
\textnormal{tr} \ \varrho(\,H_\Lambda-(E-2\varepsilon))     \big)\notag\\[2mm]
&=& -\,\int_{E-2\varepsilon}^{E+2\varepsilon}
\;\frac{\partial}{\partial \eta}\; \Big(\,
\varrho\big(H_\Lambda(V_\L)-\eta\big)\Big)\;d\eta\notag
\\ \label{eq:WBew2}
&=&\ \int_{E-2\varepsilon}^{E+2\varepsilon}\;
\sum_{j\in\L}\,\frac{\partial}{\partial V_j}\ \ \tr\, \varrho
\big(H_\Lambda(V_\L-\eta)\big)\;d\eta.
\end{eqnarray}

Therefore
\begin{align}\nonumber
&~~\E\,\big(\,N_\Lambda(E+\varepsilon)-N_\Lambda(E-\varepsilon)\,\big)\\\nonumber
\leq ~~&\E\,
\big(\sum_{j\in\Lambda}\int_{E-2\varepsilon}^{E+2\varepsilon}
\frac{\partial}{\partial V_j}\; \textnormal{tr}\,
 (\varrho\,(H_\Lambda(V_\L)-\eta)\big)\;d\eta) \\
= ~~&\sum_{j\in\Lambda}\int_{E-2\varepsilon}^{E+2\varepsilon}
\E\,\Big( \frac{\partial}{\partial V_j} \;\textnormal{tr}\,
 \big(\varrho\,(H_\Lambda(V_\L)-\eta)\big)\Big)\;d\eta.\label{eq:WBew3}
\end{align}

Since the random variables $V_\omega(i)$ are independent and have
the common distribution $d P_0(V_i)\,=\,g(V_i)\,dV_i$, the
expectation $\E$ is just integration with respect to the product of
these distributions. Moreover, since $\supp P_0$ is compact the
integral over the variable $V_i$ can be restricted to $[-M,+M]$ for
some $M$ large enough.

Hence
\begin{align}
&\E\,\Big( \frac{\partial}{\partial V_j} \;\textnormal{tr}\,
 \big(\varrho\,(H_\Lambda(V_\L)-\eta)\big)\Big)\notag\\
=~&\int_{-M}^{+M} \dots \int_{-M}^{+M}\; \frac{\partial}{\partial
V_j}\; \textnormal{tr}\,
\big(\varrho(H_\Lambda(V_\L)-\eta)\big)\;\prod_{i\in\Lambda} g(V_i) \prod_{i\in\L}\,d V_i\notag\\
=&\int_{-M}^{+M} \dots \int_{-M}^{+M}
\Big(\int_{-M}^{+M}\frac{\partial}{\partial V_j}\textnormal{tr}\,
\big(\varrho(H_\Lambda(V_\L)-\eta)\big) g(V_j)\,dV_j \Big)
\prod_{i\in\Lambda \atop i\not=j} g(V_i)\;d V_i\label{eq:WBew1} \ .
\end{align}
Since $\textnormal{tr}\, \big(\varrho(H_\Lambda(V_\L)-\eta)\big)$ is
non decreasing in $V_j\,$ we can estimate

\begin{align}
&\int_{-M}^{+M}\frac{\partial}{\partial V_j}\;\textnormal{tr}\,
\big(\varrho(H_\Lambda(V_\L)-\eta)\big)\;g(V_j)\ \;dV_j\notag\\
\leq &\;||g||_\infty\;\;\Big(
\textnormal{tr}\,\big(\varrho(H_\Lambda(V_\L, V_j=M)-\eta)\big)\,-
\,\textnormal{tr}\,\big(\varrho(H_\Lambda(V_\L, V_j=-M)-\eta)\big)
\Big)    {\label{eq:10}}
\end{align}

where $H_\L(V_\L, V_j=a)={H_0}_\L + \tilde{V}$\index{$H_\L(V_\L,
V_j=a)$} is the Anderson Hamiltonian on $\L$ with potential

\begin{align} \tilde{V}_i=\left\{
\begin{array}{ll}V_i & \textnormal{for }\quad i\not= j\\
\ a & \textnormal{for }\quad i=j.
\end{array}\right.
\end{align}

To estimate the right hand side of inequality (\ref{eq:10}) we will
use the following Lemma:
\goodbreak
\begin{lemma}
Let $A$ be a selfadjoint operator bounded below with purely discrete
spectrum and eigenvalues $E_0 \leq E_1 \leq \dots$ repeated
according to multiplicity. If $B$ is a symmetric positive rank one
operator then $\tilde{A}=A+B$ has eigenvalue $\tilde{E_n}$ with
$E_n\leq\tilde{E_n}\leq{E_{n+1}}$.
\end{lemma}

Given the Lemma we continue the proof of the theorem.

We set $A=H_\Lambda(V_\L, V_j=-M)$ and $\tilde{A}=H_\Lambda(V_\L,
V_j =+M)$. Obviously their difference is a (positive) rank one
operator

\begin{eqnarray}\nonumber
&& \textnormal{tr}\ \varrho(\tilde{A}-\eta)-\textnormal{tr}\
\varrho(A-\eta)\\\nonumber = &&\sum_n\;
\big(\varrho(\tilde{E_n}-\eta)-\varrho(E_n-\eta)\big)\\\nonumber
\leq &&\sum_n\;
\big(\varrho(E_{n+1}-\eta)-\varrho(E_n-\eta)\big)\\\nonumber
\leq && \sup_{\lambda,\mu}\ \varrho(\lambda)-\varrho(\mu)\\
= &&1 \ .
\end{eqnarray}

Thus from (\ref{eq:10}) we have

\begin{align}\notag
\int_{-M}^{+M}\frac{\partial}{\partial V_j}\;\textnormal{tr}\,
\big(\varrho(H_\Lambda(V_\L)-\eta)\big)\;g(V_j)\ \;dV_j
~\leq~||\,g||_{\,\infty} \ .
\end{align}
Since $\int_{-M}^{+M}g(v)dv=1$ we conclude from (\ref{eq:WBew1}) and
(\ref{eq:10}) that
\begin{align}
\E\,\Big( \frac{\partial}{\partial V_j} \;\textnormal{tr}\,
 \big(\varrho\,(H_\Lambda(V_\L)-\eta)\big)\Big)\notag~\leq~||\,g||_{\,\infty}.
\end{align}
So, (\ref{eq:WBew3}) implies
\begin{align}
\E\,\big(\,N_\Lambda(E+\varepsilon)-N_\Lambda(E-\varepsilon)\,\big)~\leq~4\,||\,g||_{\,\infty}\,|\,\L|\;\varepsilon
\ .
\end{align}
\end{nproof}

\begin{nproof}{Lemma}
Since $B$ is a positive symmetric rank one operator it is of the
form $B=c\,|h\rangle\langle h|$ with $c\geq 0$, i.e. $B\ \varphi =
c\ \langle h,\varphi\rangle\ h$\quad for some $h$.

By the min-max principle (Theorem \ref{th:minmax})

\begin{eqnarray}\nonumber
\tilde{E}_n &=& \sup_{\psi_1,...,\psi_{n-1}}
\inf_{\varphi\perp\psi_1,...,\psi_{n-1}\atop||\varphi||=1}
\langle\varphi,A\,\varphi\rangle+c\;|\langle\varphi,h\rangle|^2\\\nonumber
&\leq&
\sup_{\psi_1,...,\psi_{n-1}}\inf_{\varphi\perp\psi_1,...,\psi_{n-1}
\atop\varphi\perp h\;\;||\varphi||=1}
\langle\varphi,A\,\varphi\rangle+c\;|\langle\varphi,h\rangle|^2\\\nonumber
&\leq&
\sup_{\psi_1,...,\psi_{n-1}}\inf_{\varphi\perp\psi_1,...,\psi_{n-1},h
\atop||\varphi||=1} \langle\varphi,A\,\varphi\rangle\\\nonumber
&\leq&\sup_{\psi_1,...,\psi_{n-1},\psi_n}
\inf_{\varphi\perp\psi_1,...,\psi_{n-1},\psi_n\atop ||\varphi||=1}
\langle\varphi,A\,\varphi\rangle\\
&=&E_{n+1} \ .
\end{eqnarray}
\end{nproof}

By the Wegner estimate we know that any given energy $E$ is {\em
not} an eigenvalue of $(H_\omega)_{\Lambda_L}$ for almost all
$\omega$. On the other hand it is clear that for any given $\omega$
there are (as a rule $|\Lambda_L|$) eigenvalues of
$(H_\omega)_{\Lambda_L}$.

This simple fact illustrates that we are not allowed to interchange
`any given $E$' and `for $\mathbb{P}-$almost all $\omega$' in
assertions like the one above. What goes wrong is that we are trying
to take an uncountable union  of sets of measure zero. This union
may have any measure, if it is measurable at all.

In the following we demonstrate a way to overcome these difficulties
(in a sense). This idea is extremely useful when we want to prove
pure point spectrum.

\begin{theorem}\label{th:dWegner}
If $\Lambda_1, \Lambda_2$ are disjoint finite subsets of\;
$\mathbb{Z}^d$, then
\begin{align}\nonumber
\mathbb{P}\,\big(&\textnormal{ \em There is an } E \in \mathbb{R}
\textnormal{ \em such that } \textnormal{dist}( E,
\sigma(H_{\Lambda_1}))<\varepsilon \textnormal{ \em and \rm}
\textnormal{dist}(\sigma (E, H_{\Lambda_2}))) <
\varepsilon\,\big)\\[2mm]
\nonumber &\leq 2\,C\,\|\,g\,\|_\infty \; \varepsilon\;
|\,\Lambda_1| |\,\Lambda_2| \ .
\end{align}
\end{theorem}

We start the proof with the following lemma.

\begin{lemma}\label{lem:dWegner}
If $\Lambda_1, \Lambda_2$ are disjoint finite subsets of\;
$\mathbb{Z}^d$ then
\begin{displaymath}
\mathbb{P}\,\big(\,\textnormal{dist}(\sigma(H_{\Lambda_1}),
\sigma(H_{\Lambda_2})) < \varepsilon\,\big) \leq C\,
\|\,g\,\|_\infty \;\varepsilon\; |\Lambda_1| |\Lambda_2| \ .
\end{displaymath}
\end{lemma}

\begin{nproof}{Lemma}
Since $\Lambda_1 \cap \Lambda_2 = \emptyset$ the random potentials
in $\Lambda_1$ and $\Lambda_2$ are independent of each other and so
are the eigenvalues $E_0^{(1)} \leq E_1^{(1)} \leq ...$ of
$H_{\Lambda_1}$ and the eigenvalues $E_0^{(2)} \leq E_1^{(2)} \leq
...$ of $H_{\Lambda_2}$.

We denote the probability (resp. the expectation) with respect to
the random variables in $\Lambda$ by $\mathbb{P}_\Lambda$ (resp.
$\mathbb{E}_\Lambda$). Since the random variables
$\{V_\omega(n)\}_{n\in\Lambda_1}$ and
$\{V_\omega(n)\}_{n\in\Lambda_2}$ are independent for
$\Lambda_1\cap\Lambda_2=\emptyset$ we have that for such sets
$\P_{\L_1\cup\L_2}$ is the product measure
$\P_{\L_1}\otimes\P_{\L_2}$.

We compute
\begin{align}\nonumber
\mathbb{P}\,\big(\,\textnormal{dist}(\sigma(\Lambda_1),
\sigma(\Lambda_2)) < \varepsilon\,\big) =&\;
\mathbb{P}\,\big(\,\min_i\; \textnormal{dist}(E_i^{(1)},
\sigma(H_{\Lambda_2})) <
\varepsilon\,\big)\notag\\
\nonumber \leq& \;\sum_{i=1}^{|\Lambda_1|}\;
\mathbb{P}\,\big(\,\textnormal{dist}
(E_i^{(1)}, \sigma(H_{\Lambda_2})) < \varepsilon\,\big)\\
\nonumber \leq& \;\sum_{i=1}^{|\Lambda_1|}\;\;
\mathbb{P}_{\Lambda_1}\otimes\mathbb{P}_{\Lambda_2}\,\big(\,\textnormal{dist}
(E_i^{(1)}, \sigma(H_{\Lambda_2})) < \varepsilon\,\big)\\
\label{eq:dWeg} \leq& \sum_{i=1}^{|\Lambda_1|}\;
\mathbb{E}_{\Lambda_1}\,\Big(\,\mathbb{P}_{\Lambda_2}
\big(\,\textnormal{dist}(E_i^{(1)}, \sigma(H_{\Lambda_2})) <
\varepsilon\,\big)\,\Big) \ .
\end{align}
From Theorem \ref{th:Wegner} we know that
\begin{align}\notag
\mathbb{P}_{\Lambda_2} \big(\,\textnormal{dist}(E,
\sigma(H_{\Lambda_2})) < \varepsilon\,\big)\;\leq\;C\,
\|\,g\,\|_\infty \;\varepsilon\; |\Lambda_2| \ .
\end{align}
Hence, we obtain
\begin{equation} \nonumber \textnormal{(\ref{eq:dWeg})} \leq C\, \|\,g\,\|_\infty \;\varepsilon\; |\Lambda_2| |\Lambda_1| \ .
\end{equation}
\end{nproof}

The proof of the theorem is now easy.
\begin{nproof}{Theorem}
\begin{align}\nonumber
\mathbb{P}\,\big(\textnormal{ There is an } &E \in \mathbb{R}
\textnormal{  such that } \textnormal{dist}(\sigma(H_{\Lambda_1}),
E)<\varepsilon \textnormal{ and }
\textnormal{dist}(\sigma (H_{\Lambda_2}, E))) < \varepsilon\,\big)\\
\nonumber &\leq
\mathbb{P}\,\big(\,\textnormal{dist}(\sigma(H_{\Lambda_1}),
\sigma(H_{\Lambda_2})) < 2\,\varepsilon)
\\ \nonumber &\leq 2\,C\, \|\,g\,\|_\infty \;\varepsilon\; |\Lambda_1| |\Lambda_2|
\end{align}
by the lemma.
\end{nproof}
\goodbreak
$\;$\\[2mm]
{\bf Notes and Remarks }\\[2mm]  
General  references for the density of states are  \cite{p0},
\cite{km1}, \cite{av} and \cite{dels}, \cite{ksonder} \cite{simon1}
and \cite{ivan}. A thorough discussion of the geometric resolvent
equation in the context of perturbation theory can be found in
\cite{dk}, \cite{froespe} and \cite{spencer1}.

In the context of the discrete Laplacian Dirichlet and Neumann
boundary conditions were introduced and investigated in
\cite{simon1}. See also \cite{KirMuel}.

For discrete ergodic operators the integrated density of states $N$
is log-Hölder continuous, see \cite{Craig-S}. Our proof of the
continuity of $N$ is tailored after \cite{dels}.

For results concerning the Wegner estimates see \cite{wegner},
\cite{ki1}, \cite{stoll3}  \cite{CHM}, \cite{CHK},and \cite{ivan},
as well as references given there. \NeueSeite


\section{Lifshitz tails\label{ch:Lifshitz}}

\subsection{Statement of the Result}\es

Already in the 1960s, the physicist I. Lifshitz observed that the
low energy behavior of the density of states changes drastically if
one introduces disorder in a system. More precisely, Lifshitz found
that
\begin{eqnarray}\label{Liford}
N(E) \thicksim C\,(E-E_0)^{\frac{d}{2}} \qquad E \searrow E_0
\end{eqnarray}

for the ordered case (i.e. periodic potential), $E_0$ being the
infimum of the spectrum, and
\begin{eqnarray}\label{Lifdis}
N(E) \thicksim C_1 e^{-C\,(E-E_0)^{-\frac{d}{2}}} \qquad E \searrow
E_0
\end{eqnarray}

for the disordered case. The behavior (\ref{Lifdis}) of $N$ is now
called \dindex{Lifshitz behavior} or \dindex{Lifshitz tails}. We
will prove (a weak form of) Lifshitz tails for the Anderson model.
This result is an interesting and important result on its own. It is
also used as an input for the proof of Anderson
localization. 

If $P_0$ is the common distribution of the independent random
variables $V_\omega(i)$, we denote by $a_0$ the infimum of the
support $\supp P_0$ of $P_0$. From Theorem \ref{th:spAnd} we have
$E_0= \inf \sigma(H_\omega)=a_0$\;\; $\P$-almost surely. We assume
that $P_0$ is not trivial,\; i.e. is not concentrated in a single
point. Moreover, we suppose that

\begin{equation}\label{ass:PO}
P_0([a_0, a_0 + \epsilon]) \geq C \epsilon^\kappa, \quad\mbox{for
some } C,\kappa> 0 \ .
\end{equation}

Under these assumptions we prove:\\

\begin{theorem}[Lifshitz-tails]
\begin{equation}\label{eq:doublelog}
\lim_{E \searrow E_0} \frac{\ln | \ln N(E) |}{\ln (E-E_0)} =
-\frac{d}{2} \ .
\end{equation}
\end{theorem}
\vspace{4mm}
\begin{remark}
(\ref{eq:doublelog}) is a weak form of (\ref{Lifdis}). The
asymptotic formula (\ref{Lifdis}) suggests that we should expect
\emph{at least}
\begin{equation}\label{eq:Lifstrong}
\lim_{E \searrow E_0} \frac{\ln N(E)}{(E-E_0)^{-\frac{d}{2}}}
~=~-\,C \ .
\end{equation}
Lifshitz tails can be proven in the strong form (\ref{eq:Lifstrong})
for the Poisson random potential (see \cite{dv} and \cite{nakao}).
In general, however, there can be a logarithmic correction to
(\ref{eq:Lifstrong}) (see \cite{metzger}) so that we can only expect
the weak form (\ref{eq:doublelog}). This form of the asymptotics is
called the `doublelogarithmic' asymptotics.
\end{remark}

To prove the theorem, we show an upper and a lower bound.
\goodbreak

\subsection{Upper bound}\es

For the upper bound we will need \mindex{Temple's inequality}, which
we state and prove for the reader's convenience.

\begin{lemma}[Temple's inequality]
Let $A$ be a self-adjoint operator and $E_0=\inf \sigma(A)$ be an
isolated non degenerate eigenvalue. We set $E_1= \inf \big(\sigma(A)
\backslash \{E_0 \}\big)$. If $\psi \in D(A)$ with $||\psi|| =1$
satisfies
\begin{displaymath}
\langle \psi, A \psi \rangle < E_1\ ,
\end{displaymath}
then
\begin{displaymath}
E_0 \;\;\geq \;\;\langle \psi, A \psi \rangle - \frac{\langle \psi,
A^2 \psi \rangle - \langle \psi, A \psi \rangle^2}{E_1 - \langle
\psi, A \psi \rangle} \ .
\end{displaymath}
\end{lemma}

\begin{proof}
By assumption we have
\begin{displaymath}
(A-E_1)(A-E_0) \geq 0.
\end{displaymath}
Hence, for any $\psi$ with norm $1$
\begin{displaymath}
\langle \psi, A^2 \psi \rangle - E_1 \langle \psi, A \psi \rangle -
E_0 \langle \psi, A \psi \rangle + E_1 E_0 \geq 0 \ .
\end{displaymath}

This implies
\begin{displaymath}
E_1 E_0 - E_0 \langle \psi, A \psi \rangle \geq E_1 \langle \psi, A
\psi \rangle - \langle \psi, A \psi \rangle^2 - (\langle \psi, A^2
\psi \rangle - \langle \psi, A \psi \rangle^2) \ .
\end{displaymath}

Since $E_1 - \langle \psi, A \psi \rangle > 0$, we obtain
\begin{displaymath}
E_0 \geq \langle \psi, A \psi \rangle - \frac{\langle \psi, A^2 \psi
\rangle - \langle \psi, A \psi \rangle^2}{E_1 - \langle \psi, A \psi
\rangle} \ .
\end{displaymath}
\end{proof}

We proceed with the upper bound.

\begin{nproof}{upper bound}

By adding a constant to the potential we may assume that
$a_0=\,\inf\supp\,(P_0)\,=0$, so that $V_\omega(n)\geq 0$. By
(\ref{NDbrackdis}) we have that
\begin{eqnarray}\nonumber
N(E) &\leq& \frac{1}{|\Lambda_L|}\;
\mathbb{E}\,\big(N(H_{\Lambda_L}^N,E)\big)
\\ \label{ubound} &\leq& \mathbb{P}\,\big(E_0(H_{\Lambda_L}^N) < E\big)
\end{eqnarray}
for any $L$, since $N(H_{\Lambda_L}^N,E) \leq |\Lambda_L|$.

At the end of the proof, we will choose an optimal $L$.

To estimate the right hand side in (\ref{ubound}) from above we need
an estimate of $E_0(H_{\Lambda_L}^N)$ from below which will be
provided by Temple's inequality. As a test function $\psi$ for
Temple's inequality we use the ground state of
$(H_0)_{\Lambda_L}^N$, namely
\begin{displaymath}
\psi_0(n) = \frac{1}{| \Lambda_L|^{\frac{1}{2}}} \qquad
\textnormal{for all } n \in \Lambda_L \ .
\end{displaymath}

In fact $(H_0)_{\Lambda_L}^N \psi_0 =0$. We have
\begin{eqnarray}\nonumber
\langle \psi_0, H_{\Lambda_L}^N \psi_0 \rangle &=& \langle \psi_0,
V_\omega \psi_0 \rangle
\\ \label{sternlein} &=& \frac{1}{(2L+1)^d} \sum_{i \in \Lambda_L}
V_\omega (i) \ .
\end{eqnarray}

Observe that this is an arithmetic mean of independent, identically
distributed random variables. Hence, (\ref{sternlein}) converges to
$\mathbb{E}(V_\omega(0))> 0$ almost surely.

To apply Temple's inequality, we would need
\begin{displaymath}
\frac{1}{(2L+1)^d} \sum_{i \in \Lambda_L} V_\omega (i) <
E_1(H_{\Lambda_L}^N)
\end{displaymath}

which is certainly wrong for large $L$ since
$E_1(H_{\Lambda_L}^N)\rightarrow 0$. We estimate
\begin{displaymath}
E_1(H_{\Lambda_L}^N) \geq E_1((H_0)_{\Lambda_L}^N) \geq c L^{-2} \ .
\end{displaymath}

The latter inequality can be obtained by direct calculation. Now we
define
\begin{displaymath}
V_\omega^{(L)}(i) = \min \{ V_\omega(i) , \ \frac{c}{3}L^{-2} \} \ .
\end{displaymath}
\index{$V_\omega^{(L)}$}

For fixed $L$, the random variables $V_\omega^{(L)}$ are still
independent and identically distributed, but their distribution
depends on $L$. Moreover, if
\begin{displaymath}
H^{(L)}=(H_0)_{\Lambda_L}^N + V_\omega^{(L)}
\end{displaymath}
\index{$H^{(L)}$}then $E_0(H_{\Lambda_L}^N) \geq E_0(H^{(L)})$ by
the min-max principle (Theorem \ref{th:minmax}).
\newline We get
\begin{eqnarray}\label{rautchen}
\langle \psi_0, H^{(L)} \psi_0 \rangle = \frac{1}{(2L+1)^d} \sum_{i
\in \Lambda_L} V_\omega^{(L)} (i) \leq \frac{c}{3}L^{-2}
\end{eqnarray}

by definition of $V_\omega^{(L)}$, consequently
\begin{displaymath}
\langle \psi_0, H^{(L)} \psi_0 \rangle \leq \frac{c}{3}\,L^{-2} <
E_1((H_0)_{\Lambda_L}^N) \leq E_1(H^{(L)}) \ .
\end{displaymath}

Thus, we may use Temples inequality with $\psi_0$ and $H^{(L)}$:

\begin{eqnarray}\nonumber
E_0(H_{\Lambda_L}^N) &\geq& E_0(H^{(L)})\\\nonumber &\geq& \langle
\psi_0, H^{(L)} \psi_0 \rangle - \frac{\langle \psi_0, (H^{(L)})^2
\psi_0 \rangle}{cL^{-2}-\langle \psi_0, H^{(L)} \psi_0 \rangle}
\\ \nonumber &\geq& \frac{1}{(2L+1)^d} \sum_{i \in \Lambda_L} V_\omega^{(L)}
(i) - \frac{\frac{1}{(2L+1)^d} \sum_{i \in \Lambda_L} (V_\omega
(i))^2}{(c-\frac{c}{3})L^{-2}}
\\ \nonumber &\geq& \frac{1}{(2L+1)^d} \sum_{i \in \Lambda_L} V_\omega^{(L)}
(i) - \frac{1}{(2L+1)^d} \sum_{i \in \Lambda_L}
V_\omega^{(L)}\left(\frac{\frac{c}{3}L^{-2}}{\frac{2}{3}cL^{-2}}\right)
\\&\geq& \frac{1}{2} \frac{1}{(2L+1)^d} \sum_{i \in \Lambda_L}
V_\omega^{(L)}(i) \ .
\end{eqnarray}

Collecting the estimates above, we arrive at
\begin{eqnarray}\label{kreuzlein}
N(E) \leq \mathbb{P}\,\big(E_0(H_{\Lambda_L}^N) < E\big)\leq
\mathbb{P}\,\big(\,\frac{1}{(2L+1)^d} \sum_{i \in \Lambda_L}
V_\omega^{(L)} < \frac{E}{2}\,\big) \ .
\end{eqnarray}

Now we choose $L$. We try to make the right hand side of
(\ref{kreuzlein}) as small as possible.

Since $V_\omega^{(L)}\leq\frac{c}{3}L^{-2}$, the probability in
(\ref{kreuzlein}) will be one if $L$ is too big.

So we certain want to choose $L$ in such a way that
$\frac{c}{3}L^{-2}>\frac{E}{2}$.

Thus, a reasonable choice seems to be

\begin{displaymath}
L:=\lfloor\beta E^{-\frac{1}{2}}\rfloor
\end{displaymath}

with some $\beta$ small enough and $\lfloor x\rfloor$ the largest
integer not exceeding $x$.

We single out an estimate of the probability in (\ref{kreuzlein})

\begin{lemma}\label{lem:ld}
For $L=\lfloor\beta E^{-\frac{1}{2}}\rfloor$ with $\beta$ small and
$L$ large enough
\begin{displaymath}
\P\,\left(\,\frac{1}{|\LL|}\sum_{i\in\LL}V_\omega^{(L)}(i)<\frac{E}{2}\,\right)\leq
e^{-\gamma|\LL|}
\end{displaymath}
with some $\gamma > 0$.
\end{lemma}

Given the lemma, we proceed

\begin{eqnarray}\nonumber
N(E)&\leq&\P\,\left(\frac{1}{|\LL|}\sum_{i\in\LL}V_\omega^{(L)}(i)<\frac{E}{2}\,\right)\\\nonumber
&\leq& e^{-\gamma|\LL|}\\\nonumber&=& e^{-\gamma(2\lfloor\beta
E^{-\frac{1}{2}}\rfloor^d+1)}\\&\leq& e^{-\gamma'E^{-\frac{d}{2}}} \
.
\end{eqnarray}

This estimate is the desired upper bound on $N(E)$.
\end{nproof}

To finish the proof of the upper bound, it remains to prove Lemma
\ref{lem:ld}. This lemma  is a typical large deviation estimate: By
our choice of $L$ we have $\E(V_\omega^{(L)})>\frac{E}{2}$ if
$\beta$ is small enough; thus, we estimate the probability that an
arithmetic mean of independent random variables deviates from its
expectation value. What makes the problem somewhat nonstandard is
the fact that the random variables $V^{(L)}$ depend on the parameter
$L$, which is also implicit in $E$.

\begin{nproof}{Lemma}

\begin{eqnarray}\nonumber
&&\P\,\big(\frac{1}{|\LL|}\sum
V_\omega^L(i)<\frac{E}{2}\big)\\\nonumber
\leq&&\P\,\big(\frac{1}{|\LL|}\sum
V_\omega^L(i)<\frac{\beta^2}{2}L^{-2}\big)\\\label{le:stern}
\leq&&\P\left(\#\{\,i\ \mid\ V_\omega^L(i)<\frac{c}{3}L^{-2}\}
\geq(1-\frac{3\beta^2}{c})|\LL|\right) \ .
\end{eqnarray}

Indeed, if less than $(1-\frac{3\beta^2}{c})|\LL|$ of the $V(i)$ are
below $\frac{c}{3}L^{-2}$ than more than $\frac{3\beta^2}{c}|\LL|$
of them are at least $\frac{c}{3}L^{-2}$ (in fact equal to). In this
case

\begin{eqnarray}\nonumber
\frac{1}{|\LL|}\sum
V_\omega^L(i)&\geq&\frac{1}{|\LL|}\frac{3\beta^2}{c}|\LL|\frac{c}{3}L^{-2}\\\nonumber&=&\frac{\beta^2}{2}L^{-2}
\ .
\end{eqnarray}

Since $P(V(i)>0)>0$ there is a $\gamma>0$ such that
$q:=P(V(i)<\gamma)<1$.

We set $\xi_i=\left\{\begin{array}{ll}1&\textnormal{if
}V_i<\gamma,\\0&\textnormal{otherwise.}\end{array}\right.$

The random variables $\xi_i$ are independent and identically
distributed, $\E(\xi_i)=q$.

Let us set $r=1-\frac{3\beta^2}{c}$. By taking $\beta$ small we can
ensure that $q<r<1$.

Then, for $L$ sufficient large

\begin{eqnarray}\nonumber
(\ref{le:stern})&\leq&\P\left(\#\{i\ \mid\
V_\omega^L(i)<\frac{c}{3}L^{-2}\} \geq r|\LL|\right)\\\nonumber
&\leq& \P\left(\#\{i\ \mid\ V_\omega^L(i)<\gamma\} \geq
r|\LL|\right)\\\label{SLD} &\leq& \P(\frac{1}{|\LL|}\sum \xi_i\geq
r) \ .
\end{eqnarray}

Through our somewhat lengthy estimate above we finally arrived at
the standard large deviations problem (\ref{SLD}). To estimate the
probability in (\ref{SLD}) we use the inequality

\begin{displaymath}
\P\,(X>a)\leq e^{-ta}\;\E(e^{tX})\qquad\textnormal{for }t\geq0 \ .
\end{displaymath}

Indeed
\begin{eqnarray}\nonumber
\P(\,X>a)&=&
\int\chi_{\{X>a\}}(\omega)\;d\,\P(\omega)\\\nonumber&\leq&\int
e^{-ta}e^{tX}\chi_{\{X>a\}}(\omega)\;d\,\P(\omega)\\\nonumber&\leq&\int
e^{-ta}e^{tX}\;d\,\P \ .
\end{eqnarray}

We obtain

\begin{eqnarray}\nonumber
(\ref{SLD})&\leq& e^{-|\LL|\, t\,
r}\cdot\;\E(\prod_{i\in\LL}e^{t\xi_i})\\\nonumber &=&e^{-|\LL|(r t
-\ln\E(e^{t\xi_0}))} \ .
\end{eqnarray}

Set $f(t)=r t -\ln\E(e^{t\xi_0})$. If we can choose $t$ such that
$f(t)>0$, the result is proven. To see that this is possible, we
compute

\begin{displaymath}
f'(t)=r-\frac{\E(\xi_0e^{t\xi_0})}{\E(e^{t\xi_0})}
\end{displaymath}

So $f'(0)=r-q\ >0$.

Since $f(0)=0$, there is a $t>0$ with $f(t)>0$.

\end{nproof}

Thus, we have shown

\begin{equation}
\overline{\lim}_{E \searrow E_0} \frac{\ln | \ln N(E)| }{\ln
(E-E_0)} \leq - \frac{d}{2} \ .
\end{equation}

\subsection{Lower bound}\es

We proceed with the lower bound. By (\ref{NDbrackdis}) we estimate

\begin{eqnarray}\nonumber
N(E)&\geq&\frac{1}{|\LL|}\;\E\,\big(\,N(H_{\LL}^D,E)\,\big)\\
\label{lb:Brack}&\geq&\frac{1}{|\LL|}\;\P\,\big(\,E_0(H_{\LL}^D)<E\,\big).
\end{eqnarray}

As in the upper bound, the above estimate holds for any $L$.

To proceed, we have to estimate $E_0(H_{\LL}^D)$ from above.

This is easily done via the min-max principle (Theorem
\ref{th:minmax}):

\begin{eqnarray}\nonumber
E_0(H_{\LL}^D)&\leq&\langle\psi,H_{\LL}^D\psi\rangle\\\label{lb:MinMax}
&=&\langle\psi,(H_0)_{\LL}^D\psi\rangle+\sum_{i\in\LL}V_\omega(i)\;|\psi(i)|^2
\end{eqnarray}

for any $\psi$ with $||\psi||=1$. Now we try to find $\psi$ which
minimizes the right hand side of (\ref{lb:MinMax}). First we deal
with the term

\begin{equation}\label{MinMax0}
\langle\psi,(H_0)_{\LL}^D\psi\rangle \ .
\end{equation}

Since $(H_0)_{\LL}^D$ adds a positive term to $(H_0)_{\LL}^N$ at the
boundary, it seems desirable to choose $\psi(n)=0$ for $|n|=L$. On
the other hand, to keep (\ref{MinMax0}) small we don't want $\psi$
to change too abruptly.

So, we choose

\begin{displaymath}
\psi_1(n)=L-\inorm{n}\qquad ,n\in\LL
\end{displaymath}

and

\begin{displaymath}
\psi(n)=\frac{1}{||\psi_1||}\,\psi_1(n) \ .
\end{displaymath}

We have

\begin{displaymath}
\sum_{n\in\LL}|\psi_1(n)|^2\geq\sum_{n\in\Lambda_{\frac{L}{2}}}|\psi_1(n)|^2\geq|\Lambda_{\frac{L}{2}}|\;(\frac{L}{2})^2\geq
c\, L^{d+2}
\end{displaymath}

and

\begin{eqnarray}\nonumber
\langle\psi_1,(H_0)_{\LL}^D\psi_1\rangle&\leq&\sum_{(n,n')\in\LL
\atop
\gnorm{n-n'}=1}\,\big|\,\psi_1(n')-\psi_1(n)\,\big|^2\\\nonumber&\leq&\Big|\,\{\,(n,n')\in\LL\times\LL\
;\ \gnorm{n-n'}=1\}\,\Big|\\\nonumber&\leq&c_1\,L^d \ .
\end{eqnarray}

Above, we used that $|\psi_1(n)-\psi_1(n')|\leq1$ if
$\gnorm{n-n'}=1$.

Collecting these estimates, we obtain

\begin{eqnarray}\label{lb:free}
E_0\Big((H_0)_{\LL}^D\Big)&\leq&\frac{\langle\psi_1,(H_0)_{\LL}^D\psi_1\rangle}{\langle\psi_1,\psi_1\rangle}\\
\nonumber&\leq& c_0\, L^{-2} \ .
\end{eqnarray}

The bounds of (\ref{lb:Brack}), (\ref{lb:MinMax}) and
(\ref{lb:free}) give
\begin{eqnarray}\nonumber
N(E) &\geq& \frac{1}{|\Lambda_L|}\; \mathbb{P}\,\left(\,\sum_{i \in
\Lambda_L} V_\omega(i)\, | \psi (i)|^2 < E - c_0 L^{-2}\,\right)
\\ \label{lb:LD} &\geq& \frac{1}{|\Lambda_L|}\; \mathbb{P}\,\left(\,\frac{c_2}{|\Lambda_L|}
\sum_{i \in \Lambda_{L/2}} V_\omega(i) < E-c_0 L^{-2}\,\right) \ .
\end{eqnarray}

In the last estimate, we used that for $\inorm{i\,} \leq L/2$
\begin{eqnarray}\nonumber
\psi(i) &=& \frac{1}{|| \psi_1 ||}\; (L - \inorm{i})
\\ \nonumber &\geq& \overline{c} \ \frac{L - \inorm{i}}{L^{\frac{d+2}{2}}}
\\ \nonumber &\geq& \overline{c} \ \frac{L/2}{L^{\frac{d+2}{2}}}
\\ \nonumber &=& \widetilde{c}\, L^{-d/2} \ .
\end{eqnarray}

The probability in (\ref{lb:LD}) is again a large deviation
probability. As above, the $L-$independence of the right hand side
is nonstandard. We estimate (\ref{lb:LD}) in a somewhat crude way by

\begin{eqnarray}\nonumber
\textnormal{(\ref{lb:LD})} &\geq& \frac{1}{|\Lambda_L|}\;
\mathbb{P}\,\big(\,\textnormal{For all $i \in \Lambda_{L/2}$ },\quad
V_\omega(i) < \frac{1}{c_2}(E - c_0 L^{-2})\big)
\\ \label{lb:all} &=& \frac{1}{|\Lambda_L|}\; \mathbb{P}\,\Big(\,V_\omega(0) < \frac{1}{c_2}(E-
c_0\,L^{-2})\Big)^{| \Lambda_{L/2}|} \ .
\end{eqnarray}
If we take $L$ so large that $c_0L^{-2} < \frac{E}{2}$ (i.e. $L \sim
E^{-1/2}$ as for the upper bound), we obtain
\begin{displaymath}
\textnormal{(\ref{lb:all})} \geq \frac{1}{|\,\Lambda_L\,|}
P_0\Big([0, E/2) \Big)^{c_3 L^{d}} \ .
\end{displaymath}
Using assumption (\ref{ass:PO}), we finally get
\begin{eqnarray}\nonumber
N(E) &\geq& c_4 L^{-d} E^{c_3 \kappa L^d}
\\ \nonumber &=& c_4 L^{-d} e^{(\ln E) c_3 \kappa L^d} \ .
\end{eqnarray}
We remind the reader that $\kappa$ is the exponent occurring in
(\ref{ass:PO}).

So
\begin{displaymath}
N(E) \geq c_4' E^{d/2} e^{c_3' (\ln E) E^{-d/2}} \ .
\end{displaymath}

This gives the lower bound
\begin{displaymath}
\underline{\lim}_{E \searrow E_0} \frac{\ln | \ln N(E)| }{\ln
(E-E_0)} \geq - \frac{d}{2} \ .
\end{displaymath}

\goodbreak
$\;$\\[2mm]
{\bf Notes and Remarks }\\[2mm]
There are various approaches to Lifshitz tails by now. The first is
through the Donsker-Varadhan theory of large deviations, see
\cite{dv}, \cite{nakao} and \cite{p}. Related results  and further
references can be found in \cite{bk1}. For an alternative approach,
see \cite{AS}.

The results contained in these lecture and variants can be found for
example in \cite{KMlif}, \cite{ks}, \cite{ksonder}, \cite{simon1}
and \cite{stoll2}. See also \cite{KWlif}.

For an approach using periodic approximation see \cite{klopp},
\cite{klopp2}, \cite{FK1}, \cite{FK2},  and references therein. In
\cite{metzger} the probabilistic and the spectral point of view were
combined.

There are also results on other band edges than the bottom of the
spectrum, so called internal Lifshitz tails, \cite{mezincescu},
\cite{SimInt}, \cite{Klil}, \cite{KW} and \cite{Najar1}.

Magnetic fields change the Lifshitz behavior drastically, see
\cite{bhkl}, \cite{erd}, \cite{sw}.

For a recent survey about the density of states, see \cite{KirMetz}.

\NeueSeite


\setcounter{equation}{0}
\section{The spectrum and its physical interpretation\label{ch:spec}}

\subsection{Generalized Eigenfunctions and the spectrum}\es

 In this section we explore the connection between
generalized eigenfunctions of (discrete) Hamiltonians $H$ and their
spectra. A function $f$ on $\mathbb{Z}^d$ is called {\em
polynomially bounded} \index{polynomially bounded} if
\begin{equation}\label{def:polbound}
|f(n)|~\leq~ C\;\big(1+\inorm{n}\big)^k
\end{equation}
for some constants $k,C>0$. We say that $\lambda$ is a {\em
generalized eigenvalue}\index{generalized eigenvalue} if there is a
polynomially bounded solution $\psi$ of the finite difference
equation
\begin{equation}
H \psi = \lambda \psi\ .
\end{equation}

$\psi$ is called a {\em generalized eigenfunction}\index{generalized
eigenfunction}. Note that we do not require $\psi \in
\ell^2(\mathbb{Z}^d)$! We denote the set of generalized eigenvalues
of $H$ by $\ve_g(H)$.\index{$\ve_g(H)$}

We say that the sets $A,B\in\mathcal{B}(\R)$ \emph{agree up to a set
of spectral measure zero}\index{spectral measure zero} if
$\chi_{A\setminus B}(H)=\chi_{B\setminus A}(H)=0$ where $\chi_I(H)$
is the projection valued spectral measure associated with $H$ (see
Section \ref{sec:speccal}).

The goal of this section is to prove the following theorem.
\begin{theorem}{\label{th:5}}\label{th:specge}
The spectrum of a (discrete) Hamilitonian $H$ agrees up to a set of
spectral measure zero with the set $\ve_g(H)$ of all generalized
eigenvalues.
\end{theorem}

As a corollary to the proof of Theorem \ref{th:5} we obtain the
following result

\begin{corollary}\label{cor:specge}
Any generalized eigenvalue $\lambda$ of $H$ belongs to the spectrum
$\sigma(H)$, moreover
\begin{equation}
\sigma(H)~=~\overline{\;\ve_g(H)\;}
\end{equation}
\end{corollary}

\begin{remark}
The proof shows that in Theorem \ref{th:specge} as well as in
Corollary \ref{cor:specge} the set $\ve_g(H)$ can be replaced by the
set of those generalized eigenvalues with a corresponding
generalized eigenfunction satisfying
\begin{equation}\label{rem:dhalbe}
|\,\psi(n)\,|\;\leq
\,C\,{\big(1+\inorm{n}\big)}^{\frac{d}{2}+\varepsilon}
\end{equation}
for some $\ve>0$.
\end{remark}

The proof of Theorem \ref{th:specge} and Corollary \ref{cor:specge}
we present now is quite close to \cite{simon2}, but the arguments
simplify considerably in the discrete ($\ell^2$-)\,case we consider
here.

For $\Delta \subset \R$ a Borel set, let
$\mu(\Delta)=\chi_\Delta(H)$  be the projection valued measure
associated with the self adjoint operator $H$ (see Section
\ref{sec:speccal}). Thus,

\begin{displaymath}
\langle\varphi, H\psi\rangle = \int \lambda
\;d\mu_{\varphi,\psi}(\lambda)
\end{displaymath}

with

\begin{displaymath}
\mu_{\varphi,\psi}(\Delta) = \langle\varphi, \mu(\Delta)\psi\rangle\
.
\end{displaymath}

In the case of $\H=\ell^2(\Zd)$, we set

\begin{displaymath}
\mu_{n, m}(\Delta) = \langle\delta_n, \mu(\Delta)\delta_m\rangle\ .
\end{displaymath}

If $\{\alpha_n\}_{n\in \Zd}$ is a sequence of real numbers with
$\alpha_n > 0$, $\sum \alpha_n =1$, we define

\begin{equation}\label{eq:rho}
\rho(\Delta) = \sum_{n\in \Zd} \alpha_n \;\mu_{n, n}(\Delta)\ .
\end{equation}

$\rho$ is a finite positive Borel measure of total mass
$\rho(\R)=1$. We call $\rho$ a \emph{spectral
measure}\index{spectral measure!real valued} (sometimes \emph{real
valued spectral measure} to distinguish it from $\mu$, the
\emph{projection valued spectral measure})\index{spectral
measure!projection valued}. It is easy to see that

\begin{equation}
\rho(\Delta) = 0 \qquad\tn{if and only if}\qquad \mu(\Delta) = 0
\end{equation}

Thus, $A$ and $B$ agree up to a set of spectral measure zero if
$\rho(A\setminus B)=0$ and $\rho(B\setminus A)=0$. Moreover, the
support of $\rho$ is the spectrum of $H$. Although the spectral
measure is not unique (many choices for the $\alpha_n$), its measure
class and its support are uniquely defined by (\ref{eq:rho}).

We are ready to prove one half of Theorem \ref{th:specge}, namely

\begin{proposition} \label{prop:geneig}
Let $\rho$ be a spectral measure for $H=H_0+V$. Then, for
$\rho$-almost all $\lambda$ there exists a polynomially bounded
solution of the difference equation

\begin{displaymath}
H\psi=\lambda\psi\ .
\end{displaymath}
\end{proposition}

\begin{proof}
By the Cauchy-Schwarz inequality, we have
\begin{displaymath}
|\mu_{n, m}(\Delta)| \leq \mu_{n, n}(\Delta)^{\frac{1}{2}}\ \mu_{m,
m}(\Delta)^{\frac{1}{2}}
\end{displaymath}

Consequently, the $\mu_{n,m}$ are absolutely continuous with respect
to $\rho$, i.e.

\begin{equation}
  \quad \quad\rho(\Delta)=0\quad
\Longrightarrow \quad \mu_{n,m}(\Delta)=0 \ .
\end{equation}

Hence, the Radon-Nikodym theorem tells us that there exist
measurable functions $F_{n,m}$ (densities) such that

\begin{equation}\label{eq:Fnm}
\mu_{n, m}(\Delta)=\int_{\Delta} F_{n,m}(\lambda)\;d\rho(\lambda)\ .
\end{equation}

The functions $F_{n,m}$ are defined up to sets of $\rho$-measure
zero and, since $\mu_{n,n}\geq 0$ the functions $F_{n,n}$ are non
negative $\rho$-almost surely ($\rho$-a.s.). Moreover

\begin{eqnarray}\nonumber
\rho(\Delta) &=& \sum\alpha_n \;\mu_{n,n}\\\nonumber &=&
\sum\alpha_n \int_{\Delta}
F_{n,n}(\lambda)\;d\rho(\lambda)\\
&=& \int_{\Delta} \sum\alpha_n F_{n,n}(\lambda)\;d\rho(\lambda)\ .
\end{eqnarray}

Hence, $\sum\alpha_n F_{n,n}(\lambda)=1$ \quad ($\rho$-a.s.). In
particular

\begin{equation}\label{est:Fnn}
F_{n,n}(\lambda) \leq \frac{1}{\alpha_n}\ .
\end{equation}

It follows

\begin{eqnarray}\nonumber
\left|\int_{\Delta}F_{n,m}(\lambda)\;d\rho(\lambda)\right| &=&
|\mu_{n, m}(\Delta)|\\\nonumber &\leq& \mu_{n,
n}(\Delta)^{\frac{1}{2}}\ \mu_{m,
m}(\Delta)^{\frac{1}{2}}\\\nonumber &=&
 \left(\int_{\Delta}F_{n,n}(\lambda)\;d\rho(\lambda)\right)^{\frac{1}{2}}
 \left(\int_{\Delta}F_{m,m}(\lambda)\;d\rho(\lambda)\right)^{\frac{1}{2}}\\
&\leq& \alpha_n^{-\frac{1}{2}} \alpha_m^{-\frac{1}{2}}\rho(\Delta)\
.
\end{eqnarray}

Thus

\begin{equation}\label{est:Fnm}
|F_{n,m}| \leq \alpha_n^{-\frac{1}{2}} \alpha_m^{-\frac{1}{2}}\ .
\end{equation}

Equation (\ref{eq:Fnm}) implies that for any bounded measurable
function $f$

\begin{displaymath}
\langle\delta_n, f(H)\delta_m\rangle = \int f(\lambda)
F_{n,m}(\lambda)\,d\rho(\lambda)\ .
\end{displaymath}

In particular, for $f(\lambda)=\lambda\, g(\lambda)$ ($g$ of compact
support)

\begin{align}\nonumber
& \int \lambda\;g(\lambda)\, F_{n,m}(\lambda)\,d\rho(\lambda)\\
\nonumber =&\; \langle\delta_n,H\,g(H)\,\delta_m\rangle  \\\nonumber
  =&\; \langle H\,\delta_n, g(H)\,\delta_m \rangle\\
  \nonumber =&\; \sum_{|e|=1}\,\bigl(-\langle \delta_{n+e},g(H)\,\delta_m \rangle\bigr)\;+\bigl(V(n)+2d\bigr)\,\langle\delta_n,g(H)\,\delta_m \rangle\\
\nonumber =&\;
\sum_{|e|=1}\,\bigl(-\int\,g(\lambda)\,F_{n+e,m}(\lambda)\,d\rho(\lambda)\,\bigr)\;
+\,\int g(\lambda)\,(V(n)+2d) F_{n,m}(\lambda)\,d\rho(\lambda)\\
=&\; \int g(\lambda)\; H^{(n)} F_{n,m}(\lambda)\,d\rho(\lambda)
\end{align}

where $H^{(n)}F_{n,m}(\lambda)$ is the operator $H$ applied to the
function $n\mapsto F_{n,m}(\lambda)$. Thus,

\begin{displaymath}
\int g(\lambda)\,\lambda\, F_{n,m}(\lambda)\;d\rho(\lambda)=\int
g(\lambda)\,H^{(n)}F_{n,m}(\lambda)\;d\rho(\lambda)
\end{displaymath}
for any bounded measurable function $g$ with compact support.

It follows that for $\rho$-almost all $\lambda$ and for any fixed $m
\in \Zd$, the function $\psi(n)=F_{n,m}(\lambda)$ is a solution of
$H\psi=\lambda\psi$. By (\ref{est:Fnm}) the function $\psi$
satisfies

\begin{equation}
|\,\psi(n)|\; \leq \; C_0\;\alpha_n^{-\frac{1}{2}}\ .
\end{equation}

So far, the sequence $\alpha_n$ has only to fulfill $\alpha_n>0$ and
$\sum \alpha_n=1$. Now, we choose
$\alpha_n=c\,(1+\inorm{n})^{-\beta}$ for an arbitrary $\beta > d$,
hence

\begin{displaymath}
|\,\psi(n)| \leq C\,(1+\inorm{n})^{\frac{d}{2}+\varepsilon}
\end{displaymath}

for an $\varepsilon > 0$.

This proves the proposition as well as the estimate
(\ref{rem:dhalbe}).
\end{proof}

\vspace{2mm} We turn to the proof of the opposite direction of
Theorem \ref{th:5}. As usual, \ we equip \ $\mathbb{Z}^d$ with the
norm $\inorm{n}=\max_{i=1,...,d}\, |n_i|$. So $\Lambda_L = \{|n|\le
L\}$ is a cube of side length $2L+1$.

For a subset $S$ of $\mathbb{Z}^d$ we denote by $||\psi||_S$ the
$l^2$- norm \index{$||\psi||_S$} of $\psi$ over the set $S$.

We begin with a lemma:
\begin{lemma}{\label{le:delta}}
If $\psi$ is polynomially bounded $(\neq0)$ and $l$ is a positive
integer, then there is a sequence $L_n \rightarrow \infty$ such that
\begin{displaymath}
\frac{\|\psi \|_{\Lambda_{L_n+l}}}{\|\psi \|_{\Lambda_{L_n}}}
\rightarrow 1\ .
\end{displaymath}
\end{lemma}

\begin{proof}
\ Suppose the assertion of the Lemma is wrong. Then there exists
$a>1$ and $L_0$ such that for all $L\geq L_0$
\begin{displaymath}
\| \psi \|_{\Lambda_{L+l}}\geq a \| \psi \|_{\Lambda_{L}}\ .
\end{displaymath}
So
\begin{equation}\label{est:psilower}
\| \psi \|_{\Lambda_{L_0+lk}}\geq a^k \| \psi \|_{\Lambda_{L_0}}
\end{equation}
but by the polynomial boundedness of $\psi$ we have
\begin{equation}
\| \psi \|_{\Lambda_{L_0+lk}}\leq C_1\, (L_0+l\,k)^M \leq C\, k^M
\end{equation}
for some $C,M>0$ which contradicts (\ref{est:psilower}).
\end{proof}



We are now in position to prove the second half of Theorem
\ref{th:5}.

\begin{proposition} \label{krueppel}
If the difference equation $H\psi=\lambda\psi$ admits a polynomially
bounded solution $\psi$, then $\lambda$ belongs to the spectrum
$\sigma(H)$ of $H$.
\end{proposition}

\begin{proof}
We set $\psi_L(n)=\left\{\begin{array}{ll} \psi(n) & \textnormal{for
} |n|\leq L, \\ 0 & \textnormal{otherwise\ .}
\end{array} \right.$

Set $\varphi_L=\frac{1}{||\psi_L||}\psi_L$. Then $\psi_L$ is
`almost' a solution of $H\psi=\lambda\psi$, more precisely

\begin{equation*}
(H-\lambda)\psi_L(n) = 0
\end{equation*}
\vspace{1mm}

as long as $n\not\in S_L:=\{m|\,L-1\leq|m|\leq L+1\}$ \quad and

\begin{eqnarray}\nonumber
\|(H-\lambda)\,\psi_L||^2&\leq&||\,\psi||_{S_L}^2\\\nonumber &=& \sum_{m\in S_L}|\,\psi(m)|^2 \\
&=&||\,\psi||^2_{\Lambda_{L+1}}-||\psi||^2_{\Lambda_{L-2}}\ .
\end{eqnarray}

By Lemma (\ref{le:delta}) there is a sequence $L_n\rightarrow\infty$
such that

\begin{displaymath}
\frac{||\psi||^2_{\Lambda_{L_n+1}}}{||\psi||^2_{\Lambda_{L_n-2}}}
\rightarrow 1\ ,
\end{displaymath}

so

\begin{displaymath}
||(H-\lambda)\varphi_{L_n}||^2\leq
\frac{||\psi||^2_{\Lambda_{L_n+1}}-||\psi||^2_{\Lambda_{L_n-2}}}{||\psi||_{\Lambda_{L_n-2}}}\rightarrow
0\ .
\end{displaymath}

Thus, $\varphi_{L_n}$ is a Weyl sequence for $H$ and $\lambda$ and
$\lambda\in\sigma(H)$.
\end{proof}

\begin{nproof}{Corollary}
We have already seen in Proposition \ref{krueppel} that
\begin{displaymath}
\ve_g(H)\subset \sigma(H)\ .
\end{displaymath}

Since $\sigma(H)$ is closed, it follows
\begin{displaymath}
\overline{\,\ve_g(H)\,}\subset \sigma(H)\ .
\end{displaymath}
By the Theorem \ref{th:specge}, we know that
\begin{displaymath}
\rho\left( \complement\;\ve_g(H) \right) = 0
\end{displaymath}
hence
\begin{displaymath}
\complement\;\overline{\,\ve_g(H)\,} \cap \sigma(H)
 =\complement\;\overline{\,\ve_g(H)\,}\cap \supp\rho
 =\emptyset\ .
\end{displaymath}
So,
\begin{displaymath}
\sigma(H) \subset \overline{\,\ve_g(H)\,}\ .
\end{displaymath}
\end{nproof}

\subsection{The measure theoretical decomposition of the spectrum\label{sec:measspec}}\es

The spectrum gives the physically possible energies of the system
described by the Hamiltonian $H$. Hence, if $E\notin\sigma(H)$, no
(pure) state of the system can have energy $E$. It turns out that
the fine structure of the spectrum gives important information on
the dynamical behavior of the system, more precisely on the long
time behavior of the state $\psi(t)=e^{-itH}\psi_0$.

To investigate this fine structure we have to give a little
background in measure theory. By the term {\em bounded Borel
measure} \index{bounded Borel measure} (or bounded measure, for
short) we mean in what follows a complex-valued $\sigma $-additive
function $\nu$ on the Borel sets $\mathcal{B}(\R)$ such that the
total variation
$$\|\nu\|=\sup\;\{\,\sum_i^N\,|\nu(A_i)|;\;
A_i\in\mathcal{B}(\R)\quad \mbox{pairwise disjoint }\}$$ is finite.
By a {\em positive Borel measure}\index{positive Borel measure} we
mean a non-negative $\sigma $-additive function $m$ on the Borel
sets such that $m(A)$ is finite for any bounded Borel set $A$.

A bounded Borel measure $\nu$ on $\R$ is called a \em pure point
measure \rm if $\nu$ is concentrated on a countable set, i.e. if
there is a countable set $A\in\mathcal{B}(\R)$ such that
$\nu(\R\backslash A)=0$. The points $x_i\in\R$  with
$\nu(\{x_i\})\neq0$ are called the atoms of $\nu$. A pure point
measure $\nu$ can be written as $\nu=\sum\alpha_i\delta_{x_i}$,
where $\delta_{x_i}$ is the Dirac measure at the point $x_i$ and
$\alpha_i=\nu(\{x_i\})$.

A measure $\nu$ is called \em continuous \rm if $\nu$ has no
atoms,\index{measure!continuous} \index{measure!absolutely
continuous} \index{measure! pure point} \index{measure!bounded
Borel} \index{measure!positive} \index{measure!singular continuous}
i.e. $\nu(\{x\})=0$ for all $x\in\R$.

A bounded measure $\nu$ is called \em absolutely continuous \rm with
respect to a positive measure $m$ (in short $\nu\ll m$) if there is
a measurable function $\varphi\in L^1(\nu)$ such that
$\nu=\varphi(m)$, i.e. $\nu(A)=\int_A\varphi(x)dm(x)$.

The Theorem of Radon-Nikodym \index{Radon-Nikodym} asserts that
$\nu$ is absolutely continuous with respect to $m$ if (and only if)
for any Borel set $A$, \ $m(A)=0$ implies $\nu(A)=0$.

By saying $\nu$ is \em absolutely continuous \rm we always mean
$\nu$ is absolutely continuous with respect to Lebesgue measure $L$.

A measure is called \em singular continuous \rm if it is continuous
and it lives on a set $N$ of Lebesgue measure zero, i.e.
$\nu(\{x\})=0$ for all $x\in\R$, $\nu(\R\backslash N)=0$ and
$L(N)=0$.

The Lebesgue-decomposition theorem tells us that any bounded Borel
measure $\nu$ on $\R$ admits a unique decomposition

\begin{displaymath}
\nu=\nu_{pp}+\nu_{sc}+\nu_{ac}
\end{displaymath}

where $\nu_{pp}$ \index{$\nu_{pp}$} is a pure point measure,
$\nu_{sc}$\index{$\nu_{sc}$}  a singular continuous measure and
$\nu_{ac}$ \index{$\nu_{ac}$} is absolutely continuous (with respect
to Lebesgue measure). We call $\nu_{pp}$ the pure point part of
$\nu$ etc.

Let $H$ be a self adjoint operator on a Hilbert space $\H$ with
domain $D(H)$ and $\mu$ be the corresponding \emph{projection valued
spectral measure}\index{spectral measure!projection valued} (see
Section \ref{sec:speccal}). So, for any Borel set $A\subset\R$
,\quad $\mu(A)$ is a projection operator,

\begin{displaymath}
 \langle\varphi,\mu(A)\psi\rangle=\mu_{\varphi,\psi}(A)
\end{displaymath}

is a complex valued measure and

\begin{displaymath}
 \langle\varphi, H\psi\rangle=\int\lambda\; d\mu_{\varphi,\psi}(\lambda)\ .
\end{displaymath}

We also set $\mu_\varphi=\mu_{\varphi,\varphi }$ which is a positive
measure. Note that

\begin{eqnarray}\nonumber
|\,\mu_{\psi,\varphi}(A)\,|\;&=&\;|\,\langle \psi, \mu(A)
\varphi\rangle\,|
\\\nonumber
&=&\;|\,\langle \mu(A)\psi, \mu(A) \varphi\rangle\,| \\\nonumber
&\le&\;\left(\langle \mu(A)\psi, \mu(A)
\psi\rangle\right)^{\eh}\;\left(\langle \mu(A)\varphi, \mu(A)
\varphi\rangle\,\right)^{\eh}\\\label{eq:CS}
&=&\;\mu_\psi(A)^\eh\;\mu_\varphi(A)^\eh\ .
\end{eqnarray}

We define $\H_{pp}=\{\varphi\in\H\mid\mu_\varphi\ \textnormal{is
pure point}\}$ \index{$\H_{pp}$} and analogously $\H_{sc}$
\index{$\H_{sc}$} and $\H_{ac}$ \index{$\H_{ac}$}. These sets are
closed subspaces of $\H$ which are mutually orthogonal and

\begin{displaymath}
\H=\H_{pp}\oplus\H_{sc}\oplus\H_{ac}\ .
\end{displaymath}

The operator $H$ maps each of these spaces into itself (see e.g.
\cite{rs1}). We set \quad $H_{pp}=H\mid_{\H_{pp}\cap
D(H)}$\index{$H_{pp}$}, \quad $H_{sc}=H\mid_{\H_{sc}\cap D(H)}$
\index{$H_{sc}$} , \quad $H_{ac}=H\mid_{\H_{ac}\cap D(H)}$
\index{$H_{ac}$}. We define the \em pure point spectrum
\index{spectrum!pure point } \rm $\sigma_{pp}(H)$
\index{$\sigma_{pp}(H)$} of $H$ to be the spectrum $\sigma(H_{pp})$
of $H_{pp}$, analogously the \emph{singular continuous spectrum}
$\sigma_{sc}(H)$ \index{spectrum!singular continuous}
\index{$\sigma_{sc}(H)$} of $H$ to be $\sigma(H_{sc})$ and the
\emph{absolutely continuous spectrum} $\sigma_{ac}(H)$
\index{spectrum!absolutely continuous} to be $\sigma(H_{ac})$
\index{$\sigma_{ac}(H)$}. It is clear that
\begin{displaymath}
\sigma(H)\;=\;\sigma_{pp}(H) \cup \sigma_{sc}(H) \cup \sigma_{ac}(H)
\end{displaymath}
but this decomposition of the spectrum is not a {\em disjoint} union
in general.

This measure theoretic decomposition of the spectrum is defined in a
rather abstract way and we should ask: Is there any \emph{physical}
meaning of the decomposition? The answer is \emph{YES} and will be
given in the next section.

\subsection{Physical meaning of the spectral decomposition\label{sec:RAGE}}\es

The measure theoretic decomposition of the Hilbert space and the
spectrum may look more like a mathematical subtleness than like a
physically relevant classification. In fact, in physics one is
primarily interested in long time behavior of wave packets. For
example, one distinguishes bound states and scattering states. It
turns out, there is an intimate connection between the
classification of states by their long time behavior and the measure
theoretic decomposition of the spectrum. We explore this connection
in the present section.

The circle of results we present here was dubbed `RAGE-theorem'
\index{RAGE-theorem} in \cite{cfks} after the pioneering works by
Ruelle \cite{ruelle}, Amrein, Georgescu \cite{amrein} and Enss
\cite{enss} on this topic.

If $E$ is an eigenvalue of $H$ (in the $\ell^2$-sense) and $\varphi$
a corresponding eigenfunction, then the spectral measure $\mu$ has
an atom at $E$, and $\mu_\varphi$ is a pure point measure
concentrated at the point $E$. Thus, all eigenfunctions and the
closed subspace generated by them belong to the pure point subspace
$\H_{pp}$. The converse is also true, i.e. the space $\H_{pp}$ is
exactly the closure of the linear span of all eigenvectors.

It follows that the set $\varepsilon(H)$ \index{$\ve(H)$} of all
eigenvalues of $H$ is always contained in the pure point spectrum
$\sigma_{pp}(H)$ and that $\varepsilon(H)$ is dense in
$\sigma_{pp}(H)$. The set $\varepsilon(H)$ is countable (as our
Hilbert space is always assumed to be separable), it may have
accumulation points, in fact $\varepsilon(H)$ may be dense in a
whole interval.

Let us look at the time evolution of a function $\psi \in \H_{pp}$.
To start with, suppose $\psi$ is an eigenfunction of $H$ with
eigenvalue $E$. Then

\begin{displaymath}
e^{-itH}\psi\;= \;e^{-itE}\psi
\end{displaymath}

so that $|e^{-itH}\psi(x)|^2$ is independent of the time $t$. We may
say that the particle, if starting in an eigenstate, stays where it
is for all $t$. It is easy to see that for general $\psi$ in
$\H_{pp}$ the function $e^{-itH}\psi(x)$ is almost periodic in $t$.
A particle in a state $\psi$ in $\H_{pp}$ will stay inside a compact
set with high probability for arbitrary long time, in the following
sense:

\begin{theorem}\label{th:pp}
Let $H$ be a self adjoint operator on $\ell^2(\Z^d)$, take
$\psi\in\H_{pp}$ and let $\Lambda_L$ denote a cube in $\Z^d$
centered at the origin with side length $2L+1$.

Then
\begin{align}\label{eq:pp}
\lim_{L\rightarrow\infty}\;\sup_{t\ge
0}\;\left(\,\sum_{x\in\Lambda_L}\,|\,e^{-itH}\psi(x)\,|^{\,2}\right)\;\;&=\;\|{\psi}\|^2\\
\intertext{and} \label{eq:pp1} \lim_{L\rightarrow\infty}\;\sup_{t\ge
0}\;\left(\,\sum_{x\not\in\Lambda_L}\,|\,e^{-itH}\psi(x)\,|^{\,2}\right)\;\;&=\;0\
.
\end{align}

\end{theorem}

\begin{remark}
Equations (\ref{eq:pp}), (\ref{eq:pp1}) can be summarized in the
following way: Given any error bound $\varepsilon > 0$ there is a
cube $\Lambda_L$ such that for arbitrary time $t$ we find the
particle inside $\Lambda_L$ with probability $1-\varepsilon$. In
other words, the particle will not escape to infinity. Thus a state
$\psi\in\H_{pp}$ can be called a {\em bound state} \index{bound
state}.
\end{remark}

\begin{proof}
Since $\ut$ is unitary, we have for all $t$
\begin{align}
\notag {\no{\psi}\,}^2\quad~=&~\quad{\no{\ut\psi}\,}^2\\
=&~\sum_{x\in\L}\;|\,\ut\psi\,(x)\,|^2~+~\sum_{x\not\in\L}\;|\,\ut\psi\,(x)\,|^2\
. \label{eq:unitary}
\end{align}
Consequently, (\ref{eq:pp}) follows from (\ref{eq:pp1}).

Above we saw that (\ref{eq:pp1}) is valid for eigenfunctions $\psi$.
To prove it for other vectors in $\H_{pp}$, we introduce the
following notation: By $P_L$ we denote the projection onto
$\complement \Lambda_L$. Then equation (\ref{eq:pp1}) claims that

\begin{displaymath}
\|\,P_L\,e^{-itH}\psi\|\,\rightarrow\,0
\end{displaymath}

uniformly in $t$ as $L\rightarrow\,\infty$. If $\psi$ is a (finite)
linear combination of eigenfunctions, say
$\psi=\sum_{m=1}^M\,\alpha_k \psi_k$, $H\psi_k=E_k\psi_k$, then

\begin{eqnarray}\nonumber
\|\,P_L\,e^{-itH}\psi\|\;&=&\;\|\,\sum_{m=1}^M\,\alpha_k\,P_L\,e^{-itH}\psi_k\| \\
                         &\le&\;\sum_{m=1}^M\,|\alpha_k|\,\|\,
                         P_L\,e^{-itH}\psi_k\|
                         =\;\sum_{m=1}^M\,|\alpha_k|\,\|\,
                         P_L\,e^{-itE_k}\psi_k\| \nonumber\\
                         &=&\; \sum_{m=1}^M\,|\alpha_k|\,\|\,
                         P_L\,\psi_k\|\ .
\end{eqnarray}

By taking $L$ large enough, each term in the sum above can be made
smaller then
$\left(\sum_{m=1}^M\,|\alpha_k|\,\right)^{-1}\varepsilon$ \ .

If now $\psi$ is an arbitrary element of $\H_{pp}$, there is a
linear combination of eigenfunctions
$\psi^{(M)}=\sum_{m=1}^M\,\alpha_k \psi_k$ such that $\|\psi -
\psi^{(M)}\|<\varepsilon$. We conclude

\begin{eqnarray}\nonumber
\|\,P_L\,e^{-itH}\psi\|\;&\le &\|\,P_L\,e^{-itH}\psi^{(M)}\|\,+\,\|\,P_L\,e^{-itH}(\psi -\psi^{(M)})\| \\
                         &\le & \|\,P_L\,e^{-itH}\psi^{(M)}\|\,+\,\|\psi -
                         \psi^{(M)}\|\ .
\end{eqnarray}

By taking $M$ large enough the second term of the right hand side
can be made arbitrarily small. By choosing $L$ large, we can finally
make the first term small as well.
\end{proof}
\goodbreak
We turn to the interpretation of the continuous spectrum. Let us
start with a vector $\psi\in\H_{ac}$. Then, by definition the
spectral measure $\mu_\psi$ is absolutely continuous. From estimate
(\ref{eq:CS}) we learn that $\mu_{\varphi,\psi}$ is absolutely
continuous for any $\varphi\in\H$ as well. It follows that the
measure $\mu_{\varphi,\psi}$ has a density $h$ with respect to
Lebesgue measure, in fact $h\in L^1$. Hence, for any $\phi\in\H$ and
$\psi\in\H_{ac}$

\begin{eqnarray}\nonumber
\langle \varphi, \ut \psi \rangle &=&\; \int\,e^{-it\lambda}\;d\mu_{\varphi,\psi}(\lambda) \\
                               &=&\;
                               \int\,e^{-it\lambda}\,h(\lambda)\;d\,\lambda\ .
\end{eqnarray}

The latter expression is the Fourier transform of the
($L^1$-)function $h$. Thus, by the Riemann-Lebesgue-Lemma (see e.g.
\cite{rs1}), it converges to $0$ as $t$ goes to infinity.

We warn the reader that the decay of the Fourier transform of a
measure does {\em not} imply that the measure is absolutely
continuous. There are examples of singular continuous measures whose
Fourier transforms decay.

If the underlying Hilbert space is $\ell^2(\Z^d)$ we may choose
$\varphi=\delta_x$ for any $x\in\Z^d$, then $\langle \varphi, \ut
\psi \rangle = \ut \psi (x)$, thus we have immediately

\begin{theorem}\label{th:ac}
Let $H$ be a self adjoint operator on $\ell^2(\Z^d)$, take
$\psi\in\H_{ac}$ and let $\Lambda$ denote a finite subset of $\Z^d$.
Then

\begin{equation}\label{eq:ac}
\lim_{t\rightarrow\infty}\left(\;\sum_{x\in\Lambda}\,|\,e^{-itH}\psi(x)\,|^{\,2}\right)\;=\;0
\end{equation}

or equivalently
\begin{equation}\label{eq:ac1}
\lim_{t\rightarrow\infty}\left(\;\sum_{x\not\in\Lambda}\,|\,e^{-itH}\psi(x)\,|^{\,2}\right)\;=\;\|\,\psi\,\|^2\
.
\end{equation}

\end{theorem}

\begin{remark}
As $\psi\in\H_{pp}$ may be interpreted as a particle staying
(essentially) in a finite region for all time, a particle
$\psi\in\H_{ac}$ runs out to infinity as time evolves (and came out
from infinity as $t$ goes to $-\infty$). So, in contrast to the
bound states ($\psi\in\H_{pp}$), we might call the states in
$\H_{ac}$ {\em scattering states} \index{scattering states}.
Observe, however, that this term is used in scattering theory in a
more restrictive sense.
\end{remark}

In the light of these results, states in the pure point subspace are
interpreted as bound states with low mobility. Consequently,
electrons in such a state should not contribute to the electrical
conductivity of the system. In contrast, states in the absolutely
continuous subspace are highly mobile. They are the carrier of
transport phenomena like conductivity.

A (relatively) simple example of a quantum mechanical system is a
Hydrogen atom. After removal of the center of mass motion it
consists of one particle moving under the influence of a Coulomb
potential $V(x)=-\frac{Z}{|x|}$. The spectrum of the corresponding
Schrödinger operator consists of infinitely many eigenvalues
$\left(-\frac{C}{n^2}\right)$ which accumulate at $0$ and the
interval $[0,\infty)$ representing the absolutely continuous
spectrum. The eigenfunction corresponding to the negative
eigenvalues represent electrons in bound states, the orbitals. The
states of the a.c.-spectrum correspond to electrons coming from
infinity being scattered at the nucleus and going off to infinity
again.

The Hydrogen atom is typical for the classical picture of a quantum
system: Above an energy threshold there is purely absolutely
continuous spectrum due to scattering states, below the threshold
there is a finite or countable set of eigenvalues accumulating at
most at the threshold. For the harmonic oscillator there is a purely
discrete spectrum, for periodic potentials the spectrum consists of
bands with purely absolutely continuous spectrum. Until a few
decades ago almost all physicists believed that all quantum systems
belonged to one of the above spectral types.

We have seen above that there may be pure point spectrum which is
dense in a whole interval and we will see this is in fact typically
the case for random operators.

So far, we have not discussed the long time behavior for states in
the singular continuous spectrum. Singularly continuous spectrum
seems to be particularly exotic and unnatural. In fact, one might
tend to believe it is only a mathematical sophistication which never
occurs in physics. This point of view is proved to be wrong. In
fact, singularly continuous spectrum is typical for systems with
aperiodic long range order, such as quasicrystals.

The definition of singularly continuous measures is a quite indirect
one. Indeed, we have not defined them by what they are but rather by
what they are not. In other words: Singular continuous measures are
those that remain if we remove pure point and absolutely continuous
measures. There is a characterization of {\em continuous} measures
(i.e. those without atoms) by their Fourier transform which goes
back to Wiener.

\begin{theorem}[Wiener]\label{Wiener}\index{Wiener's Theorem}
Let $\mu$ be a bounded Borel measure on $\R$ and denote its Fourier
transform by $\hat{\mu}(t) = \int e^{-it\lambda}\,d\mu(\lambda)$.
Then

\begin{displaymath}
\lim_{T\to\infty}\,\frac{1}{T}\int_0^T\,|\,\hat{\mu}(t)|^2\,dt =
\sum_{x\in\R}\,|\,\mu(\{x\})|^2\ .
\end{displaymath}

\end{theorem}
\vspace{0.5cm}
\begin{corollary}
$\mu$ is a continuous measure if and only if
$$\lim_{T\to\infty}\,\frac{1}{T}\int_0^T\,|\,\hat{\mu}(t)|^2\,dt\;=\;0\ .$$
\end{corollary}

\begin{nproof}{Theorem}
\begin{eqnarray}\nonumber
&&\frac{1}{T}\int_0^T|\hat{\mu}(t)|^2dt\\\nonumber&&=\frac{1}{T}\int_0^T\;\left(\int_\R
e^{-it\lambda}\,d\mu(\lambda)\ \int_\R
e^{it\varrho}\,d\bar{\mu}(\varrho)\right)\,dt\\&&=\int_\R\int_\R\;\left(\frac{1}{T}\int_0^T
e^{it(\varrho - \lambda)}dt\right)\; d\bar{\mu}(\varrho)\;
d\mu(\lambda)\ .
\end{eqnarray}

Here $\bar{\mu}$ denotes the complex conjugate of the measure $\mu$.
The functions

\begin{displaymath}
f_T(\varrho, \lambda)=\frac{1}{T}\int_0^T e^{it(\varrho -
\lambda)}dt
\end{displaymath}

are bounded by one. Moreover for $\varrho\neq\lambda$

\begin{displaymath}
f_T(\varrho, \lambda)=\frac{1}{i(\varrho -
\lambda)T}(e^{iT(\varrho-\lambda)}-1)\rightarrow\ 0 \quad
\textnormal{as}\quad T \rightarrow\infty
\end{displaymath}

and

\begin{displaymath}
f_T(\varrho, \varrho)=1\ .
\end{displaymath}

\vspace{0.5cm}

Thus, $f_T(\varrho, \lambda)\rightarrow\chi_D(\varrho, \lambda)$
with $D=\{(x,y)\mid x=y\}$. By Lebesgue's dominated convergence
theorem, it follows that

\begin{eqnarray}\nonumber
&&\frac{1}{T}\int_0^T |\hat{\mu}(t)|^2dt\\\nonumber &\rightarrow&
\int\int\chi_D(\varrho,
\lambda)\,d\bar{\mu}(\varrho)d\mu(\lambda)\\&=&\int\bar{\mu}(\{\lambda\})\,d\mu(\lambda)=\sum|\mu(\{\lambda\})|^2\
.
\end{eqnarray}

\end{nproof}

This enables us to prove an analog of Theorem \ref{th:pp} and
Theorem \ref{th:ac} for continuous measures. It says that states in
the singularly continuous subspace represent particles which go off
to infinity (at least) in the time average.
\goodbreak
\begin{theorem}\label{th:c}
Let $H$ be a self adjoint operator on $\ell^2(\Z^d)$, take
$\psi\in\H_c$ and let $\Lambda$ be a finite subset of $\Z^d$.

Then

\begin{equation}\label{eq:thc1}
\lim_{T\rightarrow\infty}\;\frac{1}{T}\int_0^T\left(\sum_{x\not{\in}\Lambda}|e^{-itH}\psi(x)|^2\right)dt\;=\;\|\,\psi\,\|^2
\end{equation}

or equivalently

\begin{equation}\label{eq:thc2}
\lim_{T\rightarrow\infty}\;\frac{1}{T}\int_0^T\left(\,\sum_{x\in\Lambda}|e^{-itH}\psi(x)|^2\right)\,dt\;=\;0\
.
\end{equation}

\end{theorem}

\begin{proof}
The equivalence of (\ref{eq:thc1}) and (\ref{eq:thc2}) follows from
(\ref{eq:unitary}).

We prove (\ref{eq:thc2}). Let $\psi$ be in $\H_{c}$. From estimate
(\ref{eq:CS}), we learn that for any $x\in\Z^d$ the measure
$\mu_{\delta_x,\psi}$ is continuous. We have

\begin{displaymath}
\frac{1}{T}\int_0^T\sum_{x\in\Lambda}|e^{-itH}\psi(x)|^2\,dt\,=\;\sum_{x\in\Lambda}\;\;\frac{1}{T}\int_0^T\,|\hat{\mu}_{\delta_x,\psi}|^2\,dt\
.
\end{displaymath}

The latter term converges to $0$ by Theorem (\ref{Wiener}).
\end{proof}

We close this section with a result which allows us to express the
projections onto the pure point subspace and the absolutely
continuous subspace as dynamical quantities.
\goodbreak

\begin{theorem}\label{th:dyn}
Let $H$ be a self adjoint operator on $\ell^2(\Z^d)$, let $P_c$ and
$P_{pp}$ be the orthogonal projection onto $\H_c$ and $\H_{pp}$
respectively, and let $\Lambda_L$ denote a cube in $\Z^d$ centered
at the origin with side length $2L+1$. Then, for any
$\psi\in\ell^2(\Z^d)$

\begin{equation}\label{eq:dyn1}
\|{P_c\,\psi}\|^2\;\;=\;\lim_{L\rightarrow\infty}\;\lim_{T\to\infty}\,
\frac{1}{T}\,\int_0^T\;\left(\,\sum_{x\not\in\Lambda_L}\,|\,e^{-itH}\psi(x)\,|^{\,2}\right)\;dt
\end{equation}
and
\begin{equation}\label{eq:dyn2}
\|{P_{pp}\,\psi}\|^2\;\;=\;\lim_{L\rightarrow\infty}\;\lim_{T\to\infty}\,\frac{1}{T}\,\int_0^T\;
\left(\,\sum_{x\in\Lambda_L}\,|\,e^{-itH}\psi(x)\,|^{\,2}\right)\;dt\
.
\end{equation}

\end{theorem}
\goodbreak
\begin{proof}
As in (\ref{eq:unitary}) we have

\begin{align}
\notag ||\,P_c\,\psi\,||^2~=&~\frac{1}{T}\,\int_0^T\;\big(\,\sum_{x\not\in\LL}\,|\,\ut\psi\,(x)\,|^2\,\big)\;dt\\
&-\frac{1}{T}\,\int_0^T\;\big(\,\sum_{x\not\in\LL}\,|\,\ut P_{pp}\psi\,(x)\,|^2\,\big)\;dt\notag\\
&+\frac{1}{T}\,\int_0^T\;\big(\,\sum_{x\in\LL}\,|\,\ut
P_c\psi\,(x)\,|^2\,\big)\;dt\ .\label{eq:cpp}
\end{align}
By Theorem \ref{th:pp} and Theorem \ref{th:c} the second and the
third term in (\ref{eq:cpp}) tend to zero as $T$ and (then) $L$ go
to infinity. This proves (\ref{eq:dyn1}). Assertion (\ref{eq:dyn2})
is proved in a similar way.
\end{proof}
\goodbreak
$\;$\\[2mm]
{\bf Notes and Remarks }\\[2mm]
Most of the material in this chapter is based on \cite{cfks},
\cite{simon2} and the lecture notes \cite{Teschl} by Gerald Teschl.
Teschl's excellent notes are only available on the internet. For
further reading we recommend \cite{amrein}, \cite{bere},
\cite{enss}, \cite{poerscke}, \cite{ruelle} and \cite{weid}.
\NeueSeite


\setcounter{equation}{0}
\section{Anderson localization\label{ch:localization}}


\subsection{What physicists know}\es

Since the ground breaking work of P. Anderson in the late fifties,
physicists like Mott, Lifshitz, Thouless and many others have
developed a fairly good knowledge about the measure theoretic nature
of the spectrum of random Schrödinger operators, i.e. about the
dynamical properties of wave packets.

By Theorem \ref{th:spAnd} (see also Theorem \ref{th:specconst}) we
know that the (almost surely non random) spectrum $\Sigma$ of
$H_\omega$ is given by $\supp(P_0)+[0, 4d]$ where $P_0$ is the
probability distribution of $V_\omega(0)$. Thus if $\supp(P_0)$
consists of finitely many points or intervals the spectrum $\Sigma$
has a band structure in the sense that it is a union of (closed)
intervals.

In the following we report on the picture physicists developed about
the measure theoretic structure of the spectrum of $H_\omega$. This
picture is supported by convincing physical arguments and is
generally accepted among theoretical physicists. Only a part of it
can be shown with mathematical rigor up to now. We will discuss this
issue in the subsequent sections.

There is a qualitative difference between one dimensional disordered
systems \linebreak($d=1$) and higher dimensional ones ($d\geq 3$).
For one dimensional (disordered) systems one expects that the whole
spectrum is pure point. Thus, there is a complete system of
eigenfunctions. The corresponding (countably many) eigenvalues form
a dense set in $\Sigma$ ($= \cup [a_i, b_i]$). The eigenfunctions
decay exponentially at infinity. This phenomenon is called
\emph{Anderson localization} \index{Anderson localization} or
\emph{exponential localization.} \index{exponential localization} In
the light of our discussion in section \ref{ch:spec}, we conclude
that Anderson localization corresponds to low mobility of the
electrons in our system.
 Thus, one dimensional disordered systems (`thin wires with impurities') should
have low or even vanishing conductivity.

In arbitrary dimension, an ordered quantum mechanical system should
have purely absolutely continuous spectrum. This is known for
periodic potentials in any dimension. Thus, in one dimension, an
\emph{arbitrarily small disorder} will change the total spectrum
from absolutely continuous to pure point and hence a conductor to an
insulator. Anderson localization in the one dimensional case can be
proved with mathematical rigor for a huge class of disordered
systems. We will not discuss the one dimensional case in detail in
this paper.

In dimension $d\geq 3$ the physics of disordered systems is much
richer (and consequently more complicated). As long as the
randomness is not too strong Anderson localization occurs only near
the band edges of the spectrum. Thus near any band edge $a$ there is
an interval $[a, a+\delta]$ (resp. $[a-\delta, a]$) of pure point
spectrum and the corresponding eigenfunctions are `exponentially
localized' in the sense that they decay exponentially fast at
infinity.

Well inside the bands, the spectrum is expected to be absolutely
continuous at small disorder ($d\geq 3$).  Since the corresponding
(generalized) eigenfunctions are certainly not square integrable,
one speaks of \emph{extended states}\index{extended states}  or
\emph{Anderson delocalization} \index{Anderson delocalization} in
this regime. If the randomness of the system increases the pure
point spectrum will expand and the absolutely continuous part of the
spectrum will shrink correspondingly. So, according to physical
intuition, there is a phase transition from an insulating phase to a
conducting phase. A transition point between these phases is called
a \emph{mobility edge}.\index{mobility edge}

At a certain degree of randomness, the a.c. spectrum should be
`eaten up' by the pure point spectrum. The physical implications of
the above picture are that we expect an energy region for which the
corresponding states do not contribute to the conductance of the
system (pure point spectrum) and an energy region corresponding to
states with good mobility which constitute the conductivity of the
system (a.c. spectrum).

In the above discussion we have deliberately avoided the case of
space dimension $d=2$. The situation in two dimensions was under
debate in the theoretical physics community until a few years ago.
At present, the general believe seems to be that we have complete
Anderson localization for $d=2$ similar to the case
 $d=1$. However, the pure point spectrum is expected to be less stable for $d=2$, for example
a magnetic field might be able to destroy it.

\bigskip

\subsection{What mathematicians prove}\es

For more than 25 years, mathematicians have been working on random
Schrödinger operators. Despite of this, the mathematically rigorous
knowledge about these operators is far from being complete.

As mentioned above, the results on the one dimensional case are
fairly satisfactory. One can prove Anderson localization for all
energies for a huge class of one dimensional random quantum
mechanical systems.

For quite a number of models in $d \geq 2$ we also have proofs of
Anderson localization, even in the sense of dynamical localization
(see Section \ref{sec:further}), at low energies or high disorder.
There are also results about localization at spectral edges (other
than the bottom of the spectrum).

The model which is best understood in the continuous case is the
alloy-type model with potential (\ref{alloy})

\begin{equation}\label{alloy7}
V_\omega(x)= \sum q_i(\omega) f(x-i)\ .
\end{equation}

The $q_i$ are assumed to be independent with common distribution
$P_0$. Until very recently, all known localization proof (for $d
\geq 2$) required some kind of regularity of the probability measure
$P_0$, for example the existence of a bounded density with respect
to Lebesgue measure. In any case, these assumptions {\em exclude}
the case when $P_0$ is concentrated in finitely many points. From a
physical point of view such measures with a finite support are
pretty natural. They model a random alloy with finitely many
constituents. A few years ago, Bourgain and Kenig \cite{bourgain}
proved localization for the Bernoulli alloy type model, i.e. a
potential as in (\ref{alloy7}) with $P_0$ concentrated on $\{0,1\}$.

Their proof works in the continuous case, but it does \emph{not} for
the (discrete) Anderson model. In the continuous case Bourgain and
Kenig strongly use that eigenfunctions of a Schr\"{o}dinger operator
on $\Rd$ can not decay faster than a certain exponential bound. This
is a strong quantitative version of the unique continuation theorem
which says that a solution of the Schr\"odinger equation which is
zero on an open set vanishes everywhere.

Such a unique continuation theorem is wrong on the lattice, so a
fortiori the lower bound on eigenfunctions is not valid on $\Zd$.
This is the main reason why the proof by Bourgain-Kenig does not
extend to the discrete case.

Using ideas from Bourgain-Kenig \cite{bourgain}, Germinet, Hislop
and Klein \cite{ghk} proved Anderson localization for the Poisson
model (\ref{poisson}). Until their paper nothing was known about
Anderson localization for the Poisson model in dimension $d\ge 2$.
(For $d=1$ see \cite{stolzp}).

It is certainly fair to say that by now mathematicians know quite a
bit about  Anderson {\em localization}, i.e. about the insulating
phase.

The contrary is true for Anderson {\em delocalization}. There is
{\em no} proof of existence of absolutely continuous spectrum for
{\em any} of the models we have discussed so far. In particular it
is not known whether there \emph{is} a conducting phase or a
mobility edge at all.

Existence of absolutely continuous spectrum is known, however, for
the so called Bethe lattice (or Cayley tree). This is a graph
("lattice") without loops (hence a tree) with a fixed number of
edges at every site. One considers the graph Laplacian on the Bethe
lattice, which is analogously defined to the Laplacian on the graph
$\mathbb{Z}^d$ (see \cite{Klein1}, \cite{Klein2}, \cite{Klein3}) and
an independent identically distributed potential on the sites of the
graph.

There are also `toy'-models similar to the Anderson model but with
non identically distributed $V_\omega(i)$ which are more and more
diluted (or `weak') as $\inorm{i}$ becomes large. For these models,
the mobility can be determined. (see \cite{KKO}, \cite{Kirscat} and
\cite{HundKi}).
\bigskip

\subsection{Localization results}\label{sec:locres}\es

We state the localization result we are going to prove in the next
chapters. For convenience, we repeat our assumptions. They are
stronger than necessary but allow for an easier, we hope more
transparent, proof.

{\em Assumptions:}
\begin{enumerate}
    \item $H_0$ is the finite difference Laplacian  on
    $\ell^2(\mathbb{Z}^d)$.

    \item $V_\omega(i)$, $i \in \mathbb{Z}^d$ are independent random
    variables with a common distribution $P_0$.

    \item $P_0$ has a bounded density $g$, i.e. $\mathbb{P}(V_\omega(i) \in A)=P_0(A)=\int_A g(\lambda) d\lambda$ and $||\,g\,||_\infty < \infty$.

    \item $\supp P_0$ is compact.
\end{enumerate}

\begin{definition}\label{def:specloc}
We say that the random operator $H_\omega$ exhibits \emph{spectral
localization}\index{spectral localization} in an energy interval $I$
(with $I\cap\sigma(H_\omega)\not=\emptyset$) if for $\P$-almost all
$\omega$
\begin{equation}\label{eq:specloc}
\sigma_{c}(H_\omega) \cap I~=~\emptyset\ .
\end{equation}
\end{definition}

We will show spectral localization for low energies and for strong
disorder. To measure the degree of disorder of $P_0=g\, d\lambda$,
we introduce the `disorder parameter' $\delta(g):=
||g||^{-1}_\infty$. If $\delta(g)$ is large, i.e. $||g||_\infty$ is
small, then the probability density $g$ (recall $\int g =1$) is
rather extended. So one may, in deed, say that $\delta(g)$ large is
an indicator for large disorder. (If $\delta(g)$ is small then $g$
might be concentrated near a small number of points. This, however,
is not a convincing indicator of small disorder.) Let us denote by
$E_0$ the bottom of the (almost surely constant) spectrum of
$H_\omega=H_0+V_\omega$.

In the following chapters we will prove:
\goodbreak
\begin{theorem} \label{AndersonLocalization}
There exists $E_1 > E_0 = \inf(\sigma(H_\omega))$ such that the
spectrum of $H_\omega$ exhibits spectral decomposition in the
interval $I=[E_0,E_1]$.

In particular, the spectrum inside $I$ is pure point almost surely
and the corresponding eigenfunctions decay exponentially.
\end{theorem}

\begin{theorem} \label{AndersonLocalization2}
For any interval $I\neq \emptyset$, there is a $\delta_0$ such that
for any $\delta(g)\geq \delta_0$ the operator of $H_\omega$ exhibits
spectral localization in $I$.

The spectrum inside $I$ is pure point almost surely and the
corresponding eigenfunctions decay exponentially.
\end{theorem}

\subsection{Further Results}\label{sec:further}\es

As we discussed in the previous chapter, physicists are not
primarily interested in spectral properties of random Hamiltonians
but rather in dynamical properties, i.e. in the longtime behavior of
$\utr$. Consequently Anderson localization should have dynamical
consequences, as we might expect from the considerations in section
\ref{sec:RAGE}.

It seems reasonable to expect that the following property holds in
the localization regime.

\begin{definition}\label{def:dynloc}
We say that the random operator $H_\omega$ exhibits \emph{dynamical
localization}\index{dynamical localization} in an energy interval
$I$ (with $I\cap\sigma(H_\omega)\not=\emptyset$) if for all
$\varphi$ in the Hilbert space and all $p\geq 0$
\begin{equation}\label{eq:dynloc}
\sup_{t\in\R}\;\norm{\;|X|^p\,\utr\,\chi_I(H_\omega)\,\varphi\,}~<~\infty
\end{equation}
for $\P$-almost all $\omega$.
\end{definition}
Above, $\chi_I(H_\omega)$ denotes the spectral projection for
$H_\omega$ onto the interval $I$  (see Section \ref{sec:speccal})
and $|X|$ is the multiplication operator defined by
$|X|\,\psi(n)=\inorm{n}\, \psi(n)$.

Intuitively, dynamical localization tells us that the particle is
concentrated near the origin uniformly for all times. We will not
prove dynamical localization here. We refer to the references given
in the notes and in particular to the review \cite{KleinSurv}.

We turn to the question of the relation between spectral and
dynamical localization.

\begin{theorem}
Dynamical localization implies spectral localization.
\end{theorem}

\begin{proof}
From Theorem \ref{th:dyn} we know
\begin{equation}\label{eq:dyn12}
\|{P_c\,\psi}\|^2\;\;=\;\lim_{L\rightarrow\infty}\;\lim_{T\to\infty}\,
\frac{1}{T}\,\int_0^T\;\left(\,\sum_{j\,\not\in\Lambda_L}\,|\,e^{-itH}\psi(j)\,|^{\,2}\right)\;dt\
.
\end{equation}

For $\;\psi=P_I(H_\omega)\,\varphi$, we have
\begin{align}
\sum_{j\,\not\in\Lambda_L}\,|\,e^{-itH}\psi(j)\,|^{\,2}~&=~\sum_{\inorm{j\,}>L}\,\frac{1}{\inorm{j}^{2p}}
\;\Big|\;|X|^p\;e^{-itH}\psi(j)\,\Big|^{\,2}\notag\\
&\leq~\big|\big|\,|X|^p\;\utr\psi\,\big|\big|\;\sum_{\inorm{j\,}>L}\,\frac{1}{\inorm{j}^{2p}}
\end{align}
By (\ref{eq:dynloc}) we have that
\begin{align}
\lim_{T\to\infty}\,
\frac{1}{T}\,\int_0^T\;\big|\big|\,|X|^p\;\utr\psi\,\big|\big|\;dt~\leq~C~<~\infty\
.
\end{align}

Thus, for $p$ large enough,
\begin{align}
\|{P_c\,\psi}\|^2\;\;\leq\;\;C\;\lim_{L\rightarrow\infty}\;\sum_{\inorm{j\,}>L}\,\frac{1}{\inorm{j}^{2p}}~~=0\
.
\end{align}
Hence, there is only pure point spectrum inside the interval $I$.
\end{proof}

It turns out that the converse is not true, in general. There are
examples of operators with pure point spectrum without dynamical
localization \cite{DelRioJLS}.

$\;$\\[2mm]\goodbreak
{\bf Notes and Remarks }\\[2mm]
For an overview on the physics of Anderson localization /
delocalization we refer to the papers \cite{an}, \cite{lifshitz1}
and \cite{thou1}, \cite{thou2}. For the mathematical aspects we
refer to \cite{cl}, \cite{pf} and \cite{stoll1}.

In this lecture notes we have to omit many important results about
the one dimensional case. We just mention a few of the most
important papers about one dimensional localization here:
\cite{gmp}, \cite{m}, \cite{p1}, \cite{kotani}, as well as
\cite{ckm}, \cite{carmona}, \cite{dss}.

In the multidimensional case there exist two quite different
approaches to localization. The  first (in chronological order) is
the multiscale analysis based on the fundamental paper
\cite{froespe}. This is the method we are going to present in the
following chapters. For further references see the literature cited
there.

The second method, the method of fractional moments, is also called
the Aizenman-Molchanov method after the basic paper
\cite{AizenMolch}. At least for the lattice case, this method is in
many ways easier than the multiscale analysis. Moreover, it gives a
number of additional results. On the other hand its adaptation to
the continuous case is rather involved. We refer to \cite{aizenm},
\cite{Graf}, \cite{ag}, \cite{asfh}, \cite{amcont} for further
developments. We will not discuss this method here due to the lack
of space and time.

It was realized by Martinelli and Scoppolla \cite{MartScop} that the
result of multiscale analysis implies absence of a.c. spectrum. The
first proofs of spectral localization were given independently in
\cite{FMSS} (see also \cite{dk}), \cite{DLS} and \cite{SimWol} . The
latter papers develop the method of spectral averaging which goes
partly back to \cite{Kotsa}.

For delocalization on the Bethe lattice see: \cite{Klein1},
\cite{Klein2}, \cite{Klein3}. See also \cite{AizenSW1},
\cite{AizenSW} and \cite{FroeseHS} for new proofs and further
developments.

\emph{Dynamical delocalization} was shown for a random dimer model
in \cite{jss} and for a random Landau Hamiltonian in \cite{gks}.
Dynamical delocalization means that dynamical localization is
violated in some sense. It does not imply delocalization in the
sense of a.c. spectrum. Moreover, in the above cited papers
dynamical localization is only shown at special energies of Lebesgue
measure zero.

Delocalization for potentials with randomness decaying at infinity
was investigated in \cite{Krishna}, \cite{Krishna1}, \cite{Kirscat},
\cite{bour}, \cite{rs}. A localization / delocalization transition
was proved for such potentials in \cite{KKO}, \cite{HundKi}.

The result, that dynamical localization implies spectral
localization was proved in \cite{cfks}, partly following
\cite{kuso}. An example with spectral localization which fails to
exhibit dynamical localization was given in \cite{DelRioJLS}.

De Bi\`{e}vre and Germinet \cite{bg} proved dynamical localization
for the (multidimensional) Anderson model (with the same assumptions
as in section \ref{sec:locres}. Damanik and Stollmann \cite{ds}
proved that the multiscale analysis actually implies dynamical
localization. They proved a version of dynamical localization
(strong dynamical localization) which is stronger than ours.

Dynamical localization in the framework of the fractional moment
method is investigated in the work \cite{aizenm}.

There are various even stronger versions of dynamical localization,
we just mention strong Hilbert-Schmidt dynamical localization which
was proven by Germinet and Klein \cite{gk1}. We refer to the survey
\cite{KleinSurv} by Abel Klein for this kind of questions.

In theoretical physics, the theory of conductivity goes much beyond
a characterization of the spectral type of the Hamiltonian. One of
the main topics is the linear response theory and the Kubo-formula.
This approach is investigated from a mathematical point of view in
\cite{ag}, \cite{BouclGKS}, \cite{KLM} (see also \cite{KLP}).

\NeueSeite


\setcounter{equation}{0}
\section{The Green's function and the spectrum}

\subsection{Generalized eigenfunctions and the decay of the Green's function}\es

Here we start to prove Anderson localization via the multiscale
method. The proof will require the whole rest of this text. For the
reader who might get lost while trying to understand the proof, we
provided a roadmap through these chapters in chapter
\ref{ch:append}.

We begin our discuss of multiscale analysis. This method is used to
show exponential decay of Green's functions. In this section we
investigate some of the consequences of that estimate on the
spectral properties of $H_\omega$. The multiscale estimates are
discussed in the next chapter.

Let us start by defining what we mean by exponential decay of
Green's functions .

We recall some of the notations introduced in previous chapters.
$\Lambda_L(n)$ is the cube of side length $(2L+1)$ centered at
$n\in\Zd$ (see (\ref{eq:cube})), and $\Lambda_L$ denotes a cube
around the origin. $\inorm{m} =\sup_{i=1, \ldots, d}\;|m_i|$.

The inner boundary $\partial^-\Lambda_L(n)$ of $\L_L(n)$ consists of
the outermost layer of lattice points in $\Lambda_L(n)$, namely (see
\ref{eq:inbound})

\begin{eqnarray}
\partial^-\Lambda_L(n)&=&  \{m \in \mathbb{Z}^d\; | \ m \in \Lambda_L(n), \ \exists \ m' \not \in
\Lambda_L(n) \ \ (m,m') \in  \partial \Lambda_L(n) \} \quad \nonumber \\
 &=& \{m\in\Zd\ \mid\ \inorm{m-n} =L\}\ .
\end{eqnarray}

Similarly, the outer boundary of $\Lambda_L(n)$ is defined by
\begin{eqnarray}
\partial^+\Lambda_L(n)&=&  \{m \in \mathbb{Z}^d\; | \ m \not\in \Lambda_L(n), \ \exists \ m' \in
\Lambda_L(n) \ \ (m,m') \in  \partial \Lambda_L(n) \} \quad \nonumber \\
 &=& \{m\in\Zd\ \mid\ \inorm{m-n}=L+1\}\ .
\end{eqnarray}

For $A\subset\Zd$ we denote the number of lattice points inside $A$
by $|A|$\index{$"|A"|$}. So, $|\LL|=(2L+1)^d$ and
$|\partial^-\Lambda_L|=\,2d\,(2L)^{d-1}$. By $A_m\nearrow\Z^d$ we
mean: $A_m\subset A_{m+1}\subset\Z^d$ and $\bigcup A_m=\Z^d$.

The Green's function \mindex{$G_E^{\,\L}(n,m)$} is the kernel of the
resolvent of $H_\L$ given by
\begin{align}
G_E^{\,\Lambda}(n,m)~=~ (H_{\L}-E)^{-1}(n,m)~
=~\langle\,\delta_n,(H_{\L}-E)^{-1}\,\delta_m\,\rangle\ .
\end{align}

\vspace{0.2cm}
\begin{definition}\label{def:good}{ \quad}\\ \vspace{-0.5cm}
\rm
\begin{enumerate}
\item We will say that the Green's functions $G_E^{\Lambda_L(n_0)}(n,m)$
for energy $E$ and potential $V$ {\em decays
exponentially}\index{exponential decay} on  $\Lambda_L(n_0)$ with
rate $\gamma$ $(\gamma>0)$ if $E$ is not an eigenvalue for
$H_{\LL(n_0)}=(H_0+V)_{\LL(n_0)}$ and

\begin{equation}\label{stern}
|G_E^{\Lambda_L(n_0)}(n,m)|=|(H_{\LL}-E)^{-1}(n,m)\ |\ \leq
e^{-\gamma L}
\end{equation}

for all $n\in\Lambda_{L^{1/2}}(n_0)$ and all
$m\in\partial^-\LL(n_0)$. \vspace{0.3cm}
\item If the Green's function $G_E^{\Lambda_L(n_0)}$
decays exponentially with rate $\gamma>0$ we call the cube
$\LL(n_0)$\quad
\emph{$(\gamma,E)$-good}\index{$(\gamma,E)$-good}\quad for $V$.
\item We call an energy $E\quad$
\emph{$\gamma$-good}\quad (for $V_\omega$) if the there is a
sequence of cubes $\Lambda_{\ell_m} \nearrow \Z^d$ such that all
$\Lambda_{\ell_m}$ are $(\gamma, E)$-good\index{$\gamma$-good}.
(Note, that $\gamma$ is independent of $\Lambda_{\ell_m}$!)
\end{enumerate}

\end{definition}

Note that, by definition, $E \not \in \sigma(H_{\LL(n_0)})$ if
$\LL(n_0)$ is $(\gamma,E)$-good.

The behavior of the Green's function has important consequences for
the behavior of (generalized) eigenfunctions. Suppose that the
function $\psi$ is a solution of the difference equation

\begin{equation}
H \psi = E\psi\ .
\end{equation}
Then (see equations (\ref{def:Gamma}) and (\ref{eq:splitH}))
\begin{eqnarray}
0=(H- E)\psi = (H_\Lambda \oplus
H_{\complement\Lambda}+\Gamma_\Lambda-E)\psi\ ,
\end{eqnarray}
hence
\begin{eqnarray}
((H_\Lambda \oplus H_{\complement\Lambda})- E)\psi =-\Gamma_\Lambda
\psi\ .
\end{eqnarray}

So, for any $n_0 \in \Lambda$ we have

\begin{eqnarray}
(H_\Lambda- E)\psi(n_0) = (-\Gamma_\Lambda \psi)(n_0)\ .
\end{eqnarray}

Suppose that $E$ is not an eigenvalue of $H_\Lambda$, then ($n_0 \in
\Lambda$)

\begin{eqnarray}
\psi(n_0)=-[(H_\Lambda- E)^{-1} \Gamma_\Lambda \psi ](n_0)\ .
\end{eqnarray}

So

\begin{eqnarray} \label{rautegekippt}
\psi(n_0)=-\sum_{(k,m) \in
\partial\Lambda \atop k\in\partial^-\Lambda, m\in\partial^+\Lambda} G_E ^\Lambda (n_0,k)
\psi(m)\ .
\end{eqnarray}

This enables us to prove a crucial observation.

\begin{theorem}\label{th:noge}
If $E$ is $\gamma$-good for $V$ then E is not a generalized
eigenvalue of $H=H_0+V$.
\end{theorem}

\begin{proof}
Suppose $\psi$ is a polynomially bounded eigenfunction of $H$ with
(generalized) eigenvalue $E$, hence
$$H\psi\,=\,E\,\psi\quad\mbox{and}\quad |\psi(m)|\leq c\;|m|^r \qquad\textnormal{for }\; m\not=0\ .$$

Take any $n\in\Zd$, then $n\in\Lambda_{L_k^{1/2}}(n_0)$ for $k$
large enough. Thus by (\ref{rautegekippt})

\begin{eqnarray}\nonumber
|\psi(n)|&\leq&\left|\sum_{(m',m)\in\,\partial\Lambda_{L_k} \atop
m'\in \,\Lambda_{L_k}}G_E^{\Lambda_{L_k}}(n,m')\psi(m)\;\right|\\
&\leq& c_1\,L_k^{d-1}\;e^{-\gamma L_k}\;\sup_{m\in \partial^+\Lambda_{L_k}}|\psi(m)| \label{est:exp1}\\
&\leq& c_2\,L_k^{d-1+r}\;e^{-\gamma L_k}\\
&\rightarrow&\ 0\qquad\qquad \textnormal{as }\quad
k\rightarrow\infty\ .
\end{eqnarray}

Hence $\psi\equiv0$. Consequently, there are no non zero
polynomially bounded eigensolutions.
\end{proof}


There are two immediate yet remarkable consequences of Theorem
(\ref{th:noge}).

\begin{corollary}\label{cor:nospec}
If every $E\in[E_1,E_2]$ is $\gamma$-good for $V$ then\\
$\sigma(H_0+V)\cap(E_1,E_2)=\emptyset$.
\end{corollary}

\begin{corollary}\label{cor:noacspec}
If Lebesgue-almost all $E\in[E_1,E_2]$ are $\gamma$-good for $V$
then
\begin{displaymath}
\sigma_{ac}(H_0+V)\cap(E_1,E_2)=\emptyset\ .
\end{displaymath}
\end{corollary}

\begin{proof}
The assumption of Corollary \ref{cor:nospec} implies by Theorem
\ref{th:noge} that there are no generalized eigenvalues in $(E_1,
E_2)$. By Theorem \ref{th:specge} (or Proposition \ref{prop:geneig})
it follows that there is no spectrum there.

If there is any absolutely continuous spectrum in $(E_1, E_2)$ the
spectral measure restricted to that interval must have an absolutely
continuous component. Hence, by Theorem \ref{th:specge}, there must
be a set of generalized eigenvalues of {\em positive} Lebesgue
measure. However, this is not possible by the assumption of
Corollary \ref{cor:noacspec} and Theorem \ref{th:noge}.
\end{proof}
\goodbreak

\subsection{From multiscale analysis to absence of a.c. spectrum\label{sec:msa2noac}}\es

The results of the previous section indicate a close relation
between the existence of $(\gamma, E)$-good cubes and the spectrum
of the (discrete) Schrödinger operator. The following theorem gives
first hints to a probabilistic analysis of this connection.

\begin{theorem}\label{th:msanac}
If there is a sequence $R_k\rightarrow\infty$ of integers such that
for every $k$, every $E\in I=[E_1, E_2]$ and a constant $\gamma>0$
\begin{equation}
\mathbb{P}\,(\;\Lambda_{R_k} \textnormal{ is not $(\gamma,
E)$-good}\;)\;\;\rightarrow\;0 \label{msavw}
\end{equation}

then with probability one
\begin{equation}
\sigma_{ac}(H_\omega) \cap (E_1, E_2) = \varnothing\ .
\end{equation}
\end{theorem}

\begin{proof}
Set $p_k=\mathbb{P}\,(\;\Lambda_{R_k} \textnormal{ is not $(\gamma,
E)$-good}\;)$. By passing to a subsequence, if necessary, we may
assume that the $R_k$ are increasing and that $\sum p_k < \infty$.

Consequently, from the Borel-Cantelli-Lemma (see Theorem
\ref{th:BorelCantelli})) we learn that with probability one, there
is a $k_0$ such that all $\Lambda_{R_k}$ are $(\gamma, E)-$good for
$k\ge k_0$.

Hence for $\mathbb{P}-$almost every $\omega$ any given $E \in [E_1,
E_2]$ is $\gamma-$good.

We set
\begin{eqnarray}
&\mathcal{N}\, =&\; \big\{ (E, \omega) \in [E_1, E_2] \times \Omega
\;\big|\; E
\textnormal{ is not $\gamma-$good for $V_\omega$} \big\} \\
&\mathcal{N}_E\,=&\;\big\{\; \omega \in \Omega \;\big|\; E
\textnormal{ is not $\gamma-$good for $V_\omega$} \}\\
&\mathcal{N}_\omega\,=&\;\big\{ E \in [E_1, E_2] \;\big|\; E
\textnormal{ is not $\gamma-$good for $V_\omega$} \big\}\ .
\end{eqnarray}

Above we proved $\;\mathbb{P}(\mathcal{N}_E)=0$ for any $E \in [E_1,
E_2]$.

Denoting the Lebesgue measure on $\R$ by $\lambda$ we have by
Fubini's theorem
\begin{eqnarray}\nonumber
\lambda \otimes \mathbb{P}\:(\mathcal{N}) &=& \int_{E_1}^{E_2}
\mathbb{P}(\mathcal{N}_E) \;d\lambda(E)
\\ \nonumber &=& \int \lambda(\mathcal{N}_\omega) \;d\,
\mathbb{P}(\omega) \ .
\end{eqnarray}
Since $\mathbb{P}(\mathcal{N}_E)=0$ for all $E \in [E_1, E_2]$ we
conclude that
\begin{displaymath}
0 = \int_{E_1}^{E_2} \mathbb{P}(\mathcal{N}_E) \;d\lambda(E) = \int
\lambda(\mathcal{N}_\omega) \;d\, \mathbb{P}(\omega)\ .
\end{displaymath}
Thus, for almost all $\omega$ we have
$\lambda(\mathcal{N}_\omega)=0$. Consequently, by Corollary
\ref{cor:noacspec} there is no absolutely continuous spectrum in
$(E_1, E_2)$ for these $\omega$.
\end{proof}

One might be tempted to think the assumption that all $E \in [E_1,
E_2]$ are $\gamma-$good $\mathbb{P}$-almost surely would imply that
there are \emph{no} generalized eigenvalues in $[E_1, E_2]$. This
would exclude \emph{any} spectrum
 inside $(E_1, E_2)$, not only absolutely continuous one.
This reasoning is \emph{wrong}. The problem with the argument is the
following: Under this assumption, we know that for any \emph{given}
energy $E$, there are no generalized eigenvalues with probability
one, i.e. the set $\mathcal{N}_E$ is a set of probability zero.
Thus, for $\omega\in\Omega_0:=\bigcup_{E\in[E_1,
E_2]}\,\mathcal{N}_E$ there are no generalized eigenvalues in the
interval $[E_1, E_2]$. However, the set $\Omega_0$ is an
\emph{uncountable} union of sets of measure zero, therefore, we
cannot conclude that it has zero measure.

Theorem \ref{th:msanac} immediately triggers two kind of questions:
First, is (\ref{msavw}) true under certain assumptions, and how can
we prove it? This is exactly what the multiscale analysis does. We
will discuss this result in the following section \ref{sec:msares}
and prove it in  chapter \ref{ch:msa}.

The other question raised by the theorem is whether or not `good'
cubes might help to prove even \emph{pure point} spectrum, not only
the absence of absolutely continuous spectrum.

It turns out that the condition (\ref{msavw}) alone is not
sufficient to prove pure point spectrum. There are examples of
operators with (almost periodic) potential $V$ satisfying condition
(\ref{msavw}) inside their spectrum, having no
($\ell^2$-)eigenvalues at all (see e.g. \cite{cfks}). So, these
operators have purely singular continuous spectrum in the region
where (\ref{msavw}) holds. This effect is due to some kind of `long
range' order of almost periodic potentials.

In the situation of the Anderson model, we have the independence of
the random variables $V_\omega(i)$. This assumption, we may hope,
prevents the potential from `conspiring' against pure point spectrum
through long range correlations. However, the above example of an
almost periodic potential makes clear that some extra work is
required to go beyond the absence of a.c. spectrum and prove pure
point spectrum.

This question will be addressed in section \ref{sec:pp} after some
preparation in section \ref{sec:iterate}.

\subsection{The results of multiscale analysis \label{sec:msares}}\es

We define a length scale\index{length scale} $L_k$ \index{$L_k$}
inductively. The initial length\index{initial length} $L_0$ will be
defined later depending on the specific parameters (disorder, energy
region, etc.) of the problem considered. The length $L_{k+1}$ is
defined by ${L_{k}}^\alpha$ for an $\alpha$ with $1<\alpha<2$ to be
further specified later. The constant $\alpha$ will only depend on
some
 general parameters like the dimension $d$. The condition $\alpha>1$ ensures that $L_k\to\infty$,
while $\alpha<2$ makes the estimates to come easier. Finally, we
will have to choose $\alpha$ close to one. Observe, that the length
scale $L_k$ is growing very fast, in fact superexponentially.

A main result of multiscale analysis will be the following
probabilistic estimate, which holds for certain intervals $[E_1,
E_2]$.

\begin{result}[multiscale analysis - weak form]\label{res:msa}\index{multiscale analysis - weak form}
For some $\alpha>1$,  $p>2d$ and a $\gamma>0$ and for all $E\in
I=[E_1, E_2]$
\begin{equation}\label{est:msasoft}
\mathbb{P}\,(\;\Lambda_{L_k} \textnormal{ is not $(\gamma,E)-$good
for $V_\omega$}\,)\ \leq \ \frac{1}{L_k^{p}} \ .
\end{equation}
\end{result}
\goodbreak
\begin{remarks}\label{rem:ggood}\platz
\begin{enumerate}
\item We will prove this result in the next two chapters.
\item To prove Result \ref{res:msa}, we need to assume that the probability distribution
$P_0$ of the random variables $V_\omega(i)$ has a bounded density.
This ensures that we can apply Wegner's estimate (Theorem
\ref{th:Wegner}) which is a key tool in our proof. Recently,
Bourgain and  Kenig \cite{bourgain} were able to do the multiscale
analysis for some $V_\omega$ without a density for $P_0$.
\item We will proof Result \ref{res:msa} for $I=[E_1, E_2]$ when $I$ is
close to the bottom of the spectrum or for given $I$ if the disorder
is sufficiently strong.

\item As the proof shows we have to take $\alpha<\frac{2p}{2d+p}$ which is bigger than $1$ since $p>2d$.

\end{enumerate}
\end{remarks}

The proof of Result \ref{res:msa} and its variants (see below) will
take two chapters. We prove the result by induction, i. e. we prove
(\ref{est:msasoft}) for the initial scale $L_0$ and then prove the
induction step, namely: If (\ref{est:msasoft}) holds for a certain
$k$, it holds for $k+1$ as well.

The initial scale estimate will be done in chapter \ref{ch:ise}. It
is only here where we need assumptions about the energy interval $I$
(e.g. $I$ is close to the bottom of the spectrum or to an other band
edge) or about the strength of the disorder. Thus, the specific
parameters of the model enter only here.

In contrast to this, the induction step can be done under quite
general conditions for all energies and any degree of disorder. This
step will be presented in chapter~\ref{ch:msa}. \vspace{0.05cm}

The multiscale estimate Result \ref{res:msa} obviously implies the
absence of absolutely continuous spectrum inside $I$ via Theorem
\ref{th:msanac}. The estimate (\ref{est:msasoft}) per se does not
imply pure point spectrum (see the discussion at the end of the
previous section). However, for the \emph{Anderson model}  one can
use Result \ref{res:msa} to deduce pure point spectrum, provided
$P_0$ has a bounded density. This can be done using a technique
known as \emph{spectral averaging}. The basic idea goes back to
Kotani \cite{Kotsa} and was further developed and applied to the
Anderson model by various authors (see e.g. \cite{DLS, KoSi, SimWol,
CHM}). The paper \cite{SimWol} triggered also the development of the
theory of rank one perturbations \cite{SimR1}. We will not discuss
this method here and refer to the papers cited.

Instead, we will present another proof of pure point spectrum which
goes back to \cite{FMSS} and \cite{dk}. It consists in a version of
the estimate (\ref{est:msasoft}) which is `uniform' in energy $E$.
Taken literally a uniform version of (\ref{est:msasoft}) would be
\vspace{-0.05cm}
\begin{equation}\label{est:msa2str}
\mathbb{P}\,(\, \textnormal{There is an } E\in I \textnormal{ such
that } \Lambda_{L_k} \textnormal{ is not $(\gamma,E)-$good for
$V_\omega$}) \leq L_k^{-p} \ .
\end{equation}

However, it is easy to see by inspecting the proof of Theorem
\ref{th:msanac} that (\ref{est:msa2str}) implies that any $E\in I$
is $\gamma$-good, thus there is \emph{no} spectrum inside $I$ by
Corollary \ref{cor:nospec} above. In other words: condition
(\ref{est:msa2str}) is `too strong' to imply pure point spectrum.

A way out of this dilemma is indicated by the `uniform' version of
Wegner's estimate (Theorem \ref{th:dWegner}). There, uniformity in
energy is required only for \emph{pairs} of disjoint cubes. This
leads us to a uniform version of Result \ref{res:msa} for pairs of
cubes.

\vspace{0.2cm}
\goodbreak

\begin{result}[multiscale analysis - strong form]\label{res:msastrong}\index{multiscale analysis - strong form}
For some $p>2d$, an $\alpha$ with $1<\alpha<\frac{2p}{p+2d}$ and a
$\gamma>0$ we have: For any disjoint cubes
$\Lambda_1=\Lambda_{L_k}(n)$ and $\Lambda_2=\Lambda_{L_k}(m)$
\begin{eqnarray}\label{est:msastrong}
\P\,\left(\,\textnormal{For some } E\in I\textnormal{
 both } \Lambda_1\textnormal{ \emph{and}
$\Lambda_2$ are  not $(\gamma,E)-$good}\,\right)\leq L_k^{-2p}\ .
\end{eqnarray}

\end{result}

The proof of this result is an induction procedure analogous to the
one discussed above. In fact, the initial step will be the same as
for Result \ref{res:msa}, see Chapter \ref{ch:ise}.

In the induction step we assume the validity of estimate
(\ref{est:msastrong}) for $k$ and deduce the assertion for $k+1$
from this assumption. The general idea of this step is quite close
to the induction step for the weaker version (\ref{est:msasoft}),
but it is technically more involved. Therefore, we present the proof
of the weak version first and then discuss the necessary changes for
the strong (`uniform') version.


\subsection{An iteration procedure\label{sec:iterate}}\es

One of the crucial ingredients of multiscale analysis is the
observation that the estimate (\ref{est:exp1}) in the proof of
Theorem \ref{th:noge} can be iterated.

A first version of this procedure is the contents of the following
result.

We say that a subset $A\subset \Zd$ is {\em well inside}\index{well
inside} a set $\Lambda$ ($A\Subset\Lambda$)\index{$\Subset$} if
$A\subset\Lambda$ and $A\cap\partial^-\Lambda=\emptyset$. For any
set $\Lambda\subset\Zd$ we define the \emph{collection of $L-$cubes
inside $\Lambda$} \index{collection of $L-$cubes inside $\Lambda$}
by
\begin{equation}\label{def:CLLambda}\index{$\mathcal{C}_L(\Lambda)$}
\mathcal{C}_L(\Lambda)=\{\Lambda_L(n)\,|\,
\Lambda_L(n)\Subset\Lambda\}\ .
\end{equation}\index{$\mathcal{C}_L(\Lambda)$}

We also set\index{$\partial_L^-\Lambda$}
$\partial_L^-\Lambda=\{m\in\Lambda\mid\textnormal{dist}(m,\partial^-\Lambda)\leq
L\}$ \quad (where $\textnormal{dist}(m,A)=$\newline$\inf_{k\in A}
\inorm{m-k})$.\index{dist}

\begin{theorem} \label{th:expdec}
Suppose that each cube in $\mathcal{C}_M(A)$, $A\subset\Zd$ finite,
is $(\gamma,E)$-good and $M$ is large enough. If $\psi$ is a
solution of $H\psi=E\psi$ in $A$ and $n_0\in A$ with

\begin{equation}\label{mu}
\textnormal{dist}(n_0,\partial^-A)\,\geq\ \,k(M+1)\ ,
\end{equation}

then

\begin{equation}\label{partial}
|\,\psi(n_0)|~\leq~ e^{-\gamma'
kM}\;\sup_{m\,\in\partial^-_{M}\Lambda}\,|\,\psi(m)|
\end{equation}

for some  $\gamma' > 0$.
\end{theorem}
\goodbreak
\begin{remark}
Let us set
\begin{align} r~&=~2d\;(2M+1)^{d-1}\,e^{-\gamma M}\\
\intertext{and} \gamma'~&=~\gamma-\frac{1}{M}\,\ln\Big(2d\,(2M+1)^{d-1}\Big)\label{valuegp}\\
\intertext{such that} r~&=~e^{-\gamma' M}\ .
\end{align}
Then the phrase `$M$ large enough' in the theorem means that $r < 1$
and the theorem holds with $\gamma'$ as in (\ref{valuegp}). Note
that $\gamma'<\gamma$, but the `error' term
$\gamma-\gamma'=\frac{1}{M}\big((d-1) \ln(2M+1)+\ln2d)$ decreases in
$M$ and goes to zero if $M$ tends to infinity.
\end{remark}

The theorem may look a bit clumsy at first sight. Nevertheless, it
contains some of the main ideas of multiscale analysis. The estimate
(\ref{partial}) says that any solution $\psi$ decays exponentially
in regions which are filled with good cubes. In other words: The
tunneling probability of a quantum particle through such a region is
exponentially small. This will finally lead to the induction step in
multiscale analysis.

To illustrate Theorem \ref{th:expdec}, we state the following
Corollary which is essentially a reformulation of the theorem. The
Corollary follows immediately from the Theorem.

\begin{corollary}
Suppose each cube in $\mathcal{C}_M(A)$, $A\subset\Zd$ finite, is
$(\gamma,E)$-good and $M\geq C$ is large enough. Take $n_0\in A$
with $d(n_0)=\textnormal{dist}(n_0,\partial^-A)$ so large that
$\frac{d(n_0)}{M}\geq D$. If $\psi$ is a solution of $H\psi=E\psi$
in $A$, then
\begin{equation}
|\,\psi(n_0)|~\leq~ e^{-\gamma''
d(n_0)}\;\sup_{m\,\in\partial^-_{M}\Lambda}\,|\,\psi(m)|
\end{equation}
with
\begin{equation}
\gamma''~=~\gamma-\frac{1}{M}\,\ln\big(2d\,(2M+1)^{d-1}\big)\,
 \big(1-\frac{1}{C}-\frac{1}{D}\big)\ .
\end{equation}
\end{corollary}

Observe, that the error term
$\frac{1}{M}\,\ln\big(2d\,(2M+1)^{d-1}\big)\,\big(1-\frac{1}{C}-\frac{1}{D}\big)$
is small if both $M$ and the ratio of $d(n_0)$ and $M$ are big.

\begin{nproof}{Theorem}
Since $n_0\in A$ and $\textnormal{
dist}(n_0,\partial^-A)\,\geq\,(M+1)$ we have
$\Lambda_M(n_0)\in\mathcal{C}_M(A)$.

Thus by (\ref{rautegekippt}), we have
\begin{eqnarray}\label{e}\nonumber
|\psi(n_0)|&\le&\sum_{(q,q')\in\partial\L_M(n_0)\atop q\in
\,\Lambda_{M}(n_0)}|G_E^{\L_M(n_0)}(n_0,q)||\psi(q')|\\\nonumber
&\le&|\,\partial\L_M(n_0)|\;\;e^{-\gamma
M}\sup_{q'\in\partial^{+}\L_M(n_0)}|\psi(q')|\\\nonumber
&\le&2d\,(2M+1)^{d-1}\;e^{-\gamma M}\,|\psi(n_1)|\\
&=&r\,|\psi(n_1)|
\end{eqnarray}
for some $n_1\in\partial^+\L_M(n_0)$.
\goodbreak

If $n_1\in\partial^-_M A$, this is estimate (\ref{partial}) for
$k=1$. Note that $\textnormal{
dist}(n_1,\partial_M^-A)\geq\textnormal{
dist}(n_0,\partial_M^-A)-(M+1)$.

So $n_1\in\partial_M^-A$ can only happen if $k=1$.

If $n_1\not\in\partial_M^-A$, we have $\L_M(n_1)\in\mathcal{C}_M(A)$
and we can iterate the estimate (\ref{e}) to obtain

\begin{equation}
|\psi(n_1)|\leq r|\psi(n_2)|
\end{equation}
with some $n_2\in\partial^+\Lambda_M(n_1)$, so

\begin{equation}
|\psi(n_0)|\leq r^2|\psi(n_2)|\ .
\end{equation}

(Note, that for $n_1\in\partial_M^-A$ the iteration might get us out
of $A$!)

For $n_2$ we have

\begin{eqnarray}\nonumber
\textnormal{dist}(n_2,\partial_M^-A)&\geq&\textnormal{
dist}(n_1,\partial_M^-A)-(M+1)\\\nonumber&\geq&\textnormal{
dist}(n_0,\partial_M^-A)-2(M+1)\ .
\end{eqnarray}

So $n_2\in\partial_M^-\Lambda$ can happen only if $k\leq2$.

If $n_2\not\in\partial_M^-\Lambda$, then we may iterate (\ref{e})
again. We obtain
$$|\psi(n_0)|\leq r|\psi(n_1)|\leq
r^2|\psi(n_2)|\leq r^3|\psi(n_3)|\leq \ldots\leq
r^\ell|\psi(n_\ell)|\ .$$

This iteration process works fine as long as the new point
$n_\ell\not\in\partial_M^-A$. Consequently, by the assumption on
$n_0$, we can iterate at least $k$ times.

Thus, we obtain

\begin{equation}
|\psi(n_0)|\leq r^{k'}\sup_{q\in\partial_M^-\Lambda}|\psi(q)|
\end{equation}

with some $k'\geq k$.

We conclude
\begin{equation}
|\psi(n_0)|\leq
e^{-\gamma'kM}\sup_{q\in\partial_M^-\Lambda}|\psi(q)|\ .
\end{equation}

\end{nproof}

\begin{remark}
For $\psi(n_0)\neq0$, the above iteration procedure must finally
reach $\partial_L^-$. Otherwise, we have

\begin{displaymath}
|\psi(n_0)|\leq r^\ell\sup_{q\in\Lambda}|\psi(q)|
\end{displaymath}

for \em any \rm $\ell\in\N$ which implies $\psi(n_0)=0$. For
$\psi(n_0)=0$ the theorem is trivially fulfilled.

\end{remark}

\goodbreak

\subsection{From multiscale analysis to pure point spectrum\label{sec:pp}}\es

In this section we prove that the strong version (Result
\ref{res:msastrong}) of the multiscale estimate implies pure point
spectrum inside the interval where the estimate holds.

\begin{theorem}\label{th:msapp}
If Result \ref{res:msastrong} holds for an interval $I=[E_1, E_2]$,
then with probability one
\begin{displaymath}
\sigma_c(H_\omega) \cap (E_1, E_2) = \emptyset \ .
\end{displaymath}
The spectrum of $H_\omega$ inside $(E_1, E_2)$ consists of pure
point spectrum, the corresponding eigenfunctions decay exponentially
at infinity.
\end{theorem}

\begin{remark}
The theorem includes the case $(E_1, E_2)\cap
\sigma(H_\omega)=\emptyset$ but we will choose $E_1, E_2$ such that
there is some spectrum inside $(E_1, E_2)$ when we apply the
theorem.

\end{remark}

\begin{proof}\;\\[1.5mm]
\centerline{\textbf{Step 1}}\\[1mm]
We begin with a little geometry. As before we choose a sequence
$L_k$ by setting $L_k = L_{k-1}^\alpha$ with an $\alpha>1$ and $L_0$
to be determined later. We consider the cubes $\Lambda_{L_k} =
\Lambda_{L_k}(0)$ and annuli \mindex{$A_k$} which cover the region
between the boundaries of $\Lambda_{L_{k}}$ and $\Lambda_{L_{k+1}}$,
more precisely
\begin{equation}
A_k= \Lambda_{6 L_{k+1}} \setminus \Lambda_{3L_k}\ .
\end{equation}

So, $n\in A_k$ if $\inorm{n}\leq 6L_{k+1}$ and
$||\,n||_\infty>3L_{k}$. It is clear that
\begin{align}A_k\cap A_{k+1}\not=\emptyset\\
\intertext{and} \bigcup A_k=\Z^d\setminus\L_{3L_0}\ .
\end{align}
We will need also an enlarged version \mindex{$A_k^+$} of the $A_k$
namely
\begin{equation}
A_k^+ = \Lambda_{8 L_{k+1}} \setminus \Lambda_{2L_k}\ .
\end{equation}

Obviously, $A_k\subset A_k^+$ and any $n\in A_k$ has a certain
`security' distance from $\partial\,A_k^+$, in fact we have:

\begin{lemma}\label{le:geom}
For each $n \in A_k$
\begin{displaymath}
\textnormal{dist}(n, \partial A_k^+) \geq \frac{1}{3} \inorm{n} \ .
\end{displaymath}
\end{lemma}

\begin{nproof}{Lemma}
If $\inorm{n} \geq 3L_k$ we have
\begin{eqnarray}\nonumber
\textnormal{dist}(n, \partial \Lambda_{2L_k}) &=& \inorm{n}-2L_k
\\ \nonumber &\geq& \inorm{n}- \frac{2}{3}\inorm{n}
\\ \nonumber &=& \frac{1}{3}\inorm{n} \ .
\end{eqnarray}
If $\inorm{n} \leq 6L_{k+1}$
\begin{eqnarray}\nonumber
\textnormal{dist}(n, \partial \Lambda_{8L_{k+1}}) &=& 8 L_{k+1} -
\inorm{n}
\\ \nonumber &\geq& \frac{8}{6} \inorm{n}- \inorm{n}
\\ \nonumber &=& \frac{1}{3}\inorm{n} \ .
\end{eqnarray}
If $n \in A_k$ we have $3L_k \leq \inorm{n} \leq 6L_{k+1}$ , so
\begin{eqnarray}\nonumber
\textnormal{dist}(n, \partial A_k^+) &=& \min \{
\textnormal{dist}(n, \partial \Lambda_{6L_{k+1}}),
\textnormal{dist}(n, \partial \Lambda_{3L_k}) \}
\\ \nonumber &\geq& \frac{1}{3} \inorm{n} \ .
\end{eqnarray}
\end{nproof}
\\[1.5mm]
\centerline{\textbf{Step 2}}\\[1mm]

Now, we investigate the probability that $\L_{L_k}$ is \emph{not}
$(E,\gamma)$-good and, at the same time, one of the $L_k$-cubes in
$A_k$ is also \emph{not} $(E,\gamma)$-good.

Let us abbreviate \index{$\mathcal{C}_k^+$}
\begin{displaymath}
\mathcal{C}_k^+ =\mathcal{C}_{L_k}(A_k^+) = \{ \Lambda_{L_k}(m) |
\Lambda_{L_k}(m) \Subset A_k^+ \} \ .
\end{displaymath}

For a given $k$, define $p_k$ to be the probability of the event

\begin{align}\notag
B_k~=~\big\{\omega\,\big|\,\textnormal{For some }E\in[E_1,E_2]~,~
\L_{L_k} \textnormal{ and at least one cube in } \mathcal{C}_k^+
\textnormal{ are \emph{not} } (E,\gamma)\textnormal{-good}\,\big\}\
.
\end{align}
\goodbreak
We will prove

\begin{lemma}\label{lem:pk} If Result \ref{res:msastrong} holds for $I=[E_1,E_2]$, then there is a constant $C$ such that
for all $k$
\begin{align}
p_k~\leq~\frac{C}{\,L_k^{2p-\alpha d}\,}\ .
\end{align}
\end{lemma}
\begin{remark} The constants $\alpha$ and $p$ are given in Result \ref{res:msastrong}.
\end{remark}

\begin{nproof}{Lemma}
If $\L_{L_k}(m)$ is a fixed cube in $\mathcal{C}_k^+$ then
\begin{align}
\P\,\big(\,\textnormal{For some } &E\in[E_1,E_2] \textnormal{ both
}\L_{L_k}(m) \textnormal{ and } \L_{L_k}
\textnormal{ are \emph{not} $(E,\gamma)$-good }\,\big)\notag\\
~&\leq~\frac{1}{L_k^{2p}}\ .
\end{align}
Hence
\begin{align}
\P\,(B_k)~&\leq~|\,\mathcal{C}_k^+|\;\frac{1}{L_k^{2p}}
~~~~~~\leq~~~C\,L^d_{k+1}\;\frac{1}{L_k^{2p}}\notag\\
&\leq~ C\,{(L_{k}^{\alpha})}^d\;\frac{1}{L_k^{2p}}~ \leq~~
\frac{C}{\,L_k^{2p-\alpha d}}\ .
\end{align}
\end{nproof}

Since $\alpha<\frac{2p}{d}$ (by Result \ref{res:msastrong}) we have
$2p-\alpha d>0$. Thus
\begin{align}
\sum_k\;\P\,(B_k)~<~\infty\ .
\end{align}
Hence, by the Borel-Cantelli-Lemma (Theorem \ref{th:BorelCantelli}),
we have
\begin{align}
\P\,\big(\{\omega\,|\,\omega\in B_k \textnormal{ for infinitely many
$k$ }\}\big)~=~0\ .
\end{align}
Thus we have shown
\begin{proposition}\label{prop:Akgood} If Result \ref{res:msastrong} holds for $I=[E_1,E_2]$, then for
$\P$-almost all $\omega$, there is a $k_0=k_0(\omega)$ such that for all $k\geq k_0$:\\
For any $E\in[E_1,E_2]$ either $\L_{L_k}$ is $(E,\gamma)$-good or
\emph{all} cubes $\L_{L_k}(m)$ in $\mathcal{C}_{L_k}^+$ are
$(E,\gamma)$-good.
\end{proposition}

\centerline{\textbf{Step 3}}

\vspace{2mm} In this final step, we take $\omega$ such that the
assertion of Proposition \ref{prop:Akgood} is true.

Suppose now that $E\in[E_1,E_2]$ is a generalized eigenvalue. It
follows from Theorem \ref{th:noge} that there is no sequence $L_k'$
(with $L_k'\to\infty$) such that all $\L_{L_k'}$ are
$(E,\gamma)$-good. Hence by Proposition \ref{prop:Akgood}, we
conclude that for \emph{all} $k>k_1$, \emph{all} cubes in
$\mathcal{C}_k^+$ are $(E,\gamma)$-good.

Let $\psi$ be a generalized eigenfunction corresponding to the
generalized eigenvalue $E$. Take any $n \in \mathbb{Z}^d$ with
$\inorm{n}$ large enough. Then there is a $k$, $k \geq k_1$, so that
$n \in A_k$ (hence $ 3 L_k\le \inorm{n} < 6 L_{k+1}$). It follows
from Lemma \ref{le:geom} that $\textnormal{dist}(n, \partial A_k^+)
\geq \frac{1}{3} \inorm{n}$. Thus we may apply theorem
\ref{th:expdec} to conclude
\begin{equation}
|\,\psi(n)| ~\leq~ e^{-\gamma^{\prime\prime} \inorm{n}}\; \sup_{m\in
A_k^+}\,|\,\psi(m)|\ .
\end{equation}
Since $\psi$ is polynomially bounded by assumption, we have for
$m\in A_k^+$ and for some $r$
\begin{eqnarray}\nonumber
|\,\psi(m)| &\leq& C_0\,(8L_{k+1})^r \\
\nonumber &\leq& C_1\, L_k^{\alpha r}\\
\nonumber &\leq& C_2\; \inorm{n}^{\alpha r}\ .
\end{eqnarray}

Thus
\begin{align}
|\,\psi(n)| ~\leq~ e^{-\tilde{\gamma}\inorm{n}}\ .
\end{align}

We have therefore shown that any generalized eigenfunction of
$H_\omega$ with eigenvalues in $[E_1, E_2]$ decays exponentially
fast. A fortiori, any generalized eigenfunction is $\ell^2$, so the
corresponding generalized eigenvalue is a bona fide eigenvalue.
Thus, the spectrum in $(E_1, E_2)$ is pure point.

\end{proof}

\begin{remark}
Observe that eigenfunctions $\psi_1$, $\psi_2$ to different
eigenvalues are orthogonal to each other. Since the Hilbert space
$\ell^2(\Zd)$ is separable, there are only countably many $E \in
[E_1, E_2]$ with exponentially decaying eigensolutions.
\end{remark}

$\;$\\[2mm]
{\bf Notes and Remarks }\\[2mm]
multiscale analysis is based on the ground breaking paper by
Fröhlich and Spencer \cite{froespe}. That the MSA result implies
absence of a.c. spectrum was realized by Martinelli and Scoppolla
\cite{MartScop}. An alternative appoach to exclude a.c. spectrum can
be found in \cite{ss}.

The first proofs of Anderson localization were given independently
in \cite{FMSS}, \cite{DLS}, \cite{SimWol}. The latter papers develop
the method of spectral averaging which goes partly back to
\cite{Kotsa}.

The method to prove Anderson localization we present above is due to
\cite{dk} which is related to \cite{FMSS}. Germinet and Klein
\cite{gk2} investigate the relation between Localization and
multiscale analysis in great detail. They characterize a certain
version of localization in terms of the multiscale estimate.

For the literature on the continuous case, i.e. for Schr\"odinger
operators on $L^2(\R^d)$, we refer to the Notes at the end of the
next chapter.

\NeueSeite


\setcounter{equation}{0}
\section{Multiscale analysis\label{ch:msa}}

\subsection{Strategy}\es

We turn to the proof of the multiscale analysis result.

Multiscale analysis (MSA) is an {\em induction} procedure which
starts with a certain length scale $L_0$ and then proves the
validity of the multiscale estimate (\ref{res:msa} and
\ref{res:msastrong}) for $L_{k+1}=L_k^\alpha$ assuming the estimate
holds for $L_k$. The value of $\alpha$ will be fixed later. To get
an increasing sequence $L_k$ we obviously need $\alpha > 1$. We will
also choose $\alpha < 2$ for reasons that will become clear later.
In fact, later we will have to choose $\alpha$ close to one.

In this chapter we will  present the induction step (from $L_k$ to
$L_{k+1}$) deferring the initial step (for $L_0$) to the next
chapter. The induction step can be done for all energies $E$ and for
arbitrary degree of disorder (provided there is \emph{some}
disorder, of course). Thus, it is the \emph{initial step} which
distinguishes between energy regions with pure point spectrum and
those energies where we might have (absolutely) continuous spectrum.
As explained in chapter \ref{ch:localization}, we expect certain
energy regions with absolutely continuous spectrum, but are not
(yet) able to prove it.

The proof of the induction step consists of an analytical and a
probabilistic part. We start with analytic estimates.

For the rest of this chapter, we set for brevity $l=L_k$ and
$L=L_{k+1}$, so we do the induction step from $l$ to $L=l^\alpha$.
By taking $L_0$ sufficiently large we can always assume that $l$
and, a fortiori, $L$ is big enough, i. e. bigger than a certain
constant. Since $\alpha
> 1$ we have $L \gg l$. Below, we will need that both $l$ and $L$ are integers.
To ensure this we should actually choose $L$ to be the smallest
integer bigger or equal to $l^\alpha$. We will neglect this point,
it would complicate the notation. However, the reasoning of the
proof remains the same.

The analytic estimate is a puzzle with different types of cubes.
There are (small) cubes $\Lambda_l(r)$ of size $l=L_k$ and (big)
cubes $\Lambda_L(m)$ of size $L=L_{k+1}=l^\alpha$. The goal is to
prove that that the Green's function $(H_{\Lambda_L}-E)^{-1}(m,n)$
decays exponentially.

By induction hypothesis the probability that a small cube (of size
$l$) is $(\gamma,E)$-good is very high. Thus, we expect that most of
the small cubes $\Lambda_l(n)$ inside $\Lambda_L$ are $(\gamma,
E)-$good. Let us suppose for the moment, that actually \emph{all}
cubes of size $l$ inside $\Lambda_L$ are $(\gamma,E)$-good. Then,
using the geometric resolvent identity  (\ref{geoRes}) and iterating
it just as we did in the proof of Theorem \ref{th:expdec} will give
us an estimate for the Green's function $G_E^{\LL}$ of the form

\begin{equation}\label{est:goal0}
|{G_E^{\LL}(n,m)}|\;\le\;e^{-\tilde{\gamma} k
l}\,|{G_E^{\LL}(n_k,m)}|.
\end{equation}

This estimate results from applying the geometric resolvent equation
$k$ times. This step can be iterated as long as the point $n_k$ is
not too close to the boundary of $\LL$ (so that the cube of size $l$
around $n_k$ belongs to $\LL$) and the cube $\Lambda_l(n_k)$ is a
$(\gamma,E)$-good cube. If all cubes of size $l$ inside $\LL$ are
good, we expect that we can iterate roughly $\frac{L}{l}$ times
before we reach the boundary and conclude

\begin{equation}\label{est:goal}
|{G_E^{\LL}(n,m)}|\;\le\;e^{-\tilde{\gamma} L}\,|{G_E^{\LL}(n',m)}|.
\end{equation}

We may \emph{hope} that we can obtain an estimate of the type
(\ref{est:goal}) even if not all $l$-cubes in $\Lambda_L$ are good
but, at least, an overwhelming majority of them is.

Once we have (\ref{est:goal}) we need a {\em rough} a priori bound
on $G_E^{\LL}(n',m)$ to obtain the desired exponential estimate for
$G_E^{\LL}(n,m)$, i. e. we need to know that $\Lambda_L$ is not an
{\em extremely} bad cube. We say that a cube is extremely bad, if it
is {\em resonant}\index{resonant} in the sense of the following
definition.

\begin{definition}\label{def:resonant}
We call a cube $\Lambda_L(n)$ {\em $E$-resonant}\index{E-resonant}
if $\textnormal{ dist}(E, \sigma(H_{\Lambda_L(n)})) < e^{-
\sqrt{L}}$.
\end{definition}

From Wegner's estimate (Theorem \ref{th:Wegner}) we immediately
learn that it is very unlikely (at least for large $L$) that a cube
is $E$-resonant, in fact

\begin{proposition}\label{est:res}
If the (single-site) measure $P_0$ has a bounded density, then
\begin{equation}
\P(\LL(n) \textnormal{ is $E$-resonant})\,\le \;C\, (2L+1)^d\,e^{-
\sqrt{L}}\ .
\end{equation}

\end{proposition}
\medskip
If $\Lambda_L(n)$ is not $E$-resonant, we know that the Green's
function $G_E^{\LL(n)}$ exists, because $E$ is not in the spectrum.
We even have a rough estimate on the Green's function which tells us
that $\LL$ is not `extremely bad'.

\begin{proposition}\label{prop:nonres}
If the cube $\LL(n)$ is not E-resonant, then for all \goodbreak $m,
m'\in\LL(n)$
\begin{equation}
|{G_E^{\LL(n)}(m,m')}| \le e^{\sqrt{L}}\ .
\end{equation}
\end{proposition}
\medskip

\begin{proof}
If $\Lambda_L$ is not $E$-resonant then
\begin{eqnarray}
|{G_E^{\LL}(m,m')}| &=& |(H_{\Lambda_L}-E)^{-1}(m,m')| \nonumber\\
 &\leq& ||\,(H_{\Lambda_L}-E)^{-1}||\nonumber
\\  &\leq& \frac{1}{\textnormal{dist}(E,
\sigma(H_{\Lambda_L}))} \nonumber
\\  &\leq& e^{\sqrt{L}} \ \label{est:nonres}\ .
\end{eqnarray}

\end{proof}
\medskip

Thus, if the cube $\LL$ is not resonant and if we have
(\ref{est:goal}), we get an estimate of the form

\begin{eqnarray}
|{G_E^{\LL}(n,m)}|\;&\le&\;e^{-\tilde{\gamma} L}\,e^{\sqrt{L}}\\
&\le& \;e^{-\gamma' L}\ .
\end{eqnarray}

What we finally shall prove in (the analytical part of) the
induction step is:

If an overwhelming majority of the cubes $\Lambda_l(m)$ in
$\Lambda_L$ is $(\gamma, E)-$good and $\Lambda_L$ itself is not
$E$-resonant, then $\Lambda_L$ is $(\gamma ', E)-$good.

Note that the exponential rates differ. In fact, $\gamma ' <
\gamma$. That is to say, we can not avoid to {\em decrease} the
decay rate in each and every induction step. As a result we get a
sequence of rates $\gamma_0, \gamma_1,\ldots$ (for induction step 0,
1, \ldots) . Of course, if $\gamma_n \rightarrow 0$ (or becomes
negative) the whole result is pretty useless. So, we have to prove
that $\gamma_n \searrow \gamma_\infty >0$.

Once we have an analytic estimate of the above type, the induction
step will be completed by a \emph{probabilistic} estimate. We have
to prove that with high probability most cubes $\Ll(j)$ inside of
$\LL$ are $(\gamma,E)$-good and $\LL$ is not $E$-resonant. This
probability has to be bigger than $1-L^{-p}$. To prove that most
cubes $\Ll(j)$ are good, we use the induction hypothesis. That $\LL$
is not resonant with high probability follows from the Wegner
estimate Theorem \ref{th:Wegner}.

We have deliberately used the vague terms `most cubes' and `an
overwhelming majority'. What they exactly mean is yet to be defined.

\subsection{Analytic estimate - first try \label{sec:msaft} }\es

We start with a first attempt to do the analytic part of the
induction step. This first try assumes that {\em all} cubes of size
$l$ inside $\Lambda_L$ are $(\gamma, E)-$good. We recall that
\index{$\mathcal{C}_l(\Lambda_L)$} $\mathcal{C}_l(\Lambda_L) = \{
\Lambda_l(m) \,|\, \Lambda_l(m) \Subset \Lambda_L \}$.

The main idea of the approach is already contained in the proof of
Theorem \ref{th:expdec}.

\begin{proposition}\label{pr:msaft}
Suppose all cubes in $\mathcal{C}_l(\Lambda_L)$ are $(\gamma,
E)-$good. Then for any $\bar{\gamma} < \gamma$ there is an $l_0$
such that for  $\,l \geq l_0$
\begin{eqnarray}\label{est:msaft}
|{G_E^{\LL}(m,n)}| =\,|(H_{\Lambda_L}-E)^{-1}(m,n)| \leq
\frac{1}{\textnormal{dist}(E, \sigma(H_{\Lambda_L}))}\;
e^{-\bar{\gamma} L}
\end{eqnarray}
for any $m\in\Lambda_{L^{1/2}}$ and any $n
\in\,\partial^-\Lambda_L$.
\end{proposition}

\begin{proof}
Take $m \in \Lambda_{L^{1/2}}$. Since $\textnormal{dist}(m,
\partial^-\Lambda_L) \geq l+1$ if $l_0$ and hence $l$ is large enough, we have $\Lambda_l(m) \in
\mathcal{C}_l(\Lambda_L)$ and we may apply the geometric resolvent
equation (\ref{geoRes}). Thus, we have

\begin{eqnarray}\label{est:step}
|{G_E^{\LL}(m,n)}| &\leq& \sum_{(q,q')\in\partial\Ll(m)\atop q\in
\,\Ll(m)}
|{G_E^{\Ll(m)}(m,q)}|\;\; |{G_E^{\LL}(q',n)}|\\
&\leq& 2d\,(2l+1)^{d-1}\, e^{- \gamma l}\;  |{G_E^{\LL}(n_1,n)}| \\
&\leq&  e^{- \tilde{\gamma}
l}\;\;|{G_E^{\LL}(n_1,n)}|\label{est:exp}
\end{eqnarray}
with
\begin{equation}
\tilde\gamma = \gamma - \frac{(d-1)\ln(2l+1)}{l} - \frac{\ln 2d}{l}
\end{equation}
for some $n_1 \in \partial^+\Lambda_l(m)$.

If $\textnormal{dist}(n_1, \partial^-\Lambda_L) \geq l+1$, we may
repeat this estimate with $\Lambda_l(m)$ replaced by
$\Lambda_l(n_1)$ and obtain
\begin{displaymath}
|{G_E^{\LL}(m,n)}| \leq e^{- \tilde\gamma\, 2l}\;
|{G_E^{\LL}(n_2,n)}|
\end{displaymath}
with $n_2 \in \partial^+ \Lambda_l(n_1)$.

Note that $\textnormal{dist}(n_1, \partial^-\Lambda_L) \geq
L-\sqrt{L}-(l+1)$, since $n_1 \in \partial^+\Lambda_l(m)$.

So, the second estimation step is certainly possible if
$L-\sqrt{L}-(l+1) \geq l+1$. If this is so, we may try to iterate
(\ref{est:step}) a second time. This is possible if
\begin{displaymath}
L-\sqrt{L}-2(l+1) \geq l+1
\end{displaymath}
and the result is
\begin{displaymath}
|{G_E^{\LL}(m,n)}| \leq e^{- \tilde\gamma \,3l}\;
|{G_E^{\LL}(n_3,n)}|\ .
\end{displaymath}
We may apply this procedure $k$ times as long as $L-\sqrt{L}-k(l+1)
\geq l+1$, i. e. for
\begin{eqnarray}\label{est:k1}
k \leq  \frac{L}{l+1} - \frac{\sqrt{L}}{l+1} - 1 \ .
\end{eqnarray}
The largest integer $k_0$ satisfying (\ref{est:k1}) is at least
\begin{eqnarray}\label{est:k2}
k_0 \geq  \frac{L}{l+1} - \frac{\sqrt{L}}{l+1} - 2 \ .
\end{eqnarray}
Consequently, we obtain
\begin{eqnarray}\nonumber
|{G_E^{\LL}(m,n)}|&\leq& e^{- \tilde\gamma\, k_0 l}\;
|{G_E^{\LL}(n_{k_0},n)}|
\\ \nonumber &\leq& ||(H_{\Lambda_L}-E)^{-1}||\;\; e^{- \tilde\gamma\, k_0 l}
\\  &=& \frac{1}{\textnormal{dist}(E,\sigma(H_\Lambda))} e^{- \tilde\gamma\;
k_0 l} \ .
\end{eqnarray}
As long as $\tilde\gamma>0$, we have
\begin{eqnarray}\nonumber
e^{- \tilde\gamma k_0 l} &\leq& e^{- \tilde\gamma (L\frac{l}{l+1} -
\sqrt{L} \frac{l}{l+1} - 2l)}
\\ \nonumber &=& e^{- \tilde\gamma (\frac{l}{l+1} -
\frac{1}{\sqrt{L}} \frac{l}{l+1} - 2\frac{l}{L})L}
\\ \nonumber &\leq& e^{- \tilde\gamma (1- \frac{1}{l+1} -
\frac{1}{\sqrt{L}} \frac{l}{l+1} - 2\frac{l}{L})L}
\\ &\leq& e^{- \tilde\gamma (1- \frac{1}{l} -
\frac{1}{l^{\alpha/2}} - 2\frac{1}{l^{\alpha-1}})L} \ .
\end{eqnarray}
So, estimate (\ref{est:msaft}) holds if
\begin{eqnarray}\label{est:gamma1}
\left(\gamma - \frac{(d-1)\ln(2l+1)}{l} -
\frac{2d}{l}\right)\left(1- \frac{1}{l} - \frac{1}{l^{\alpha/2}} -
2\frac{1}{l^{\alpha-1}}\right) \geq \bar{\gamma} \ .
\end{eqnarray}
By taking $l$ large enough we can assure that (\ref{est:gamma1})
holds.
\end{proof}

If we assume that $\Lambda_L$ is not $E$-resonant (see Definition
\ref{def:resonant}), we can further estimate expression
(\ref{est:msaft}).

\begin{theorem}\label{th:msaft}
If the cube $\Lambda_L$ is not E-resonant and if all the cubes in
$\mathcal{C}_l(\Lambda_L)$ are $(\gamma, E)-$good and $\gamma ' <
\gamma$, then
\begin{displaymath}
\Lambda_L \textnormal{ is $(\gamma', E)-$good}
\end{displaymath}
if $l$ is large enough.
\end{theorem}

\begin{proof}
By (\ref{est:msaft}) and the assumption that $\Lambda_L$ is not
resonant (see \ref{est:nonres}) we obtain
\begin{eqnarray}\nonumber
|{G_E^{\LL}(m,n)}| &\leq& e^{- \bar{\gamma}\, L}\; e^{L^{1/2}}
\\ &\le& e^{- \gamma' L}\label{est:gamma2}
\end{eqnarray}
with $\gamma' =  \bar{\gamma}-\frac{1}{l^{\alpha/2}}$.
\end{proof}

\bigskip
\begin{corollary}\label{cor:gamma}
If the cube $\Lambda_L$ is not E-resonant and if all the cubes in
$\mathcal{C}_l(\Lambda_L)$ are $(\gamma, E)-$good, then $\LL$ is
$(\gamma', E)-$good with
\begin{equation}
\gamma'\ \ge\ \gamma\;\big(1-\frac{4}{l^{\alpha-1}}\big)\;-
\;\big(\frac{3d\,\ln(2l+1)}{l}+\frac{1}{l^{\alpha/2}}\big)\
.\label{est:gamma}
\end{equation}
Moreover, for $l\ge C_0$, with $C_0$ depending only on $\alpha$ and
the dimension $d$, we have
\begin{equation}
\gamma'\ \ge\ \gamma\;\big(1-\frac{4}{l^{\alpha-1}}\big)\;-
\;\frac{2}{l^{\alpha/2}}\ .\label{est:gammag}
\end{equation}
\end{corollary}

\begin{proof}
Estimate (\ref{est:gamma}) follows from  (\ref{est:gamma1}),
(\ref{est:gamma2}) and the observation that
\begin{equation}
\alpha\,-\,1\ <\ \alpha/2\ <\ 1\label{est:alpha}
\end{equation}
since $1<\alpha<2$.

Moreover, there is a constant $C_0 = C_0(\alpha, d)$ such that for
$l\ge C_0$ we have $\frac{3d\,\ln(2l+1)}{l}\le
\frac{1}{l^{\alpha/2}}$ which implies (\ref{est:gammag}).
\end{proof}


An obvious problem with the above result is the fact that we have to
decrease the rate $\gamma$ of the exponential decay in each
induction step. Suppose we start with a rate $\gamma_0$ for length
scale $L_0$. Let us assume $L_0\ge C_0$, the constant appearing
before (\ref{est:gammag}). We call $\gamma_k \le \gamma_0$
\index{$\gamma_k$} the decay rate we obtain from Theorem
\ref{th:msaft} and Corollary \ref{cor:gamma} in the $k^{th}$ step,
i. e. for $L_k=(L_{k-1})^\alpha$.

We get the lower bound
\begin{eqnarray}
\gamma_{k+1}\ &\geq&\ \gamma_{k}\; -
\gamma_{k}\,\frac{4}{{L_k}^{\alpha-1}}\;
-\;\frac{2}{{L_k}^{\alpha/2}}\label{est:gammak1}\\
&\geq&\ \gamma_{k} - \gamma_0 \,\frac{4}{{L_k}^{\alpha-1}}\;
-\;\frac{2}{{L_k}^{\alpha/2}}\ . \label{est:gammak2}
\end{eqnarray}

Thus \index{$\gamma_\infty$}
\begin{equation}
\gamma_\infty\;=\;\liminf\,\gamma_k\ \ge
\gamma_0\;-\;\gamma_0\:\sum_{k=0}^\infty\,\frac{4}{{L_k}^{\alpha-1}}
\;-\;\sum_{k=0}^\infty\,\frac{2}{{L_k}^{\alpha/2}}\label{est:ginfty}\
.
\end{equation}

\bigskip
To estimate the right hand side of (\ref{est:ginfty}), we use the
following lemma.

\begin{lemma}\label{lem:sumLk} For $\beta>0$ and $L_0$ large enough we have
\begin{equation}\label{est:sumLk}
\sum_{k=0}^\infty\;\frac{1}{{L_k}^\beta}\ \le\
\frac{2}{{L_0}^\beta}\ .
\end{equation}
\end{lemma}

\begin{remark} In the lemma $L_0$ large means: ${L_0}^{\beta(\alpha-1)}\ge 2$.
\end{remark}

\begin{proof}
\begin{eqnarray}
r_k&:=&\frac{1}{L_k^{\beta}}\;\le\,\frac{1}{{({L_0}^{\alpha^k})}^{\beta}}
\le \,\frac{1}{{(L_0^{\beta})}^{\alpha^k}}\nonumber \\
&\le&\,\frac{1}{{(L_0^{\beta})}^{1+k(\alpha-1)}}
\le\,\frac{1}{L_0^{\beta}}\ \
\big(\frac{1}{\:{L_0^{{\beta(\alpha-1)}}\:}}\big)^{k}\ .
\end{eqnarray}

Above, we used $\alpha^k\ge 1+k(\alpha-1)$.

From these estimates we obtain for ${L_0}^{\beta(\alpha-1)}\ge 2$
\begin{eqnarray}
r:=\sum_{k=0}^\infty\,r_k \le \
\frac{1}{\,1-{L_0}^{-\beta(\alpha-1)}\,}\; \frac{1}{L_0^{\beta}}\
\le\ \frac{2}{L_0^{\beta}}\ .
\end{eqnarray}

\end{proof}

From this lemma we learn that the `final' decay rate $\gamma_\infty$
is positive if $L_0$ and $\gamma_0$ are not too small, more
precisely:

\begin{proposition}
If $L_0$ is big enough and
\begin{equation} \label{gamma0a}
\gamma_0\ge\,\frac{16}{{L_0}^{\alpha/2}}
\end{equation}
then
\begin{equation}\label{gammainf}
\gamma_\infty = \inf\,\gamma_k\;\ge\;\frac{1}{2}\,\gamma_0\ .
\end{equation}
\end{proposition}

\begin{remark}
$L_0$ big enough means
\begin{equation}\label{L0large}
{L_0}^{\alpha-1}\ge 32 \qquad\textnormal{ and }\qquad
{L_0}^{(\alpha-1)^2}\ge 2\ .
\end{equation}
\end{remark}

\begin{proof}
Since $\alpha<2$, we know $\frac{\alpha}{2}\geq(\alpha-1)$. So, if
${L_0}^{(\alpha-1)^2}\ge 2$, by Lemma \ref{lem:sumLk} we have
\begin{align}
\sum_{k=0}^\infty\;\frac{1}{{L_k}^{\alpha/2}}\ &\le\ \frac{2}{{L_0}^{\alpha/2}}\\
\intertext{and} \sum_{k=0}^\infty\;\frac{1}{{L_k}^{\alpha-1}}\ &\le\
\frac{2}{{L_0}^{\alpha-1}}\ .
\end{align}
Thus, (\ref{gamma0a}) and (\ref{L0large}) inserted in
(\ref{est:ginfty}) give
\begin{align}
    \gamma_\infty\ &\ge\ \gamma_0\;-\;\frac{8}{{L_0}^{\alpha-1}}\,\gamma_0\;-\;\frac{4}{{L_0}^{\alpha/2}}\\
    &\ge\ \frac{3}{4}\:\gamma_0\ -\ \frac{1}{4}\:\gamma_0\ =\
    \frac{1}{2}\,\gamma_0\ .
\end{align}

\end{proof}
\goodbreak

Let us pause to summarize what we have done so far.

\bigskip
\begin{theorem}\label{th:MSAft}
Define the length scale $L_{k+1}={L_k}^\alpha$ with $1<\alpha<2$ and
a suitable $L_0$,
which is not too small.\\
If for a certain $k$
\begin{enumerate}
\item all the cubes in $\mathcal{C}_{L_k}(\Lambda_{L_{k+1}})$ are $(\gamma_{k}, E)$- good and
\item the cube $\Lambda_{L_{k+1}}$ is not $E$-resonant
\end{enumerate}
then the cube $\Lambda_{L_{k+1}}$ is $(\gamma_{k+1}, E)$- good with
a rate $\gamma_{k+1}$ satisfying
\begin{equation} \label{est:rate}\gamma_{k+1}\ \geq\ \gamma_{k}\; - \gamma_{k}\,\frac{4}{{L_k}^{\alpha-1}}\;
-\;\frac{2}{{L_k}^{\alpha/2}}\ .
\end{equation}
\end{theorem}

Moreover, we have some control on the sequence $\gamma_k$.

\begin{corollary}
If the initial rate $\gamma_0$ satisfies $\gamma_0 \ge
\frac{16}{{L_0}^{\alpha/2}}$ and $L_0$ is large enough, then the
$\gamma_k$ (as in (\ref{est:rate})) satisfy $\gamma_k \ge
\frac{\gamma_0}{2}$ for all $k$.
\end{corollary}

\bigskip
Thus, we have done a first version of the analytic part of the
MSA-proof. So far for the good news about Theorem \ref{th:MSAft}.

\bigskip
We are left with the probabilistic estimates, namely:

Prove that if $\Lambda_{L_k}$ is good with high probability then the
hypothesis' (1) and (2) in Theorem \ref{th:MSAft} above are true
with high probability. More precisely, we \emph{would like} to
prove:

If
\begin{eqnarray}\nonumber
\mathbb{P}(\Lambda_{l} \textnormal{ is not $(\gamma, E)-$good}) \leq
\frac{1}{l^p}
\end{eqnarray}
then
\begin{eqnarray}\label{est:PL}
\mathbb{P}(\Lambda_L \textnormal{ is not $(\gamma, E)-$good}) \leq
\frac{1}{L^p}
\end{eqnarray}
with $L=l^\alpha$.

\medskip
Here comes the bad news: There is {\em no chance} for such an
estimate.

In fact, Theorem \ref{th:MSAft} allows us to estimate
\begin{align}\nonumber
&\mathbb{P}(\Lambda_L \textnormal{ is not $(\gamma, E)-$good})\\
 \leq\;
&\mathbb{P}(\Lambda_L \textnormal{ is not $E$-resonant}) +\;
\mathbb{P}(\textnormal{at least one cube in
$\mathcal{C}_l(\Lambda_L)$ is not $(\gamma, E)-$good}) \
.\label{est:probft}
\end{align}
The first term in (\ref{est:probft}) can be estimated by the Wegner
estimate (\ref{th:Wegner}). However the second term is certainly
{\em bigger} than $\mathbb{P}(\Lambda_L(0) \textnormal{ is not
$(\gamma, E)-$good})$. The only estimate we have for this is
$\frac{1}{l^p}$. So the best we can possibly hope for is an estimate
like
\begin{equation}
\mathbb{P}(\Lambda_L \textnormal{ is not $(\gamma, E)-$good}) \leq
\frac{1}{l^p} = \frac{1}{L^{p/ \alpha}} \ .\label{toobad}
\end{equation}
This is much worse than estimate (\ref{est:PL}).

What goes wrong here is that the probability that \emph{all} small
cubes are good is too small. Consequently, we have to accept at
least one or even a few cubes in $\mathcal{C}_l(\Lambda_L)$ which
are not $(\gamma, E)-$good. Dealing with bad cubes in
$\mathcal{C}_l(\LL)$ requires a refined version of the above
analytic reasoning.

\subsection{Analytic estimate - second try}\label{sec:st}\es

Now, we try to do the induction step allowing  a few bad cubes in
$\mathcal{C}_l(\Lambda_L)$. We start with just \emph{one} bad cube.
More precisely, we suppose now that $\mathcal{C}_l(\Lambda_L)$ does
not contain two disjoint cubes which are not $(\gamma, E)-$good.

If two cubes overlap, events connected with these cubes are
\emph{not} independent, so probability estimates are hard in this
case. That is why we insist above on non overlapping sets.

The above assumption implies that there is an $m_0 \in \Lambda_L$
such that all the cubes $\Lambda_l(m) \in \mathcal{C}_l(\Lambda_L)$
with $||\,m-m_0\,||_\infty > 2l$ are $(\gamma, E)-$good.
Consequently, there are no bad cubes with centers outside
$\Lambda_{2l}(m_0)$. The cube $\Lambda_{2l}(m_0)$ is the `dangerous'
region which requires special care.

As in the proof of Proposition \ref{pr:msaft}, we use and iterate
the geometric resolvent equation to estimate
\begin{equation}
|G^{\LL}_E(m,n)| \ \leq\ e^{- \tilde\gamma\, l r}\;
|G^{\Lambda_L}_E(n_r,n)|
\end{equation}

as long as possible. With a bad cube inside $\LL$, this procedure
can stop not only when $n_r$ is near the boundary of $\Lambda_L$ but
also if $n_r$ reaches the problematic region around $m_0$ where
cubes $\Lambda_l(m)$ might be bad.

Let us concentrate for a moment how we can handle sites $n_r$ inside
the dangerous region $\Lambda_{2l}(m_0)$. So, suppose that
$u:=n_r\in\Lambda_{2l}(m_0)$. Hence we can\emph{not} be sure the
cube $\Lambda_l(u)$ is good. We can still try to apply the geometric
resolvent equation and obtain

\begin{equation}
|G_E^{\LL}(u,n)| \leq \sum_{(q,q')\in\partial\Ll(u)\atop q\in
\,\Ll(u)} |G_E^{\Ll(u)}(u,q)|\; |G_E^{\LL}(q',n)|\ .
\end{equation}

If we assume \emph{nothing} about the cube $\Lambda_l(u)$, there is
no chance to estimate $ G_E^{\Ll(u)}(u,q) $. In fact, this Green's
function may be arbitrarily large or even non existing. It seems
reasonable to suppose that the `trouble making' region, the cube
$\Lambda_{2l}(m_0)$, is `not completely bad' in the sense, that
$\Lambda_{2l}(m_0)$ is  not $E$-resonant. This allows us to estimate

\begin{eqnarray}
|G_E^{\LL}(u,n)| &\leq&
\sum_{(q,q')\in\partial\Lambda_{2l}(m_0)\atop q\in
\,\Lambda_{2l}(m_0)}
|G_E^{\Lambda_{2l}(m_0)}(u,q)|\; |G_E^{\LL}(q',n)|\nonumber\\
&\leq& 2d\,(4l+1)^{d-1}\,e^{\sqrt{2l}}\
|G_E^{\LL}(u',n)|\label{step:Wegner}
\end{eqnarray}
for a $\;u'\in\LL\backslash \Lambda_{2l}(m_0)$.

Observe, that the cube $\Lambda_l(u')$ is $(\gamma_l, E)-$good by
induction hypothesis since \goodbreak $u'\not\in\Lambda_{2l}(m_0)$.
Therefore, the \emph{next} iteration of the geometric resolvent
estimate will give us an \emph{exponentially decreasing} term
\begin{eqnarray}
&&|G_E^{\LL}(u,n)| \leq \ 2d\;(4l+1)^{d-1}\;e^{\sqrt{2l}}\ |G_E^{\LL}(u',n)|\nonumber\\
&\leq& (2d)^2\;(4l+1)^{d-1}\,(2l+1)^{d-1}\ e^{\sqrt{2l}}\
e^{-\gamma_l\,l}\ |G_E^{\LL}(n_{r+1},n)|\label{step:IH}\ .
\end{eqnarray}

In the double step (\ref{step:Wegner}) and (\ref{step:IH}), we pick
up a factor

\begin{equation}
\rho:=(2d)^2\;(4l+1)^{d-1}\,(2l+1)^{d-1}\ e^{\sqrt{2l}}\
e^{-\gamma_l\,l}\ .
\end{equation}

The second step (\ref{step:IH}) compensates the first one
(\ref{step:Wegner}) if $\rho\leq 1$. This is the case if
\begin{equation}
\gamma_l\ge \frac{\sqrt{2}}{\sqrt{l}}+\frac{2\,\ln{(2d)}\;+
2\,(d-1)\,\ln{(4l+1)}}{l}
\end{equation}
which is fulfilled for
\begin{equation}\label{ass:gamma}
\gamma_l\ge \frac{2}{\sqrt{l}}
\end{equation}
if $l$ is bigger than a constant depending only on the dimension.

In the proof of Theorem \ref{th:msaft}, we could choose (see
(\ref{est:gammak1}))
\begin{equation}\label{est:gammak12}
\gamma_{k+1}\;\ge\;\gamma_{k}\;-\,
\gamma_{k}\;\frac{4}{{L_k}^{\alpha-1}}\;-\,\frac{2}{{L_k}^{\alpha/2}}\
.
\end{equation}

An induction argument using (\ref{est:gammak12}) shows

\begin{lemma}
If $L_0 \ge M$, a constant depending only on $\alpha$ and $d$, and
if (\ref{est:gammak12}) holds, then $\gamma_0\ge
\frac{2}{L_0^{1/2}}$ implies
\begin{equation}\label{est:sqrt}
    \gamma_k\ \ge\ \frac{2}{L_k^{1/2}}\qquad\text{for all }k\ .
\end{equation}
\end{lemma}

This Lemma ensures that we can iterate the induction step in the
multiscale analysis even if we hit the dangerous region
$\L_{2l}(m_0)$. In fact, once we start with $\gamma_0\ge
\frac{2}{{L_0}^{1/2}}$, we can be sure that all the the rates
satisfy the condition $\gamma_k\ge \frac{2}{{L_k}^{1/2}}$.

\begin{proof}
By taking $L_0$ large enough we can ensure that:
\begin{align}
\frac{4}{L_k^{\alpha-1}}~\leq~\frac{1}{2}\quad \textnormal{ and
}\quad
\frac{4}{L_k^{\alpha/2}}~\leq~\frac{1}{L_k^{1/2}}\qquad\textnormal{for
all } k\ .
\end{align}

So, if $\gamma_k~\geq~\frac{2}{L_k^{1/2}}$, then
\begin{align}
\gamma_{k+1}~&\geq~\gamma_k\,(1-\frac{4}{L_k^{\alpha-1}})\,-\,\frac{2}{L_k^{\alpha/2}}\notag\\
&\geq~\eh\,\gamma_k\,-\,\frac{2}{L_k^{\alpha/2}}\notag\\
&\geq~\frac{1}{L_k^{1/2}}\,-\,\frac{2}{L_k^{\alpha/2}}\notag\\
&\geq~\frac{2}{L_k^{\alpha/2}}\notag\\
&\geq~\frac{2}{L_{k+1}^{1/2}}\notag
\end{align}
Thus, the Lemma follows by induction.
\end{proof}

Knowing how to deal with the cubes inside $\L_{2l}(m_0)$, we now
sketch our strategy. We use the geometric resolvent equation to
estimate the resolvent on the big cube of size $L$ in terms of the
resolvent of small cubes of size $l$. As long as the first argument
$n_r$ of the Green's function (for $\LL$) belongs to a good cube, we
use an exponential bound as in (\ref{est:exp}). If $n_r$ belongs to
the `bad' region which may contain cubes that are not good, then we
do the double step estimate (\ref{step:Wegner}) and (\ref{step:IH}).
This procedure can be repeated until we get close to the boundary of
$\LL$. The number of times we do the exponential bound in this
procedure is at least of the order $L/l$. In fact, analogously to
(\ref{est:k1}) the number $k_0$ of `good' steps is at least
\begin{eqnarray}
k_0 \ge \frac{L}{l+1} - \frac{\sqrt{L}}{l+1} - C_1 \ .
\end{eqnarray}
Consequently, the estimates of the previous section can be redone if
we allow `one' bad cube with the following changes
\begin{itemize}
\item We need $L_0\ge C_2$ with a constant $C_2$ (possibly) bigger than the previous one.
\item We have to take $\gamma_0\ge 2\, {L_0}^{-1/2}$
\item The procedure requires that all cubes of size $2l$ inside $\LL$ are non resonant. While we need this only
for the cube $\Lambda_{2l}(m_0)$ around the `bad' cube, we do not
know, where the bad cube is, so we require non resonance for all
cubes of the appropriate size.
\end{itemize}

Thus, we have shown the following improvement of Theorem
\ref{th:MSAft}.

\bigskip
\begin{theorem}\label{th:MSAst}
Suppose $L_0$ is large enough and $L_{k+1}={L_k}^\alpha$ with $1<\alpha<2$.\\
If for a certain $k$ ($l:=L_k$ and $L:=L_{k+1}$)
\begin{enumerate}
\item there do not exist two disjoint cubes in $\mathcal{C}_{l}(\Lambda_{L})$ which are
not $(\gamma_{k}, E)$- good with a rate $\gamma_k\ge
\frac{2}{l^{1/2}}$,
\item no cube $\Lambda_{2l}(m)$ in $\LL$ is $E$-resonant and
\item the cube $\Lambda_{L}$ is not $E$-resonant,
\end{enumerate}
then the cube $\Lambda_{L}$ is $(\gamma_{k+1}, E)$- good with a rate
$\gamma_{k+1}$ satisfying $\gamma_{k+1}\ge \frac{2}{L^{1/2}}$.

Moreover we can choose the rate $\gamma_{k+1}$ such that
\begin{equation} \label{est:rate1}\gamma_{k+1}\ \geq\ \gamma_{k}\; - \gamma_{k}\,\frac{C}{{L_k}^{\alpha-1}}\;
-\;\frac{C}{{L_k}^{\alpha/2}}\ .
\end{equation}
\end{theorem}

As above, we can estimate the decay rates as follows.

\begin{corollary}
If the initial rate $\gamma_0$ satisfies $\gamma_0 \ge
\frac{C}{{L_0}^{1/2}}$ and $L_0$ is large enough, then the
$\gamma_k$ in Theorem \ref{th:MSAst} satisfy $\gamma_k \ge
\frac{\gamma_0}{2}$ for all $k$.
\end{corollary}

This result allows us to prove the multiscale estimate in its weak
form (\ref{res:msa}) as we will show in the next section
\ref{sec:prob1} where we do the corresponding probabilistic
estimates.

The above analytic results (especially the counterpart of Theorem
\ref{th:MSAst}) can be shown for the strong version (Result
\ref{res:msastrong}) as well with not too much difficulties.
Unfortunately, the probabilistic estimate breaks down for the strong
form, as we will discuss below. To make the probabilistic part of
the argument work for the strong case, we have to allow more than
just one bad $l$-cube inside the $L$-cubes. In Section \ref{sec:tt},
we show how to deal with this problem.

\subsection{Probabilistic estimates - weak form\label{sec:prob1}}\es

We turn to the probablistic estimates of the induction step in
multiscale analysis. Here, we will prove the multiscale result in
its weak form (Result \ref{res:msa}).

In the whole section we assume that the probability distribution
$P_0$ of the independent, identically distributed random variables
$V_\omega(i)$ has a \emph{bounded density}, i.~e.
\begin{eqnarray}\label{ass:density}
    P_0(A)\;&:=&\; \P(\,V_\omega(i)\in A\,) = \int_A g(\lambda)d\lambda, \notag\\
&&\textnormal{with\quad}||g||_\infty = \sup_{\lambda}
|\,g(\lambda)\,|\;<\; \infty)\ .\label{ass:dens}
\end{eqnarray}
This condition is assumed throughout this section even when not
explicitly stated.

The main result is

\begin{theorem}\label{th:msasoftres}
Assume that the probability distribution $P_0$ has a bounded
density. Suppose $L_0$ is large enough, $\gamma\ge
\frac{1}{{\,L_0}^{1/2}}$, $p>2d$  and $1<\alpha<\frac{2p}{p+2d}$. If
\begin{eqnarray}\label{ind:ise}
\mathbb{P}(\Lambda_{L_0} \textnormal{ is not $(2 \gamma,E)-$good })
\leq \frac{1}{{\,L_0}^p}\ ,
\end{eqnarray}
then for all $k$
\begin{eqnarray}
\mathbb{P}(\Lambda_{L_{k}} \textnormal{ is not $(\gamma,E)-$good})
\leq \frac{1}{{\,L_{k}}^p} \ .
\end{eqnarray}
\end{theorem}

\begin{remark}\platz
\begin{itemize}
\item Note that $p>2d$ ensures that we can choose $\alpha>1$.
\item We need the assumption (\ref{ass:dens}) on $P_0$ (only) in order to have the Wegner estimate (Theorem \ref{th:Wegner}).
\end{itemize}
\end{remark}

This theorem reduces the multiscale analysis to the initial scale
estimate (\ref{ind:ise}) which we discuss in chapter \ref{ch:ise}.
As we remarked above, Theorem \ref{th:msasoftres} is proved by
induction. Thus, under the assumptions of Theorem
\ref{th:msasoftres} and with the rates $\gamma_k$ as in  Theorem
\ref{th:MSAst}, we have to prove the following theorem.

\begin{theorem}\label{th:isp}
If
\begin{eqnarray}\label{hyp:ind}
\mathbb{P}(\Lambda_{L_k} \textnormal{ is not $(\gamma_k,E)-$good})
\leq \frac{1}{L_k^p}\ ,
\end{eqnarray}
then
\begin{eqnarray}
\mathbb{P}(\Lambda_{L_{k+1}} \textnormal{ is not
$(\gamma_{k+1},E)-$good}) \leq \frac{1}{L_{k+1}^p} \ .
\end{eqnarray}
\end{theorem}

\begin{proof}
As usual, we set $l=L_k$, $L=L_{k+1}$ and $\gamma =\gamma_k$,
$\gamma' =\gamma_{k+1}$. To prove Theorem \ref{th:isp}, we use
Theorem \ref{th:MSAst} to estimate
\begin{eqnarray}
\lefteqn{\mathbb{P}\,(\;\Lambda_{L} \textnormal{ is not $(\gamma',E)-$good}\;)}\notag\\
&\leq& \mathbb{P}\,(\;\Lambda_{L} \textnormal{ is
$E-$resonant}\;)\label{pe:s1}
\\  &&+\;\mathbb{P}\,(\;\textnormal{One of the cubes $\Lambda_{2l}(m)\subset\Lambda_L$ is $E-$resonant}\;)\label{pe:s2}
\\  && +\;\mathbb{P}\,(\;\textnormal{There are two disjoint cubes in $\mathcal{C}_l(\LL)$}\nonumber\\
&& \quad\textnormal{which are not $(\gamma ,E)-$good}\;)\
.\label{pe:s3}
\end{eqnarray}

Both (\ref{pe:s1}) and (\ref{pe:s2}) can be bounded using the Wegner
estimate (Theorem \ref{th:Wegner})

\begin{eqnarray}
\mathbb{P}\,(\;\Lambda_{L} \textnormal{ is $E-$resonant}\;)&\leq& (2L+1)^d e^{-\sqrt{L}}\notag\\
&\leq& \frac{1}{3}\;\frac{1}{L^p}
\end{eqnarray}
provided $L$ is large enough, and

\begin{eqnarray}
\lefteqn{\mathbb{P}\,(\;\textnormal{One of the cubes $\Lambda_{2l}(m)\subset\Lambda_L$ is $E-$resonant}\;)}\\
&\leq& (2L+1)^d\;\; \mathbb{P}\,(\;\textnormal{The cube $\Lambda_{2l}(0)$ is $E-$resonant}\;)\\
&\leq& (2L+1)^d\; (4l+1)^d\;\; e^{-\sqrt{2l}}\notag\\
&\leq& (2L+1)^d\; (4L^{1/\alpha}+1)^d\;\; e^{-\sqrt{2}\,L^{\frac{1}{2\alpha}}}\notag\\
&\leq&\frac{1}{3}\;\frac{1}{L^p}
\end{eqnarray}
if $L$ is large enough.

Using the induction hypothesis (\ref{hyp:ind}), we can estimate the
term (\ref{pe:s3}) by

\begin{eqnarray}
\lefteqn{\sum_{i,j \in\LL\atop
\Lambda_l(i)\cap\Lambda_l(j)=\emptyset}\;\mathbb{P}\,(\Lambda_l(i)
\textnormal{ and }\Lambda_l(j)
\textnormal{ are both not $(\gamma,E)$-good}\,)}\notag\\
&\leq&\sum_{i,j \in\LL}\;\mathbb{P}\,(\Lambda_l(i) \textnormal{ is
not $(\gamma,E)$-good}\,)\: \mathbb{P}\,(\Lambda_l(j)
\textnormal{ is not $(\gamma,E)$-good}\,)\notag\\
&\leq& (2L+1)^{2d}\;\frac{1}{l^{2p}}\notag\\
&\leq& \frac{C}{L^{\frac{2p}{\alpha}-2d}}\notag\\
&\leq& \;\frac{1}{3}\: \frac{1}{L^p}
\end{eqnarray}
provided $L$ is large.

We used above that $\alpha<\frac{2p}{p+2d}$ implies
$\frac{2p}{\alpha}-2d>p$.

Summing up, we get

\begin{equation*}
\mathbb{P}\,(\;\Lambda_{L} \textnormal{ is not
$(\gamma',E)-$good}\;)\leq \frac{1}{L^p}\ .
\end{equation*}

\end{proof}

\subsection{Towards the strong form of the multiscale analyis}\label{sec:tostrong}\es

When we try to prove the `uniform' Result \ref{res:msastrong}, i.e.
the strong form of the multiscale estimate, we may proceed in the
same manner as above for awhile. Let us suppose we consider two
disjoint cubes $\Lambda_1=\LL(n)$ and $\Lambda_2=\LL(m)$. We want to
prove
\begin{eqnarray}\label{est:msastrong1}
\P\,\left(\,\textnormal{For some } E\in I \textnormal{
 both} \Lambda_1\textnormal{ \emph{and }
$\Lambda_2$ are  not $(\gamma',E)-$good}\,\right)\leq L^{-2p}\ .
\end{eqnarray}

We set
\begin{eqnarray}\label{def:sets}
A_1(E)&=&\{\,\Lambda_1\textnormal{ is  not
$(\gamma',E)-$good}\,\}\notag \\
R_1(E)&=&\{\,\Lambda_1 \textnormal{ or a cube in $\mathcal{C}_{2l}(\Lambda_1)$ is not $E$-resonant}\,\}\\
B_1(E)&=&\{\,\mathcal{C}_{l}(\Lambda_1) \textnormal{ contains two
disjoint cubes which are not $(\gamma,E)-$good}\,\}\ . \notag
\end{eqnarray}
We define $A_2(E), R_2(E), B_2(E)$ analogously for the cube
$\Lambda_2$.

The event we are interested in (see \ref{est:msastrong1}) can be
expressed through $A_1(E), A_2(E)$, namely
\begin{eqnarray}
\{\,\exists_{E\in I}\textnormal{
 such that } \Lambda_1\textnormal{ \emph{and}
$\Lambda_2$ are  not
$(\gamma',E)-$good}\,\}\\
= \bigcup_{E\in I}\:\big(\,A_1(E) \cap A_2(E)\,\big)\ .
\end{eqnarray}
Theorem \ref{th:MSAst} implies that

\begin{eqnarray}
&&\P\,(\;\bigcup_{E\in I}\;(\,A_1(E)\cap A_2(E)\,)\;)\notag\\
&\leq& \P\,(\;\bigcup_{E\in I}\;(\,R_1(E)\cap R_2(E)\,)\;)\label{RR}\\
&+& \P\,(\;\bigcup_{E\in I}\;(\,B_1(E)\cap B_2(E)\,)\;)\label{BB}\\
&+& \P\,(\;\bigcup_{E\in I}\;(\,R_1(E)\cap B_2(E)\,)\;)\label{RB}\\
&+& \P\,(\;\bigcup_{E\in I}\;(\,B_1(E)\cap R_2(E)\,)\;)\label{BR}\ .
\end{eqnarray}

The term (\ref{RR}) can be estimated using the `uniform' Wegner
estimate (Theorem \ref{th:dWegner}) and (\ref{BB}) will be handled
using the induction hypothesis. It turns out that the critical terms
are the mixed ones (\ref{RB}) and (\ref{BR}).

The only effective way we know to estimate (\ref{BR}) is
\begin{eqnarray}
\P\,(\;\bigcup_{E\in I}\;(\,B_1(E)\cap R_2(E)\,)\;)\;&\leq&\;\P\,(\;\bigcup_{E\in I}\;\,B_1(E)\;)\notag\\
&\leq& L^{2d}\; \frac{1}{\,l^{2p}\,}\notag\\
&\leq& \frac{1}{\,L^{2p/\alpha - 2d}\,}\label{BRplus}
\end{eqnarray}
where we used the induction hypothesis and the fact that there are
at most $L^{2d}$ disjoint cubes of side length $l$ in $\Lambda_1$.

Observe that the term $\bigcup_{E\in I} R_2(E)$ which we neglected
above does not have small probability as long as there is spectrum
inside $I$.

Since we need $\alpha>1$, the exponent in (\ref{BRplus}) is
certainly \emph{less} than $2p$. Consequently there is no way to do
the induction step the way we tried above. The induction step would
require that (\ref{BRplus}) is less than $\frac{1}{L^{2p}}$.

Observe that the situation is completely analogous to the one in
Section \ref{sec:msaft} (see (\ref{toobad})). There we needed to
allow more (namely one) bad cubes. The same idea remedies the
present situation: We have to accept `three' bad cubes, as will be
explained in  the next section.

\subsection{Estimates - third try \label{sec:tt}}\es

In a third round, we accept `three' bad cubes. More precisely: We
assume that the cube $\LL$ does not contain four disjoint cubes of
side length $l$ which are not $(\gamma, E)$-good. Then, there are
(at most) three cubes, $\Lambda_{2l}(m_1), \Lambda_{2l}(m_2),
\Lambda_{2l}(m_3)\subset \LL$, such that there are no bad cubes
outside $M=\bigcup_{\nu=1}^3\,\Lambda_{2l}(m_\nu)$.

As in Section \ref{sec:st}, we use the geometric resolvent equation
and an exponential bound on the Green's function as long as we do
not enter one of the $\Lambda_{2l}(m_\nu)$. Once we enter such a
set, we would like to use the geometric resolvent equation in
connection with a Wegner-type bound for $\Lambda_{2l}(m_\nu)$ as in
the following expression for $u\in \Lambda_{2l}(m_\nu)$:

\begin{eqnarray}
\nor{G_E^{\LL}(u,n)} &\leq&
\sum_{(q,q')\in\partial\Lambda_{2l}(m_\nu)\atop q\in
\,\Lambda_{2l}(m_\nu)} \nor{G_E^{\Lambda_{2l}(m_\nu)}(u,q)}
\;\nor{G_E^{\LL}(q',n)}\ .\label{eq:3bad}
\end{eqnarray}

If we assume that $\Lambda_{2l}(m_\nu)$ is not $E$-resonant we can
estimate the first term on the right hand side of (\ref{eq:3bad}) by
$e^{\sqrt{2l}}$. If the site $q'$ is the center of a good cube, we
may estimate the second term $G_E^{\LL}(q',n)$ by applying the
geometric resolvent equation for the cube $\Lambda_l(q')$ and using
the exponential bound for this cube. However, it is not guaranteed
that $\Lambda_l(q')$ is $(\gamma,E)$-good. $q'$ could belong to one
of the other `dangerous' cubes $\Lambda_{2l}(m_{\nu'})$. The problem
here is that two (or all three) of these cubes could touch or
intersect.

To get rid of this problem, we redefine the `dangerous' regions
where we use the Wegner bound instead of the exponential bound. We
say that two subsets $A$ and $B$ of $\Z^d$ \emph{touch} if $A\cap
B\not= \emptyset$ or if there are points $x\in A$ and $y\in B$ such
that $||x-y||_\infty = 1$.

As before we use the geometric resolvent equation iteratively to
estimate the Green's function $G_E^{\LL}$. We define sets $M_1, M_2,
M_3$ - the dangerous regions - where we use the Wegner estimate, i.
e. we will assume that the $M_i$ are not $E$-resonant. We construct
the $M_i$ in such a way that for all sites $x$ outside the $M_i$,
the cube $\Lambda_l(x)$ is $(\gamma,E)$-good. Moreover, any two of
the $M_i$ do not touch.

If the cubes $\Lambda_{2l}(m_\nu)$ do not touch each other, we set
$M_\nu=\Lambda_{2l}(m_\nu)$.

If two of the $\Lambda_{2l}(m_\nu)$ touch, say $\Lambda_{2l}(m_1)$
and $\Lambda_{2l}(m_2)$, we set $M'=\Lambda_{6l+1}(m_1)$. Then
$\Lambda_{2l}(m_1)\cup\Lambda_{2l}(m_2)\subset M'$. Indeed, if
$\Lambda_{2l}(m_1)$ and $\Lambda_{2l}(m_2)$ touch, there are points
$x\in\Lambda_{2l}(m_1)$ and $y\in\Lambda_{2l}(m_2)$ with
$\no{x-y}_\infty\leq 1$. If $z\in\Lambda_{2l}(m_2)$ we have
\begin{align}
||z-m_1||_\infty\;\leq\; &||z-m_2||_\infty + ||m_2-y||_\infty +
||y-x||_\infty + ||x-m_1||_\infty\notag\\\leq\;&
6\,l+1\label{constr:M}\ .
\end{align}

If $M'$ and $\Lambda_{2l}(m_3)$ do not touch we set $M_1=M'$ and
$M_2=\Lambda_{2l}(m_3)$ (The set $M_3$ is not needed, we may
formally set $M_3=\emptyset$.). If $M'$ and $\Lambda_{2l}(M_3)$ do
touch then $M',\Lambda_{2l}(m_3)\subset \Lambda_{10l+2}(m_1)$ which
is shown by a calculation analogous to (\ref{constr:M}). In this
case, we set $M_1=\Lambda_{10l+2}(m_1)$ and $M_2=M_3=\emptyset$.

We have shown

\begin{lemma}\label{lem:tt}
If there are not four disjoint cubes in $\mathcal{C}_l(\LL)$ which
are not $(\gamma,E)$-good, then either
\begin{itemize}
\item There are three cubes $M_1, M_2, M_3 \in\mathcal{C}_{2l}(\LL)$ which do not touch and such that
any cube in $\mathcal{C}_{l}(\LL)$ with center outside the $M_i$ is $(\gamma,E)$-good,\\
{\noindent\quad\quad\rm or}
\item There is a cube $M_1 \in \mathcal{C}_{6l+1}(\LL)$ and a cube $M_2\in\mathcal{C}_{2l}(\LL)$ which do not touch
such that any cube in $\mathcal{C}_{l}(\LL)$ with center outside the $M_i$ is $(\gamma,E)$-good,\\
{\quad\quad\rm or}
\item There is a cube $M_1 \in \mathcal{C}_{10l+2}(\LL)$ such that any cube in $\mathcal{C}_{l}(\LL)$
with center outside $M_1$ is $(\gamma,E)$-good.
\end{itemize}
\end{lemma}

\goodbreak
We are now in a position to prove the analytic part of the induction
step of multiscale analysis in the final form.

\begin{theorem}\label{th:MSAtt}
Suppose $L_0$ is large enough and $L_{k+1}={L_k}^\alpha$ with $1<\alpha<2$.\\
If for a certain $k$ ($l:=L_k$ and $L:=L_{k+1}$)
\begin{enumerate}
\item there do not exist four disjoint cubes in $\mathcal{C}_{l}(\Lambda_{L})$ which are
not $(\gamma_{k}, E)$- good with a rate $\gamma_k\ge
\frac{12}{l^{1/2}}$,
\item no cube in
\begin{equation}
\mathcal{C}_{2l}(\Lambda_{L})\;\cup\;\mathcal{C}_{6l+1}(\Lambda_{L})\;\cup\;\mathcal{C}_{10l+2}(\Lambda_{L})
\end{equation}
is $E$-resonant and
\item the cube $\Lambda_{L}$ is not $E$-resonant,
\end{enumerate}
then the cube $\Lambda_{L}$ is $(\gamma_{k+1}, E)$- good with a rate
$\gamma_{k+1}$ satisfying $\gamma_{k+1}\ge \frac{12}{L^{1/2}}$.

Moreover we can choose the rate $\gamma_{k+1}$ such that
\begin{equation} \label{est:rate2}\gamma_{k+1}\ \geq\ \gamma_{k}\; - \gamma_{k}\,\frac{C}{{L_k}^{\alpha-1}}\;
-\;\frac{C}{{L_k}^{\alpha/2}}\ .
\end{equation}
\end{theorem}

As above, we can estimate the decay rates as follows.

\begin{corollary}
If $L_0$ is large enough and the initial rate $\gamma_0$ satisfies
$\gamma_0 \ge \frac{12}{{L_0}^{1/2}}$ then the $\gamma_k$ in Theorem
\ref{th:MSAtt} satisfy $\gamma_k \ge \frac{\gamma_0}{2}$ for all
$k$.
\end{corollary}

\begin{proof}
We set $\gamma=\gamma_k$ and $\gamma'=\gamma_{k+1}$. From Lemma
\ref{lem:tt} we know that there are three cubes $M_1, M_2, M_3$ of
side length $2l$, $6l+1$ or $10l+2$ (or $0$ if $M_i=\emptyset$) such
that the $M_i$ contain all cubes in $\mathcal{C}_{l}(\Lambda_{L})$
which are not $(\gamma,E)$-good.

Starting with $m\in\Lambda_{\sqrt{L}}$ and $n\in\partial^-\LL$, we
use the geometric resolvent equation repeatedly.

If $m$ does not belong to one of the `dangerous' cubes $M_i$ we know
$\Lambda_l(m)$ is $(\gamma,E)$-good, so we estimate

\begin{eqnarray}
\nor{G_E^{\LL}(m,n)} &\leq& \sum_{(q,q')\in\partial\Ll(m)\atop q\in
\,\Ll(m)}
\nor{G_E^{\Ll(m)}(m,q)}\; \nor{G_E^{\LL}(q',n)}\\
&\leq& 2d\,(2l+1)^{d-1}\, e^{- \gamma l}\;  \nor{G_E^{\LL}(n_1,n)} \\
&\leq&  e^{- \tilde{\gamma}
l}\;\;\nor{G_E^{\LL}(n_1,n)}\label{est:exp2}\ .
\end{eqnarray}

We call such a step  an exponential bound. We do this repeatedly, as
long as the new point $n_1, n_2, \ldots$ neither belongs to one of
the $M_i$ nor is close to the boundary of $\LL$.

If $n_j$ belongs to one of the $M_i$, say to $M_1$, we use a
Wegner-type bound

\begin{eqnarray}
\nor{G_E^{\LL}(n_j,n)} &\leq& \sum_{(q,q')\in\partial M_1\atop q\in
\,M_1}
\nor{G_E^{M_1}(n_j,q)}\; \nor{G_E^{\LL}(q',n)}\nonumber\\
&\leq& 2d\,(20l+5)^{d-1}\,e^{\sqrt{10l+2}} \
\nor{G_E^{\LL}(n_j',n)}\label{step:Wegner2}
\end{eqnarray}
for a certain $n_j'\in\partial^+M_1$. Since the $M_i$ do not touch,
we can be sure that $\Lambda_l(n_j')$ is $(\gamma,E)$-good.
Consequently, we can always (as long as $n_j'$ is not near the
boundary of $\LL$)
 do an exponential bound after a Wegner-type bound and obtain

\begin{align}
&\nor{G_E^{\LL}(n_j,n)}\notag\\
\leq&\; 2d\,(20l+5)^{d-1}\,e^{\sqrt{10l+2}}\
\nor{G_E^{\LL}(n_j',n)}\notag\\ \label{est:double} \leq&\;
(2d)^2\,(20l+5)^{d-1}\,(2l+1)^{d-1}\;e^{\sqrt{10l+2}}\;e^{-\gamma\,l}\
\nor{G_E^{\LL}(n_{j+1},n)}\ .
\end{align}

If $l$ is larger than a certain constant and $\gamma\ge
\frac{12}{l^{1/2}}$, we have

\begin{align}
\rho~=~(2d)^2\,(20l+5)^{d-1}\,(2l+1)^{d-1}\;e^{\sqrt{10l+2}}\;e^{-\gamma\,l}~\leq~1
\end{align}
thus
\begin{equation}
(\ref{est:double})\leq \nor{G_E^{\LL}(n_{j+1},n)}\ .
\end{equation}

Whenever the point $n_j$ does not belong to one of the `dangerous'
regions $M_i$, we know that $\L_l(n_j)$ is $(\gamma,E)$-good. Hence,
we obtain an exponential bound of the Green's function and gain an
exponential factor $e^{-\gamma\,l}$. This step can be done roughly
$\frac{L}{l}$ times. Hence, we get the desired bound. The details
are as in the previous sections.
\end{proof}

Now, we do the probabilistic estimate.

\begin{theorem}\label{th:msastrongres}
Assume that the probability distribution $P_0$ has a bounded
density. Suppose $L_0$ is large enough, $\gamma\ge
\frac{12}{{\,L_0}^{1/2}}$, $p>2d$  and $1<\alpha<\frac{2p}{p+2d}$.
If for any disjoint cubes $\Lambda_{L_0}(n)$ and $\Lambda_{L_0}(m)$
\begin{equation} \begin{split}
\P\,\big(\quad&\textnormal{For some } E\in I\textnormal{ both }
\Lambda_{L_0}(n)\textnormal{ \emph{and}
$\Lambda_{L_0}(m)$} \\
&\textnormal{ are  not $(2\gamma,E)-$good}\ \big)\quad\leq
\quad\frac{1}{{L_0}^{2p}}
\end{split}\end{equation}

then for all $k$ and all disjoint cubes $\Lambda_{L_k}(n)$ and
$\Lambda_{L_k}(m)$

\begin{equation} \begin{split}
\P\,\big(\quad&\textnormal{For some } E\in I\textnormal{ both }
\Lambda_{L_k}(n)\textnormal{ \emph{and}
$\Lambda_{L_k}(m)$} \\
&\textnormal{ are  not $(\gamma,E)-$good}\ \big)\quad\leq
\quad\frac{1}{{L_k}^{2p}}\ . \label{ind:k}
\end{split}\end{equation}
\end{theorem}

\begin{proof}
The prove works by induction. So, we suppose, we know (\ref{ind:k})
already for $k$. We try to prove it for $k+1$.

As usual, we set $l=L_k$, $L=L_{k+1}$, $\gamma=\gamma_k$, and
$\gamma'=\gamma_{k+1}$.

We also abbreviate $\Lambda_1=\Lambda_{L_{k+1}}(n)$ and
$\Lambda_2=\Lambda_{L_{k+1}}(m)$.

Similar to (\ref{def:sets}) we define ($i=1, 2$)
\begin{eqnarray}
A_i(E)&=&\{\;\Lambda_i\textnormal{ is  not
$(\gamma',E)-$good}\,\}\notag \\
Q_i(E)&=&\{\;\Lambda_i \textnormal{ or a cube in
$\mathcal{C}_{2l}(\Lambda_i)
\cup\mathcal{C}_{6l+1}(\Lambda_i)\cup\mathcal{C}_{10l+2}(\Lambda_i)$
is not $E$-resonant}\,\}
\notag\\
D_i(E)&=&\{\;\mathcal{C}_l(\Lambda_i) \textnormal{ contains four
disjoint cubes which are not $(\gamma,E)-$good}\,\}\ . \notag
\end{eqnarray}
Let us denote by $\mathcal{S}(E)$ the set of all cubes of side
length $l$ which are \emph{not} $(\gamma,E)$-good. Like in Section
\ref{sec:tostrong}, we estimate
\begin{equation}\begin{split}
&\P\,\big(\ \;\exists_{E\in I} \textnormal{
 such that } \Lambda_1 \textnormal{ \emph{and}
$\Lambda_2$} \textnormal{ are  not
$(\gamma',E)-$good}\;\big)\notag\\
=\;\;\; &\P\:\bigg(\; \bigcup_{E\in I}\:\big(\,A_1(E) \cap A_2(E)\,\big)\ \;\bigg)\\
\leq \;\;\;&\P\,\bigg(\;\bigcup_{E\in I}\;(\,Q_1(E)\cap
Q_2(E)\,)\;\bigg)\;+\;
\P\,\bigg(\;\bigcup_{E\in I}\;(\,D_1(E)\cap D_2(E)\,)\;\bigg)\\
+\;& \P\,\bigg(\;\bigcup_{E\in I}\;(\,Q_1(E)\cap
D_2(E)\,)\;\bigg)\;+
\; \P\,\bigg(\;\bigcup_{E\in I}\;(\,D_1(E)\cap Q_2(E)\,)\;\bigg)\\
\leq\;\;\; &\P\,\bigg(\;\bigcup_{E\in I}\;(\,Q_1(E)\cap
Q_2(E)\,)\;\bigg)\;
+\;\P\,\bigg(\;\bigcup_{E\in I}\;\,D_1(E)\;\bigg)\;\P\,\bigg(\;\bigcup_{E\in I}\;\,D_2(E)\;\bigg)\\
+\;&\P\,\bigg(\;\bigcup_{E\in I}\;\,D_1(E)\;\bigg)\;+\;\P\,\bigg(\;\bigcup_{E\in I}\;\,D_2(E)\;\bigg)\;\\
\leq\quad&\P\,\bigg(\;\bigcup_{E\in I}\;(\,Q_1(E)\cap
Q_2(E)\,)\;\bigg)\quad +\quad 3\;\;\; \P\,\bigg(\;\bigcup_{E\in
I}\;\,D_1(E)\;\bigg)\;\ .
\end{split}\end{equation}

Let us first estimate the latter term:

\begin{align}
    &\quad\P\,\bigg(\;\bigcup_{E\in I}\;\,D_1(E)\;\bigg)\;\notag\\
    =&\quad\P\,\big(\; \oexists_{\textstyle E\in I}\;\;\;
    \oexists_{\textstyle C_1, C_2, C_3, C_4 \in\mathcal{C}_l(\LL)\atop \mbox{pairwise disjoint} }
    C_1, C_2, C_3, C_4 \in \mathcal{S}(E)\;\big)\notag\\
    \leq&\sum_{ \textstyle C_i \in\mathcal{C}_l(\LL)\atop \mbox{pairwise disjoint} }\;\;
    \P\,\big(\; \oexists_{\textstyle E\in I}\;
        C_1\in \mathcal{S}(E),\, C_2\in \mathcal{S}(E),\, C_3\in \mathcal{S}(E)\,
        \textnormal{ and }C_4 \in \mathcal{S}(E)\;\big)\notag
\end{align}
\begin{align*}
    \leq&\sum_{ \textstyle C_i \in\mathcal{C}_l(\LL)\atop \mbox{pairwise disjoint} }\;\;
    \P\;\bigg(\quad \Big(\oexists_{\textstyle E\in I}\;\;\;C_1\in \mathcal{S}(E)\; \textnormal{ and }
        C_2\in \mathcal{S}(E)\;\Big)\quad \textnormal{ and }\notag\\
        &\hspace{3.4cm}\Big(\oexists_{\textstyle E\in I}\;\;\;C_3\in \mathcal{S}(E)\;
        \textnormal{ and }C_4 \in \mathcal{S}(E)\;\Big)\qquad\bigg)\notag\\
\end{align*}
\begin{align*}
    \leq&\sum_{ \textstyle C_i \in\mathcal{C}_l(\LL)\atop \mbox{pairwise disjoint} }\;\;
    \P\;\bigg(\quad \oexists_{\textstyle E\in I}\;C_1, C_2\in \mathcal{S}(E)\;\bigg)\quad
        \P\;\bigg(\oexists_{\textstyle E\in I}\;C_3, C_4\in \mathcal{S}(E)\;
        \;\bigg)\notag\\
    \leq&\quad C\;L^{4d}\;\left(\frac{1}{l^{2p}}\right)^2\quad\leq\;\frac{C}{\,L^{4p/\alpha\, -\, 4d}\,}\quad
    \leq\;\;\frac{1}{4}\;\frac{1}{\,L^{2p}\,}\ .
\end{align*}
In the last step, we used that $p>2d$ and
$1<\alpha<\frac{2p}{p+2d}$.

We turn to the estimate of
\begin{equation*}
    \P\,\bigg(\;\bigcup_{E\in I}\;(\,Q_1(E)\cap Q_2(E)\,)\;\bigg)\;
\end{equation*}
By setting $\mathcal{Q}_i=\mathcal{C}_{2l}(\Lambda_i)
\cup\mathcal{C}_{6l+1}(\Lambda_i)\cup\mathcal{C}_{10l+2}(\Lambda_i)
\cup \{\Lambda_i\}$, we get
\begin{equation}
    \P\,\bigg(\;\bigcup_{E\in I}\;(\,Q_1(E)\cap Q_2(E)\,)\;\bigg)\;\\
    \leq \sum_{c_1\in\mathcal{Q}_1, c_2\in\mathcal{Q}_2}\;\;\P\,\Big(\;\oexists_{E\in I}\;c_1
    \textnormal{ and } c_2 \textnormal{ are $E$-resonant}\;\Big)\ .\label{dwegest}
\end{equation}

\vspace{0.8cm} Each term in the sum in (\ref{dwegest}) can be
estimated using Theorem \ref{th:dWegner} by a term of the form
$C\,L^k\,e^{L^\frac{1}{2}}$ and the sum does not have more than
$C\,L^d$ terms, thus the sum can certainly be bounded by
$\frac{1}{4}\;\frac{1}{L^{2p}}$.

This finishes the proof.
\end{proof}

$\;$\\[2mm]
{\bf Notes and Remarks }\\[2mm]
The celebrated paper by Fröhlich and Spencer \cite{froespe} laid the
foundation for multiscale analysis. This technique was further
developed and substantially simplified in the paper by Dreifus and
Klein \cite{dk}. Germinet and Klein \cite{gk1} developed a
`Bootstrap multiscale analysis' which uses the output of a
multiscale estimate as the input of a new multiscale procedure.
These authors obtain the best available estimates of this kind. In
fact, in \cite{gk2} they prove that their result characterizes the
regime of `strong localization'.

Multiscale analysis can be transferred to the continuous case as
well, see e.g. \cite{MartHold,ch1,bch,ki1,kloppLoc,fk,KSS,KSS1,
Veselic}. \NeueSeite

\setcounter{equation}{0}
\section{The initial scale estimate\label{ch:ise}}
\ua{ Large disorder} In this final chapter, we will prove an initial
scale estimate for two cases, namely for energies near the bottom of
the spectrum with arbitrary disorder and for arbitrary energies at
large disorder.

We prove the initial scale estimate first for the case of high
disorder. As usual we have to assume that the random variables
$V_\omega(n)$ are independent and identically distributed with a
bounded density $g(\lambda)$. We may say that \emph{the disorder is
high} if the norm $\|\,g\,\|_\infty$ is small. In fact, small
$\|\,g\,\|_\infty$ reflects a wide spreading of the random
variables.

\begin{theorem} Suppose the distribution $P_0$ has a bounded density $g$.

Then for any $L_0$ and any $\gamma>0$, there is a $\rho>0$ such
that:

If $\;\|\,g\,\|_\infty<\rho$ and $\Lambda_1=\Lambda_{L_0}(n),\
\Lambda_2=\Lambda_{L_0}(m)$ are disjoint, then
\begin{equation}
\P\,(\;\exists_E\;\Lambda_{1} \textnormal{ and } \Lambda_{2}
\textnormal{ are both not }(\gamma,E)\textnormal{-good}\;) \leq
\;\frac{1}{{L_0}^{2p}}\ .
\end{equation}
\end{theorem}

\begin{proof}
Since $\nor{G_E^{\Lambda_i}(m,n)}\leq
\|\,(H_{\Lambda_i}-E)^{-1}\,\|$ we have
\begin{eqnarray*}
   && \P\,(\;\exists_E\quad\Lambda_{1} \textnormal{ and } \Lambda_{2} \textnormal{ are both not }(\gamma,E)\textnormal{-good}\;) \\
  &\leq& \P\,(\;\exists_E\quad\|\,(H_{\Lambda_1}-E)^{-1}\,\| > e^{-\gamma\,L_0}\textnormal{ and } \|\,(H_{\Lambda_2}-E)^{-1}\,\| > e^{-\gamma\,L_0}\;) \\
  &\leq& \P\,(\;\exists_E\quad\mbox{dist}(E,\sigma(H_{\Lambda_1}) \leq e^{\gamma\,L_0}\textnormal{ and } \mbox{dist}(E,\sigma(H_{\Lambda_2}) \leq e^{\gamma\,L_0}\;) \\
  &\leq& 2\,C\;\|\,g\,\|_\infty\;e^{\gamma\,L_0}\,(2\,L_0\,+\,1)^{2d}
\end{eqnarray*}
where we used the `uniform' Wegner estimate, Theorem
\ref{th:dWegner}, in the final estimate. By choosing $\rho$ and,
hence, $\|\,g\,\|_\infty$ very small we obtain the desired estimate.
\end{proof}

\ua{ The Combes-Thomas estimate\label{sec:CT}}

To prove the initial scale estimate for small energies, the
following bound is crucial.

\begin{theorem}[Combes-Thomas estimate]\label{th:CT}
If $H=H_0+V$ is a discrete Schrödinger operator on $\ell^2(\Z^d)$
and $\textnormal{dist}(E,\sigma(H))=\delta\leq 1$, then for any $n,
m\in\Z^d$
\begin{equation}\label{est:CT}
\big|{(H-E)^{-1}(n,m)}\big|\;\;\leq\;\;\frac{2}{\delta}\;\;\;e^{-\,\:\frac{\delta}{12\,d}\,\gnorm{n-m}}\
.
\end{equation}
\end{theorem}

\begin{remark}Theorem \ref{th:CT} can be improved in various directions, see for example the discussion
of the Combes-Thomas estimate in \cite{stoll1}. In particular, the
condition $\delta\leq 1$ which we need for technical reasons is
rather unnatural. Our proof can easily be extended to $\delta\leq C$
for any $C<\infty$ but then the exponent $\frac{\delta}{12\,d}$ in
the right hand side of (\ref{est:CT}) has to be adjusted depending
on the value of $C$.
\end{remark}

\begin{proof}
For fixed  $\mu>0$ to be specified later and fixed $n_0$, we define
the multiplication operator $F=F_{n_0}$ on $\ell^2(\Z^d)$ by

\begin{equation}\label{def:F}
    F\,u(n)\;=\;F_{n_0}\,u(n)\;=\;e^{\,\mu\,\gnorm{n_0-n}}\;u(n)\ .
\end{equation}
Then for any operator $A$ we have
\begin{equation}\label{prop:covarF}
    \Big(\,F^{-1}_{n_0}\,A\,F_{n_0}\,\Big)\;(n,m)\;\;=
    \;e^{\,-\mu\,\gnorm{n_0-n}}\;A(n,m)\;\;e^{\,\mu\,\gnorm{n_0-m}}\ .
\end{equation}

Hence, with $F=F_n$
\begin{align}
  \big|\,(H-E)^{-1}(n,m)\,\big| &= e^{-\,\mu\,\gnorm{n-m}}\;\big|\,F^{-1}\,(H-E)^{-1}\,F\;(n,m)\big| \nonumber\\
  &= e^{-\,\mu\,\gnorm{n-m}}\;\big|\,(F^{-1}\,H\,F-E)^{-1}\;(n,m)\,\big| \nonumber \\
  &\leq e^{-\,\mu\,\gnorm{n-m}}\;\|\,(F^{-1}\,H\,F - E)^{-1}\|\ .\label{est:oper1}
\end{align}

To compute the norm of the operator  $(F^{-1}\,H\,F - E)^{-1}$, we
use the resolvent equation to conclude
\begin{align}
  &(F^{-1}\,H\,F - E)^{-1} \notag\\
 =\;& (H-E)^{-1}- (F^{-1}\,H\,F - E)^{-1}\big(F^{-1}HF - H\big) (H-E)^{-1}\ .\notag
 \end{align}

 This implies
 \begin{equation*}
(F^{-1}\,H\,F - E)^{-1} (1+(F^{-1}HF - H) (H-E)^{-1})=(H-E)^{-1}\ .
 \end{equation*}

If $\|(F^{-1}HF - H) (H-E)^{-1}\|\le 1$, we may invert $(1+(F^{-1}HF
- H) (H-E)^{-1})$ and obtain
\begin{eqnarray}
  &&  (F^{-1}H F - E)^{-1}\notag\\[0.15cm]
  &=&  (H-E)^{-1} \Big(1+(F^{-1}H F - H) (H-E)^{-1}\Big)^{-1}
  \ .\label{est:oper}
\end{eqnarray}

We compute the norm of the operator $F^{-1}HF - H$. If an operator
$A$ on $\ell^2(\Z^d)$ has matrix elements $A(u,v)$, then $A$ is
bounded if
\begin{eqnarray*}
    a_1&=&\sup_{u\in\,\Z^d}\,\sum_{v\in\,\Z^d}\;\big|{A(u,v)}\big|\;<\;\infty\qquad\mbox{and}\\
    a_2&=&\sup_{v\in\,\Z^d}\,\sum_{u\in\,\Z^d}\;\big|{A(u,v)}\big|\;<\;\infty\ .
\end{eqnarray*}

Moreover, we have
\begin{equation}\label{est:opnorm}\|\,A\,\|\;\leq\;\; {a_1}^{1/2}\; {a_2}^{1/2} \end{equation}
(see e.g. \cite{weid}). We estimate using (\ref{prop:covarF})

\begin{eqnarray}
  \sum_{v\in\,\Z^d}\;|\;\big(F_n^{-1}H F_n - H\big) (u,v)\;|\;&\leq&\;
  \sum_{v: \gnorm{v-u}=1}\;\no{e^{-\mu\gnorm{n-u}}\,e^{\,\mu\gnorm{n-v}}-1} \notag\\[0.5cm]
  &\leq& \quad 2\,d\;\;\mu\;e^{\mu}\ . \label{est:FHF}
 \end{eqnarray}
The last inequality results from an elementary calculation:

For $\gnorm{u-v}\leq 1$ we have
\begin{equation*}
\big|\,\gnorm{n-u}\,-\,\gnorm{n-v}\,\big|\;\leq\;\gnorm{u-v}\;\leq\;1\
.
\end{equation*}

Moreover, for $|a|\leq 1$ and $\mu>0$

\begin{align*}
\big|\,e^{\mu a}-1\,\big|\;&\leq\;\big|\,e^\mu-1\,\big|\;=\;e^\mu-1\\
&\leq\;\int_0^1\;\mu\,e^{\mu\,t} \leq\;\mu\,e^{\mu}
\end{align*}

which proves (\ref{est:FHF}).

Estimate (\ref{est:FHF}) and an analogous estimate with the role of
$u$ and $v$ interchanged imply using (\ref{est:opnorm})

\begin{equation}\label{est:ctexp}
\|\,F^{-1}HF - H\,\|\;\leq\; 2\,d\;\mu\,e^{\mu}\ .
\end{equation}

Now we \emph{choose} $\mu=\frac{\delta}{\,12\,d\,}$. As
$\textnormal{dist}(E,\sigma(H))=\delta\leq 1$ we conclude
\begin{align}
\|(F^{-1}HF - H) (H-E)^{-1}\|\;&\leq \;\|(F^{-1}HF - H)\|\,\|(H-E)^{-1}\|\notag\\
&\leq\;2\,d\;\mu\,e^\mu\;\frac{1}{\delta}\notag\\
&=\;2\,d\;\frac{\delta}{\,12\,d\,}\;e^{\frac{\delta}{\,12\,d\,}}\;\;\frac{1}{\delta}\notag\\
&\leq\quad\eh
\end{align}

Above we used $e^{\frac{\delta}{\,12\,d\,}} \leq e \leq 3$ since
$\delta\leq 1$.

It follows that the operator $1+ (F^{-1}HF - H) (H-E)^{-1}$ is
indeed invertible and, using the Neumann series, we conclude that

\begin{equation}
\Big|\Big|\,\Big(1+ (F^{-1}HF - H)
(H-E)^{-1}\Big)^{-1}\,\Big|\Big|\;\leq\;2
\end{equation}

Thus, by (\ref{est:oper}) we have
\begin{align}
\big|\big|(F^{-1}HF - E)^{-1}\big|\big|\;
&=\; \big|\big|(H-E)^{-1}\,\big(1+ (F^{-1}HF - H)\, (H-E)^{-1}\big)^{-1}\big|\big|\notag\\
&\leq\;\frac{2}{\delta}
\end{align}

and (\ref{est:oper1}) gives
\begin{align}
 \big|\,(H-E)^{-1}(n,m)\,\big|
  &\leq\; e^{-\,\mu\,\gnorm{n-m}}\;\|\,(F^{-1}\,H\,F - E)^{-1}\|\notag\\
  &\leq\; \frac{2}{\delta}\;e^{-\,\frac{\delta}{12\,d\,}\,\gnorm{n-m}} .
\end{align}
\end{proof}

\ua{ Energies near the bottom of the spectrum}

For energies near the bottom of the spectrum, we prove the following
estimate.

\begin{theorem} Suppose the distribution $P_0$ has a bounded support.
Denote by $E_0$ the infimum of the spectrum of $H_\omega$. Then for
arbitrary large  $L_0$, any $\,C$ and $p$ there is an energy
$E_1>E_0$ such that
\begin{equation}
\P\,(\;\Lambda_{L_0} \textnormal{ is not }
(\frac{C}{{L_0}^{1/2}},E)\textnormal{-regular for some $E\leq E_1$}
\;) \leq \;\frac{1}{{L_0}^{p}}\ .
\end{equation}
\end{theorem}

\begin{remark}
By the results of Chapter \ref{ch:msa} the above result implies pure
point spectrum for energies near the bottom of the spectrum.
\end{remark}

\begin{proof}
If $E_0(H_{\Lambda_{L}})\,\ge 2 \gamma$ then Theorem \ref{th:CT}
implies that $\Lambda_{L}$ is $(\gamma,E)$-regular for any $E\le
\gamma$. Indeed, for such an $E$
\begin{equation}
\mbox{dist}\bigg(E,\sigma(H_{\LL})\bigg)\;\ge\;E_0(H_{\LL})-\gamma\;\ge\;\gamma\
.
\end{equation}

From our study of Lifshitz tails (Chapter \ref{ch:Lifshitz}), we
have already a lower bound on some $E_0(H_{\Lambda_{L}}^N)\,\leq\,
E_0(H_{\Lambda_{L}})\,$, namely:

By (\ref{kreuzlein}) and Lemma \ref{lem:ld} there exist $\ell_0$ and
$\beta$ such that
\begin{equation}\label{est:Neumann2}
\P\,\bigg(\;E_0(H_{\Lambda_{\ell_0}}^N)\,<
\frac{1}{\,\beta\,{\ell_0}^2\,}\;\bigg)\ \leq\ e^{-c\,{\ell_0}^d}\ .
\end{equation}

This estimate tells us that for $E\leq
\gamma\,=\,\frac{1}{\,2\,\beta\,{\ell_0}^2\,}$, the cube
$\Lambda_{\ell_0}$ is $(\gamma,E)$-good with very high probability.

This sounds like it is exactly what we need for the initial scale
estimate. Unfortunately, it is \emph{not quite} what makes the
machine work.

The multiscale scheme requires for the initial step the assumption
(see Theorem \ref{th:msastrongres})
\begin{equation}\label{est:gammaLif}
\gamma\;\ge\;\frac{C}{{L_0}^{1/2}}
\end{equation}
but the $\gamma$ we obtain from (\ref{est:Neumann2}) is much smaller
than the rate required by (\ref{est:gammaLif}). On the other hand,
the right hand side of (\ref{est:Neumann2}) is much better than what
we need (exponential versus polynomial bound). So, we may hope we
can `trade probability for rate'. This is exactly what we do now.

We build a big cube $\Lambda_{L_0}$ by piling up disjoint copies of
the cube $\Lambda_{\ell_0}$, more precisely

\begin{equation}\label{bigcube}
\Lambda_{L_0}\;=\;\bigcup_{j\in R}\;\Lambda_{\ell_0}(j)\ .
\end{equation}

Indeed, for any odd integer $r$ we may take $L_0 = r\, \ell_0 +
\frac{r-1}{2}$. The set $R$ in (\ref{bigcube}) contains $r^d$
points.

By (\ref{NDbrackdis}) we have
\begin{equation}
    H_{\Lambda_{L_0}}^N\;\geq\;\bigoplus_{j\in
    R}\,H_{\Lambda_{\ell_0}}^N(j)\ ,
\end{equation}

hence
\begin{equation}
    E_0\Big(H_{\Lambda_{L_0}}^N\Big)\;\geq\;\inf_{j\in
    R}\;E_0\Big(H_{\Lambda_{\ell_0}}^N(j)\Big)\ .
\end{equation}

It follows that
\begin{eqnarray}
  &&\P\,\bigg(\; E_0\Big(H_{\Lambda_{L_0}}^N\Big)\;\leq\;2\,\gamma\;\bigg) \notag\\
  &\leq&\P\,\bigg(\;\inf_{j\in R} E_0\Big(H_{\Lambda_{\ell_0}}^N(j)\Big)\;\leq\;2\,\gamma\;
  \bigg) \notag\\
  &\leq&\P\,\bigg(\; E_0\Big(H_{\Lambda_{\ell_0}}^N(j)\Big)\;\leq\;2\,\gamma\;\mbox{ for some } j\in R\bigg) \notag\\
  &\leq& r^d\quad \P\,\bigg(\;
  E_0\Big(H_{\Lambda_{\ell_0}}^N\Big)\;\leq\;2\,\gamma\;\bigg)\label{est:manysmall}\ .
\end{eqnarray}

If we choose $\gamma\,=\,\frac{1}{\,2\,\beta\,{\ell_0}^2\,}$, we may
use (\ref{est:Neumann2}) to estimate (\ref{est:manysmall}) and
obtain

\begin{equation}\label{est:bigcube}
\P\,\bigg(\;
E_0\Big(H_{\Lambda_{L_0}}^N\Big)\;\leq\;2\,\gamma\;\bigg)\;
\leq\;r^d\;e^{-\,c\,{\ell_0}^d}\ .
\end{equation}

Now, we choose $r$ and hence $L_0$ in such a way that
$\gamma>\frac{C}{{L_0}^{1/2}}$. This leads to setting $r\sim
{\ell_0}^3$, thus $L_0\sim {\ell_0}^4$. With this choice,
(\ref{est:bigcube}) gives

\begin{equation}\label{est:bigcube2}
\P\,\bigg(\;
E_0\Big(H_{\Lambda_{L_0}}^N\Big)\;\leq\;2\,\gamma\;\bigg)\;
\leq\;C_1\,{L_0}^{d}\;e^{-\,c'\,{L_0}^{d/4}}\ .
\end{equation}

Since the right hand side of (\ref{est:bigcube2}) is smaller than
$\frac{1}{{L_0}^p}$, this proves the initial scale estimate.

\end{proof}

{\bf Notes and Remarks }\\[2mm]
Already the paper \cite{froespe} contained the proof for high
disorder localization we gave above. The idea to use Lifshitz tails
to prove localization for small energies goes back to
\cite{MartHold} and was further developed in \cite{KSS} (see also
\cite{KiWa2}), but an intimate connection between Lifshitz tails and
Anderson localization was clear to physicists for a long time (see
\cite{lifshitz1}).

The Combes-Thomas inequality was proved in \cite{ct}. It was
improved in \cite{bch}, see also \cite{stoll1}. We took the proof
above from \cite{aizenm}. \NeueSeite


\section{Appendix: Lost in Multiscalization -- A guide through the jungle\label{ch:append}}
This is a short guide to the proof of Anderson localization via
multiscale analysis given in this text.

The core of the localization proof is formed by the estimates stated
in Section \ref{sec:msares} as Result \ref{res:msa} and
\ref{res:msastrong}. The first estimate (\ref{res:msa}) says that
for a given energy $E$, exponential decay of the Green's function is
very likely on large cubes. Cubes with exponentially decaying
Green's functions will be called `good' cubes. In Section
\ref{sec:msa2noac} we prove that the estimate in Result
\ref{res:msa} implies the absence of absolutely continuous spectrum.

The strong version (Result \ref{res:msastrong}) of the multiscale
estimate considers a whole energy interval $I$ and two disjoint
cubes. The result tells us that with high probability for all
energies in $I$ at least one of the cubes has an exponentially
decaying resolvent. This result is a strong version of the former
result as it is \emph{uniform} in the energy. The price to be paid
is the consideration of a second cube. A single cube cannot be good
for all energies in $I$ if there is spectrum at all in $I$ (see
\ref{th:noge}). We show in Section \ref{sec:pp} that the strong form
of the multiscale estimate implies pure point spectrum inside $I$.
This is done using the exponential decay of eigenfunctions which we
deduce from the key Theorem \ref{th:expdec}. The connection between
spectrum and (generalized) eigenfunctions is discussed in Chapter
\ref{ch:spec}.

The proofs of the multiscale estimates (Results \ref{res:msa} and
\ref{res:msastrong}) are contained in the Chapters \ref{ch:msa} and
\ref{ch:ise}. We prove the estimates inductively for cubes of side
length $L_k, k=0, 1, 2 \dots$ . The length scale is such that
$L_{k+1}=L_k^\alpha$ for an $\alpha>1$.

The induction step from $L_k$ to $L_{k+1}$ is done in Chapter
\ref{ch:msa}. In a first attempt (Section \ref{sec:msaft}) to do
this for the weaker form we prove that if all the small cubes (of
size $L_k$) inside a big cube (of size $L_{k+1}$) are good, then the
big cube itself is good if we have a rough a priori estimate for the
big cube. This a priori bound is provided by the `Wegner estimate',
a key ingredient to our proof. We prove the Wegner estimate in
Section \ref{sec:Wegner}. Unfortunately, the probability that
\emph{all} small cubes inside the big one are good is rather small.
So, this `first try' is not appropriate to prove that the big cube
is good with high enough probability.

In the `second try' we allow \emph{one} bad small cube inside the
big cube. (For the precise formulation see Section \ref{sec:st}). To
prove that this still implies that the big cube is good requires
more work. We need again that the big cube and also the `bad' small
cube allow an a priori bound of the Wegner type. The advantage of
allowing one bad cube is that this event has a much higher
probability. In this way, we prove the induction step for the weak
form of the multiscale analysis.

The strong form of the multiscale analysis is then treated in
Section \ref{sec:tt}. Here we have to allow even a few bad cubes
among the small ones. This makes the proof yet a bit more
complicated.

So far we have done the induction step. Of course, we still have to
prove the estimate for the initial length $L_0$. This is done in
Chapter \ref{ch:ise}. We prove that the initial estimate is
satisfied if either the disorder is large or the energy is close to
the bottom of the spectrum. An important tool in this chapter is the
Combes-Thomas inequality. We prove this result in section
\ref{sec:CT}.

The strategy of proof outlined above is certainly not the fastest
one to prove localization via multiscale analysis. However, we
believe that for a first reading, it is easier to learn the subject
this way than in a streamlined turbo version.

\NeueSeite



\begin{theindex}

  \item $A(i,j)$, 15
  \item $A\leq B$, 19
  \item $A\tmspace  +\thinmuskip {.1667em}\leq \tmspace  +\thinmuskip {.1667em}B$,
		35
  \item $A_\Lambda(i,j)$, 37
  \item $A_k$, 89
  \item $A_k^+$, 90
  \item $C(K)$, 16
  \item $C_0(\mathbb{R})$, 29
  \item $C_\infty(\R)$, 19, 26, 42
  \item $C_b(\mathbb{R})$, 31
  \item $\mathcal{C}_L(\Lambda)$, 87
  \item $\mathcal{C}_k^+$, 90
  \item $\mathcal{C}_l(\Lambda_L)$, 95
  \item $E_n(H_{\Lambda _L}^X)$, 41
  \item $\mathcal{F}$, 14
  \item $G_E^{\tmspace  +\thinmuskip {.1667em}\Lambda }(n,m)$, 81
  \item $G_z^\Lambda $, 41
  \item $H^{(L)}$, 53
  \item $H_0$, 13
  \item $H_\L$, 36
  \item $H_\L(V_\L, V_j=a)$, 46
  \item $H_\Lambda ^X$, 41
  \item $H_\Lambda^D$, 39
  \item $H_\Lambda^N$, 39
  \item $H_{\L}(V_\L)$, 44
  \item $H_{ac}$, 67
  \item $H_{pp}$, 67
  \item $H_{sc}$, 67
  \item $\H_{ac}$, 67
  \item $\H_{pp}$, 67
  \item $\H_{sc}$, 67
  \item $L_k$, 85
  \item $\ell^2(\Z^d)$, 13
  \item $\ell_0^2(\Z^d)$, 21
  \item $N(E)$, 30
  \item $N(H_\Lambda ^X,E)$, 42
  \item $N_\Lambda (E)$, 43
  \item $n(\lambda)$, 43
  \item $n_\Lambda(i)$, 37
  \item $P_0$, 20
  \item $\P$-almost all, 9
  \item $\P$-almost surely, 9
  \item $\mathop {\hbox {supp}}P_0$, 20
  \item $U_i$, 25
  \item $V_\omega^{(L)}$, 53
  \item $V_{\L}$, 44

  \item $\Gamma _\Lambda $, 36
  \item $\Gamma _\Lambda ^D$, 38
  \item $\Gamma _\Lambda ^N$, 38
  \item $(\gamma,E)$-good, 81
  \item $\gamma$-good, 82
  \item $\gamma_\infty$, 97
  \item $\gamma_k$, 97
  \item $\delta _i$, 14
  \item $\ve(H)$, 18, 67
  \item $\ve_g(H)$, 61
  \item $\Lambda _L$, 13
  \item $\Lambda_L(n_0)$, 13
  \item $\nu _L$, 29
  \item $\nu_{ac}$, 66
  \item $\nu_{pp}$, 66
  \item $\nu_{sc}$, 66
  \item $\tilde{\nu}_L$, 41
  \item $\tilde{\nu}_L^D$, 41
  \item $\tilde{\nu}_L^N$, 41
  \item $\mathaccentV {tilde}07E{\nu }_L^X$, 41
  \item $\rho(A)$, 15
  \item $\sigma(A)$, 15
  \item $\sigma_{ac}(H)$, 67
  \item $\sigma_{dis}(A)$, 18
  \item $\sigma_{ess}(A)$, 18
  \item $\sigma_{pp}(H)$, 67
  \item $\sigma_{sc}(H)$, 67
  \item $\chi _\Lambda $, 29
  \item $\chi_M$, 17

  \item $\Subset$, 86
  \item $\partial \Lambda$, 36
  \item $\partial _{\Lambda _2}\Lambda _1$, 39
  \item $\partial^+ \Lambda$, 36
  \item $\partial^- \Lambda$, 36
  \item $\partial_L^-\Lambda$, 87
  \item $|A|$, 13, 81
  \item $|\Lambda_L|$, 29
  \item $\inorm{n}$, 13
  \item $\ns{n}$, 13
  \item $\norm{u}$, 13

  \indexspace

  \item adjacency matrix, 37
  \item alloy-type potential, 9
  \item Anderson delocalization, 75
  \item Anderson localization, 75
  \item Anderson model, 15

  \indexspace

  \item Borel-Cantelli lemma, 20
  \item bound state, 68
  \item boundary, 36
  \item bounded below, 18
  \item bounded Borel measure, 65

  \indexspace

  \item canonical probability space, 23
  \item characteristic function, 17, 29
  \item collection of $L-$cubes inside $\Lambda$, 87
  \item continuous case, 10
  \item coordination number, 37
  \item counting measure, 8
  \item cylinder sets, 23

  \indexspace

  \item density of states, 43
  \item density of states measure, 30
  \item Dirichlet Laplacian, 38
  \item Dirichlet-Neumann bracketing, 35
  \item discrete case, 10
  \item discrete spectrum, 18
  \item dist, 16, 87
  \item distribution, 20
  \item dynamical localization, 78

  \indexspace

  \item $E$-resonant, 94
  \item ergodic, 24
  \item ergodic operators, 25
  \item essential spectrum, 18
  \item event, 20
  \item exponential decay, 81
  \item exponential localization, 75
  \item extended states, 75

  \indexspace

  \item finitely degenerate, 18
  \item Fourier transform, 14
  \item free operator, 7

  \indexspace

  \item generalized eigenfunction, 61
  \item generalized eigenvalue, 61
  \item geometric resolvent equation, 40
  \item graph Laplacian, 14, 37
  \item Green's functions, 41

  \indexspace

  \item identically distributed, 20
  \item iid, 20
  \item independent, 20
  \item initial length, 85
  \item inner boundary, 36
  \item integrated density of states, 30
  \item invariant, 24
  \item isolated, 18

  \indexspace

  \item kernel, 15

  \indexspace

  \item Laplacian, discrete, 14
  \item length scale, 85
  \item Lifshitz behavior, 51
  \item Lifshitz tails, 51

  \indexspace

  \item matrix entry, 15
  \item measure
    \subitem  pure point, 66
    \subitem absolutely continuous, 66
    \subitem bounded Borel, 66
    \subitem continuous, 66
    \subitem positive, 66
    \subitem singular continuous, 66
  \item measure preserving transformation, 23
  \item min-max principle, 18
  \item mobility edge, 76
  \item multiplicity, 18
  \item multiscale analysis - strong form, 86
  \item multiscale analysis - weak form, 85

  \indexspace

  \item Neumann Laplacian, 37
  \item non degenerate, 18

  \indexspace

  \item outer boundary, 36

  \indexspace

  \item Poisson model, 9
  \item Poisson random measure, 9
  \item polynomially bounded, 61
  \item positive Borel measure, 66
  \item positive operator, 18, 35
  \item projection valued measure, 17

  \indexspace

  \item Radon-Nikodym, 66
  \item RAGE-theorem, 67
  \item random point measure, 8
  \item random variable, 19
  \item resolvent, 15
  \item resolvent equations, 15
  \item resolvent set, 15
  \item resonant, 94

  \indexspace

  \item scattering states, 70
  \item Schr\"odinger operator, 7
  \item simple, 18
  \item simple boundary conditions, 36
  \item single site potential, 8
  \item spectral localization, 77
  \item spectral measure
    \subitem projection valued, 17, 62, 66
    \subitem real valued, 62
  \item spectral measure zero, 61
  \item spectrum, 15
    \subitem absolutely continuous, 67
    \subitem pure point , 67
    \subitem singular continuous, 67
  \item stochastic process, 23
  \item Stone-Weierstra\IeC {\ss } Theorem, 19
  \item support, 20

  \indexspace

  \item Temple's inequality, 52
  \item thermodynamic limit, 29

  \indexspace

  \item vague convergence, 29

  \indexspace

  \item weak convergence, 31
  \item Wegner estimate, 43
  \item well inside, 86
  \item Weyl criterion, 22
  \item Weyl sequence, 22
  \item Wiener's Theorem, 71

\end{theindex}

\newpage
\printindex
\end{document}